\documentclass[final]{jfm} 
\usepackage{graphicx,amsmath,amssymb,subfigure,psfrag,natbib}

\DeclareMathSymbol{\R}{\mathalpha}{AMSb}{"52}
 
\providecommand\bcdot{\boldsymbol{\cdot}}

\newcommand\Rey{\mbox{\textit{Re}}}  

\newcommand\slsB{\mathsfbi{B}} 
\newcommand\slsC{\mathsfbi{C}} 
\newcommand\slsD{\mathsfbi{D}} 
\newcommand\slsF{\mathsfbi{F}} 
\newcommand\slsK{\mathsfbi{K}} 
\newcommand\slsL{\mathsfbi{L}} 
\newcommand\slsP{\mathsfbi{P}} 
\newcommand\slsI{\mathsfbi{I}} 
\newcommand\slsQ{\mathsfbi{Q}} 
\newcommand\slsT{\mathsfbi{T}} 
\newcommand\slsW{\mathsfbi{W}} 

\newsavebox{\astrutbox}
\sbox{\astrutbox}{\rule[-5pt]{0pt}{20pt}}

\newcommand\etc{etc.\ }
\newcommand\eg{e.g.\ }

\newfont{\figurefont}{cmr10 scaled 0800} 
\title[Cluster-based reduced-order modelling of a mixing layer]
{Cluster-based reduced-order modelling of a mixing layer}

\author[E.~Kaiser and friends]
{E\ls U\ls R\ls I\ls K\ls A\ns  K\ls A\ls I\ls S\ls E\ls R$^{1}$
\thanks{Author to whom correspondence should be addressed: eurika.kaiser@univ-poitiers.fr}, \ns
B\ls E\ls R\ls N\ls D\ns R.\ns N\ls O\ls A\ls C\ls K$^{1}$,\break
L\ls A\ls U\ls R\ls E\ls N\ls T\ns C\ls O\ls R\ls D\ls I\ls E\ls R$^{1}$,\ns
A\ls N\ls D\ls R\ls E\ls A\ls S\ns S\ls P\ls O\ls H\ls N$^{1}$,\break
M\ls A\ls R\ls C\ns S\ls E\ls G\ls O\ls N\ls D$^{2}$,\ns
M\ls A\ls R\ls K\ls U\ls S\ns A\ls B\ls E\ls L$^{2,3,4}$,\break
G\ls U\ls I\ls L\ls L\ls A\ls U\ls M\ls E\ns D\ls A\ls V\ls I\ls L\ls L\ls E\ls R$^5$,\ns 
J\ls A\ls N\ns \"O\ls S\ls T\ls H$^{6}$,\break
S\ls I\ls N\ls I\ls \v{S}\ls A\ns K\ls R\ls A\ls J\ls N\ls O\ls V\ls I\ls \'{C}$^{6}$,\ns
and\
R\ls O\ls B\ls E\ls R\ls T\ns K.\ns N\ls I\ls V\ls E\ls N$^7$\break}

\affiliation{
$^1$ Institut PPRIME, CNRS -- Universit\'e de Poitiers -- ENSMA, UPR 3346,
D\'epartement Fluides, Thermique, Combustion, CEAT, 43, rue de l'A\'erodrome,
F-86036 Poitiers Cedex, France\\[\affilskip]
$^2$ Ambrosys GmbH, Albert Einstein Str. 1-5, D-14469 Potsdam, Germany\\[\affilskip]
$^3$ LEMTA, 2 Av. de la For\^et de Haye, F-54518 Vandoeuvre-l\`{e}s-Nancy, France\\[\affilskip]
$^4$ Potsdam University, Institute for Physics and Astrophysics, 
Karl-Liebknecht Str. 24-25, D-14476 Potsdam, Germany\\[\affilskip]
$^5$ CERFACS, 42 Avenue Gaspard Coriolis, F-31057 Toulouse Cedex 01, France\\[\affilskip]
$^6$Division of Fluid Dynamics, Department of Applied Mechanics, 
Chalmers University of Technology, SE-412 96 G\"oteborg, Sweden\\[\affilskip]
$^7$ School of Engineering and Information Technology, The University of New South Wales at ADFA, 
Canberra, Australian Capital Territory, 2600, Australia\\[\affilskip]
}

\pubyear{???}
\volume{???}
\pagerange{???--???}
\date{\today\! and in revised form ??}
\usepackage[usenames,dvipsnames,svgnames,table]{xcolor}

\definecolor{Gray}{gray}{0.9}
\definecolor{LightCyan}{rgb}{0.88,1,1}

\begin{document}

\maketitle
\begin{abstract}
We propose a novel cluster-based 
reduced-order modelling  (CROM) strategy  for unsteady flows.
CROM combines the cluster
analysis pioneered in Gunzburger's group \citep{Burkardt2006cmame} 
and transition matrix models
introduced in fluid dynamics in Eckhardt's group \citep{Schneider2007pre}. 
CROM constitutes a potential alternative to POD models
and generalises the Ulam-Galerkin method classically used
in dynamical systems to determine a finite-rank approximation of the 
Perron-Frobenius operator.
The proposed strategy processes a time-resolved sequence of flow snapshots in two steps.
First,
the snapshot data are clustered into a small number
of representative states, called centroids, in the state space. 
These centroids
partition the state space in complementary 
non-overlapping regions (centroidal Voronoi cells).
Departing from the standard algorithm, 
the probabilities of the clusters are determined, 
and the states are sorted by analysis of the transition matrix.
Secondly, 
the transitions between the states are dynamically modelled 
using a Markov process.
Physical mechanisms are then distilled 
by a refined analysis of the Markov process,
e.g.\ using finite-time Lyapunov exponent and entropic methods.
This CROM framework is applied to
the Lorenz attractor (as illustrative example), 
to velocity fields of the spatially evolving incompressible mixing layer
and the three-dimensional turbulent wake of a bluff body.
For these examples,
CROM is shown to identify non-trivial quasi-attractors and transition processes
in an unsupervised manner.
CROM has numerous potential applications
for the systematic identification of physical mechanisms of complex dynamics,
for comparison of flow evolution models, 
for the identification of precursors to desirable and undesirable events,
and for flow control applications exploiting nonlinear actuation dynamics.
\end{abstract}

\section{Introduction}
\label{sec:intro}

Multi-scale phenomena, like turbulence, exhibit
a large number of degrees of freedom.
The dimensionality of corresponding detailed flow models
poses a challenge for understanding the physical mechanisms, 
for optimising realistic configurations and 
for controlling turbulent flows in real time,
even though computational power and resources are continuously increasing.
The general strategy of coarse-graining, system reduction or model reduction 
--- the term depends on the community ---
is to reduce the number of degrees of freedom through simplification of the used models, 
by keeping only those degrees of freedom that are relevant to 
the particular phenomena of interest, and also to the purpose of the model. 
Coarse-graining was introduced in the physics community more than a century ago, 
and is deeply connected to statistical physics and thermodynamics. 
Model reduction was more recently introduced for fluid flow analysis and control. 
A wide range of reduced-order models (ROMs) have been proposed and applied.
Here, focus is placed on a data-driven ROM. 
Model reduction typically 
involves two separate but related steps: 
i) selection of the coordinates comprising the reduced system, and 
ii) dynamical modelling of their 
interactions.   

A wide class of model reduction techniques are linked to 
the governing evolution equations via projection techniques. 
These methods rest on the assumption that in many cases
trajectories in the high-fidelity phase space are contained in low-dimensional subspaces. 
In the Petrov-Galerkin framework \citep{Antoulas_2005}, 
a low-dimensional 
subspace is first used to introduce a reduced state vector, 
and the reduced-order equations are then obtained by 
projection of the residual of the full-order model equations 
onto another low-dimensional subspace. 
Proper Orthogonal Decomposition (POD) models of fluid flows 
provide one elegant example of this class \cite[see, e.g.][]{Holmes2012book,Noack2011book}.
Kinematically, 
POD yields an ``optimal'' Galerkin expansion
which minimises the averaged residual in the energy norm 
with respect to all other expansions of the same order.
Dynamically, 
the POD coefficient space
can be described by an evolution equation, here called Galerkin system,
determined by the projection of the Navier-Stokes equation onto the POD modes, 
obtained via a model identification technique \citep{Navon2013book},
or a combination of the two approaches \citep{Cordier2013ef}.
For control purposes, 
an actuation term can be added to the Galerkin system
and exploited for control design \citep{Bergmann_Cordier_2008_JCP,Luchtenburg2009jfm}. 
Twenty-five years after the  pioneering work on wall turbulence by \citet{Aubry1988jfm}, 
POD models have become a widely used class of ROM, and numerous 
variations and generalisations have been suggested 
(weighted POD, POD with derivatives, adaptive POD, \dots). 
For the control of linear systems, one 
of the most important is balanced POD \citep{Rowley2005ijbc} 
where approximate balanced truncation \citep{Moore_1981} is computed through a POD 
approximation of the product of the controllability and observability Gramians. 
By construction, 
those ROMs are optimal in terms of input-output 
description of the system, and error bounds exist 
that are close to the lowest error possible from any reduced-order model. 
Another way of deriving a 
reduced-order model is to project a linearised version of the governing equations 
around some base flow onto the global and adjoint global modes. 
This approach was popularised for the control of flow 
instabilities observed in some aerodynamic configurations by \citet{Sipp2010amr}. 
Those linearised models are particularly appropriate for model-based control 
since many open-loop and closed-loop control strategies were 
developed years ago in the 
control theory community \citep[see][for a review]{Bagheri_Hoepffner_Schmid_Henningson_2009}.

When projection methods are not used to derive ROMs, system-identification techniques 
relying on Markov parameters and Hankel matrix \citep{Huang_Kim_POF_2008}, 
or auto-regressive models, e.g. ARMAX \citep{Herve2012jfm} are 
mostly employed. 
One exciting recent direction is based on Koopman or dynamic mode decomposition (DMD)
\citep{Rowley2009jfm,Schmid2010jfm} 
in which pure frequency modes are distilled.

In this study,
we propose a novel cluster-based reduced-order modelling (CROM) strategy for unsteady flows.
CROM combines a cluster analysis of an ensemble of observations 
and a Markov model for transitions between the different flow states educed by the clustering. 
Recently, clustering principles have also been applied to fluid dynamics problems in the group of Gunzburger.
In \citet{Burkardt2006cmame}, 
a centroidal Voronoi tessellation (CVT) approach is introduced for producing a  reduced-order model 
of the Navier-Stokes equation. Starting with a 
snapshot set as it is done in POD, 
CVT relies on the construction of a special Voronoi clustering of the snapshots 
for which the means of the clusters 
are also the generators of the corresponding Voronoi cells \citep{Du_Faber_Gunzburger_SIAM_Review_1999}. 
In the CVT setting, a dynamical model is then 
derived by projection of the governing equations onto the clusters' centroids. 
This approach can then be interpreted as a generalisation of POD-based 
reduced-order modelling. 
In the CROM framework, the dynamical model does not rest on a Galerkin projection. 
On the contrary, we follow
inspiring developments in the
group of Eckhardt \citep{Schneider2007pre}
and model the transition process between the identified clusters with 
a Markov process \citep{Norris1998book}.
This dynamical transition model proves to be simple and powerful.

Cluster analysis is an integral part of 
machine learning \citep[see, e.g.][for an introduction]{Bishop2007book}, 
a branch of artificial intelligence dedicated to the development of methods 
that can learn automatically from data.   
The starting point of clustering is --- like POD --- 
a sequence of snapshots, called {\em observations}, 
defined in some state space and a predefined distance measure.
The distance measure employs a {\em feature vector},
quantifying important characteristics of the snapshot space.
In our case, 
we define the original state space as feature
and use the standard Euclidean distance for defining similarity between observations.
The objective of clustering is to organise data so that the inner-cluster similarity is maximised while the inter-cluster similarity is minimised.
Clustering partitions a large amount of snapshots 
into a given small amount of geometrically close subsets,
called {\em clusters},
and determines a representative center of mass, named {\em centroid}.
The clustering approach identifies relatively homogeneous groups of observations 
according to the chosen metric.
For solving the cluster analysis problem, 
an efficient heuristic algorithm known as k-means is 
used here \citep{Steinhaus1956, Lloyd1982ieeetit, Ball1965tchrprt, MacQueen1967proc}. 
This algorithm  
is probably the most popular technique of clustering due to its simplicity of 
implementation and fast execution. It is used in numerous areas
ranging from market segmentation, 
computer vision, 
geostatistics, and 
astronomy 
to agriculture \citep{Murphy2012book}. 
Clustering is sometimes used as a preprocessing step for other algorithms, 
for example to find a starting configuration. 
The close relation between CVTs and k-means algorithm was already mentioned in \citet{Burkardt2006cmame}.

The CROM strategy can be interpreted 
as a generalisation of the Ulam-Galerkin method for 
the finite-rank approximation of the Perron-Frobenius operator.
Complex systems can be studied by the evolution of a single trajectory of a dynamical system
or by adopting a probabilistic viewpoint 
and considering a swarm of trajectories evolving in the state space.
The Perron-Frobenius operator is associated with such evolution equations, 
i.e. a Liouville equation, and describes 
the evolution of a probability density 
for a trajectory ensemble in the state space.
In contrast to the Ulam-Galerkin method, 
CROM rests on a cluster analysis which allows 
to code additional, e.g. entropic, information in the distance metric.

CROM is first applied to the Lorenz attractor
as a well-known introductory example (\citealp{Lorenz1963jas}; see \citealp{Sparrow1982book} for details).
The second application is a two-dimensional incompressible mixing layer flow
undergoing several successive vortex pairings.
The instability mechanism of spatially developing mixing layers 
and the spatial organisation of their dominant large-scale structures
have been the subject of many studies 
\citep[see, e.g.][]{Brown1974jfm, Browand1976jfm, Dimotakis1976jfm, Monkewitz1988jfm}.
The last example is the three-dimensional turbulent wake of a bluff body, the so-called Ahmed body.
The Ahmed body is a generic car model to study 
the relationship between characteristic flow structures
typically arising at passenger cars and their impact on the drag and lift forces
\citep{Ahmed1984, Lienhart2003sae, Pastoor2008jfm}.
Recently, \cite{Grandemange2013jfm} showed that 
the wake of a square-back body as in our example 
is characterised by a bi-modal behaviour, i.e.
the flow switches between two asymmetric flow states
over large time scales $T \sim 10^2 - 10^3 H/U_{\infty}$
for Reynolds numbers ranging from $Re \sim 10^2$ to $Re \sim 10^6$ \citep{Grandemange2012pre}.

These examples show complex dynamical behaviour 
associated with so-called 'quasi-attractors'.
We shall employ the term quasi-attractor 
as an isolated region in the state space 
with a long average residence time of the entering trajectory.
For instance, the two ears of the Lorenz attractor
can be considered as quasi-attractors since the switching 
between them is slow as compared to the oscillatory time scale.
Other considered examples of quasi-attractors 
are the harmonic and subharmonic vortex shedding regime of a mixing layer, and 
the bi-modal behaviour of the Ahmed body wake.
More formally, 
a quasi-attractor \citep[see, e.g.][]{Afraimovich1983book, Gonchenko1997cmm} 
is an attractive limit set of an attractor 
which contains nontrivial hyperbolic subsets 
and which may contain attractive periodic orbits of extremely long periods.
The modelling of such quasi-attractors is challenging
due to the complex interaction between long and short time scales.

This paper is organised as follows.
In \S~\ref{Sec:CROM} the methodology of the CROM is outlined.
Section \S~\ref{Sec:LorenzModel} 
presents a CROM for the Lorenz attractor  as an introductive example,
while in \S~\ref{Sec:MixingLayerModel} CROM distils physical mechanisms of the mixing layer.
An interpretation of CROM as a generalised Ulam-Galerkin method is  
presented in \S~\ref{Sec:UlamGalerkinMethod}.
The relative merits and challenges of CROM benchmarked against POD models 
are discussed in \S~\ref{Sec:Discussion}.
Finally, the main findings are summarised and 
future directions are given (\S~\ref{Sec:Conclusions}). 
In the appendix \ref{Sec:AppA:ExampleBroadbandTurbulenceSimpleStructure},
CROM is applied to a three-dimensional wake with broadband frequency dynamics.
A CROM-based control strategy is outlined in appendix~\ref{Sec:AppB:TowardsCROMbasedFlowControl}.
CROM's connection to the Perron-Frobenius and Koopman operators is elaborated 
in appendix~\ref{Sec:AppC:Comparison-KoopmanModes}.
A simple visualisation method for the cluster topology
as explained in appendix~\ref{Sec:AppD:Visualisation} 
provides additional insights on the CROM analysis.

\section{Cluster-based reduced-order modelling}
\label{Sec:CROM}

\begin{figure}
\psfrag{1}[cc][][1][0]{\raisebox{-8pt}{$1$}}
\psfrag{2}[cc][][1][0]{\raisebox{-8pt}{$2$}}
\psfrag{3}[cc][][1][0]{\raisebox{-8pt}{$3$}}
\psfrag{4}[cc][][1][0]{\raisebox{-8pt}{$4$}}
\psfrag{5}[cc][][1][0]{\raisebox{-8pt}{$5$}}
\psfrag{6}[cc][][1][0]{\raisebox{-8pt}{$6$}}
\psfrag{7}[cc][][1][0]{\raisebox{-8pt}{$7$}}
\psfrag{8}[cc][][1][0]{\raisebox{-8pt}{$8$}}
\psfrag{9}[cc][][1][0]{\raisebox{-8pt}{$9$}}
\psfrag{10}[cc][][1][0]{\raisebox{-8pt}{$10$}}
\psfrag{Standard operating procedure}[cc][][1][0]{Standard operating procedure}
\psfrag{Data}[cc][][1][90]{Data}
\psfrag{Statistics}[cc][][1][0]{Statistics}
\psfrag{Clustering}[cc][][1][0]{Clustering}
\psfrag{Kinematics}[cc][][1][90]{Kinematics}
\psfrag{Probability}[cc][][1][90]{Probability}
\psfrag{Cluster index}[cc][][1][0]{Cluster index}
\psfrag{11}[cc][][1][0]{$1$}
\psfrag{...}[cc][][1][0]{$\dots$}
\psfrag{K}[cc][][1][0]{$K$}
\psfrag{Temporal evolution}[cc][][1][0]{Temporal evolution}
\psfrag{Markov model}[cc][][1][0]{Markov model}
\psfrag{Dynamics}[cc][][1][90]{Dynamics}
\psfrag{Refined analysis}[cc][][1][0]{Refined analysis}
\psfrag{Attractor diameter}[cc][][1][0]{\hspace*{-1mm}Attractor diameter}
\psfrag{Cluster diameter}[cc][][1][0]{Cluster diameter}
\psfrag{Cluster standard}[cc][][1][0]{Cluster standard}
\psfrag{deviation}[cc][][1][0]{deviation}
\psfrag{Lyapunov}[cc][][1][0]{Lyapunov}
\psfrag{exponent}[cc][][1][0]{exponent}
\psfrag{Entropy of}[cc][][1][0]{Entropy of the}
\psfrag{transition matrix}[cc][][1][0]{transition matrix}
 \centering
 \includegraphics[scale = 1]{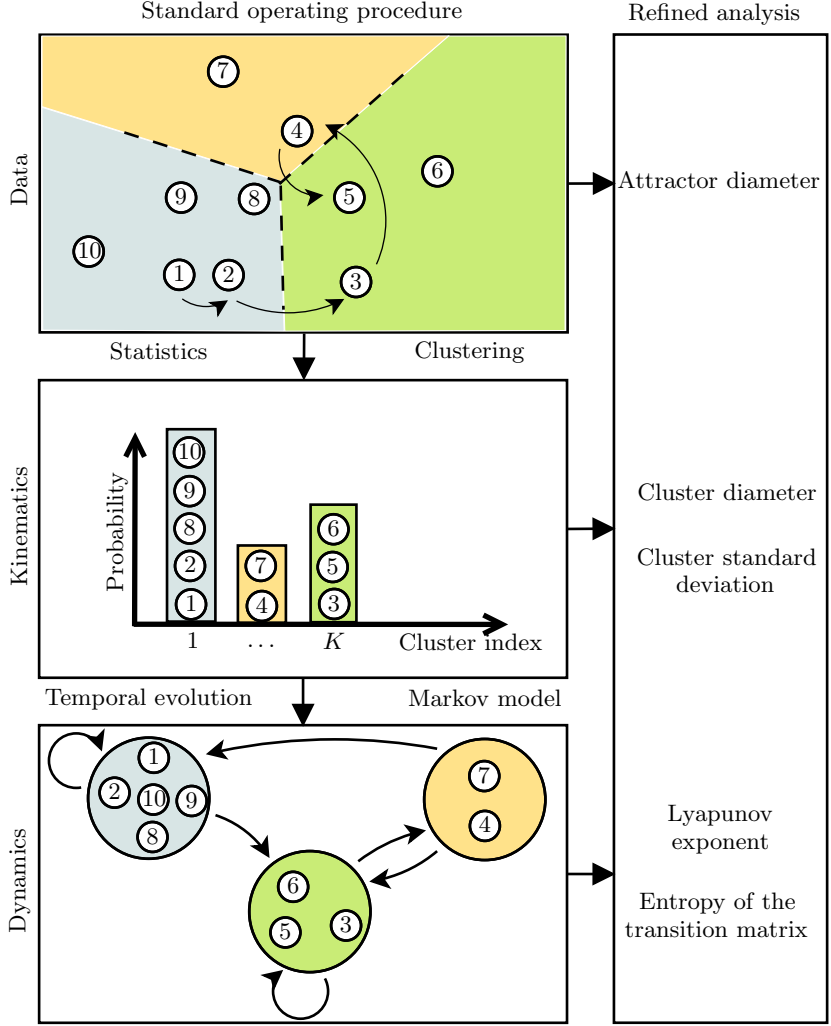}
 \caption{
 Scheme of CROM methodology applied to a sequence of observations. 
 The standard operating procedure employs:  
 (1) A partitioning of the observations into clusters of kinematically similar states (k-means algorithm). 
 (2) An estimation of the probability distribution of being in a given state is provided by a statistical analysis.
 This estimation is based on the assumption that a representative state of each cluster is supplied by its
 centroid.
 (3) An exploitation of the sequential information of the observations by assuming a Markov process for the 
 cluster transitions. This step yields to a Markov model for the evolution of the probability distribution. 
 For all steps (1)-(3), a refined analysis provides additional information on the attractor.
 For details see \S\ \protect\ref{Sec:CROM}.
 }
 \label{Fig:Figure01}
\end{figure}

In this section, 
the cluster-based reduced-order modelling (CROM) methodology is presented.
CROM combines a low-order kinematic description of the state space as obtained 
by a cluster analysis of an ensemble of snapshots (see \S~\ref{SubSec:CROM-kinematics}), 
and a dynamical model for the cluster states (see \S~\ref{SubSec:CROM-dynamics}). 
In \S~\ref{SubSec:CROM-attractor}, different criteria are defined 
which help to characterise the attractor educed by the CROM. 
These criteria will be later employed in \S~\ref{Sec:LorenzModel} and 
\S~\ref{Sec:MixingLayerModel} for the Lorenz attractor and mixing layer, respectively.
A schematic road map for the CROM is displayed in figure~\ref{Fig:Figure01}.

\subsection{Kinematic analysis}
\label{SubSec:CROM-kinematics}
CROM rests on the definition of a metric (see \S~\ref{SubSubSec:CROM-metric}) 
which measures the similarity between snapshots. 
The state space is then partitioned 
with a clustering algorithm called k-means (\S~\ref{SubSubSec:CROM-kmeans}). 
The means of the resulting clusters define  
a low-dimensional description of the state space 
and produce an optimal reduced basis in the sense that the total cluster variance is minimised.

\subsubsection{Metric}
\label{SubSubSec:CROM-metric}
We consider velocity fields in a steady domain $\Omega$.
The location is described in a Cartesian coordinate system $x$, $y$, $z$ with 
the location vector $\boldsymbol{x}=(x,y,z)$.
The velocity field $\boldsymbol{u}(\boldsymbol{x},t)$
depends on the  position $\boldsymbol{x}$ and the time $t$.
Basis for CROM is 
an ensemble of $M$ snapshots $\{\boldsymbol{u}^m(\boldsymbol{x})\}_{m=1}^M$, 
where $\boldsymbol{u}^m (\boldsymbol{x}):= \boldsymbol{u}(\boldsymbol{x},t_m)$
is the $m$-th flow realisation at discrete time $t_m$.
Thus, the employed data are similar as for reduced-order representations
via POD
\citep[see, \eg][]{Holmes2012book}
or DMD 
\citep{Rowley2009jfm,Schmid2010jfm}.
The flow may be obtained experimentally or from numerical simulations.

In principle, the snapshots may arise from transient processes 
or from a post-transient stationary state (attractor).
In the later case, the snapshots can be expected 
to approximate first and second statistical moments of the data, like for POD.
In the following, focus is placed on such attractor data,
including the applications to the Lorenz attractor and mixing-layer simulation.

For later reference, we define a Hilbert space, an inner product and an associated metric.
Let $\boldsymbol{f}$ and $\boldsymbol{g}$ be two velocity fields 
in the Hilbert space $\mathcal L^2(\Omega)$ of square-integrable vector fields in the domain $\Omega$. 
Their inner product is defined by
\begin{equation}
 \left( \boldsymbol{f},\boldsymbol{g} \right)_{\Omega} 
 := \int\limits_{\Omega}\,\mathrm d\boldsymbol{x}\,\boldsymbol{f}\bcdot\boldsymbol{g}
\end{equation}
with the corresponding norm
\begin{equation}
 \left\Vert \boldsymbol{f} \right\Vert_{\Omega}^2 
 := (\boldsymbol{f},\boldsymbol{f})_{\Omega}\,.
\end{equation}

The metric for the cluster analysis is chosen 
to be the distance between two velocity fields
according to the above norm.
In particular, the distance between 
snapshots $\boldsymbol{u}^m$ and $\boldsymbol{u}^n$
is given by
\begin{equation}
\label{Eqn:Dmn_velocity}
 \slsD_{mn}^{\scriptscriptstyle\bullet} := ||\boldsymbol{u}^m - \boldsymbol{u}^n||_{\Omega}.
\end{equation}
Here and in the following, 
the superscript '$\bullet$' refers to the raw snapshot data.

\citet{Burkardt2006cmame} showed that the k-means clustering technique leads naturally to the definition of 
basis functions that can be used to derive reduced-order models as classically done with the POD modes. 
For this reason, 
we decide to approximate the velocity fields by POD to illustrate in the results (see \S~\ref{Sec:MixingLayerModel}) 
the relation between the modes obtained by POD and by clustering. 
Moreover, 
neglecting the preprocessing costs of POD for a moment, the computational 
and input load of the cluster analysis is significantly reduced in the POD subspace 
since the evaluation of $\slsD_{mn}^{\scriptscriptstyle\bullet}$ by \eqref{Eqn:Dmn_velocity}
implies an expensive volume integration over all grid points.

POD rests on a time or ensemble average denoted by $\langle .\rangle$.
For the considered snapshot ensemble, the conventional average is employed.
POD generalises the Reynolds decomposition of a flow 
into a mean flow
$\boldsymbol{u}_0(\boldsymbol{x}) := \left\langle \boldsymbol{u}(\boldsymbol{x},t) \right\rangle$
and a fluctuation 
$\boldsymbol{v}(\boldsymbol{x},t)$.
The POD expansion \citep{Noack2003jfm, Holmes2012book} reads 
\begin{equation}
\label{Eqn:GE}
 \boldsymbol{v}(\boldsymbol{x},t) := \boldsymbol{u}(\boldsymbol{x},t) - \boldsymbol{u}_0(\boldsymbol{x}) \approx \sum\limits_{i=1}^N\, 
a_i(t)\,\boldsymbol{u}_i(\boldsymbol{x})\,,
\end{equation}
where the $\boldsymbol{u}_i$'s are the POD modes, 
and $a_i:= \left (\boldsymbol{v},\boldsymbol{u}_i \right)_{\Omega}$, 
$i=1,\,\ldots,\,N$, are the mode coefficients. 
Note that $M$ snapshots can span a space of maximum dimension $M-1$.
Hence, the number of modes in the expansion is restricted to $N \le M-1$.

The POD expansion enables the the cluster analysis of the flow snapshots
in terms of the mode coefficients.
Each snapshot is represented 
by a POD coefficient vector $\boldsymbol{a}^m = [a_1(t_m),\,a_2(t_m),\,\ldots,\,a_N(t_m)]^T$, 
where the superscript $T$ denotes the transpose. 
If the Galerkin expansion \eqref{Eqn:GE} were exact,
the distance between two snapshots would be expressed
by the Euclidean distance between the two corresponding POD coefficient vectors, i.e.
\begin{equation}
\label{Eqn:Dmn}
\slsD_{mn}^{\scriptscriptstyle\bullet} := 
  \left\Vert\boldsymbol{v}^m-\boldsymbol{v}^n\right\Vert_{\Omega}
\textcolor{black}{= \left\Vert\boldsymbol{a}^m - \boldsymbol{a}^n\right\Vert\,.}
\end{equation}
The relation \eqref{Eqn:Dmn} is also true for any orthonormal basis.
The evaluation of the right-hand side requires 
about $M$ multiplications, subtractions and additions. 
Hence, the computational savings by employing the mode coefficient metric
for the following CROM operations  are enormous.
However, the preprocessing cost of the POD 
requires the computation of $M (M+1)/2$ correlation integrals,
which have a similar computational count as the distance integrals.
For comparison,
one iteration of the clustering algorithm
will effect  $K M$ distance integrals 
between  $M$ snapshots and $K$ clusters.
The total cost of the cluster algorithm scales with $I K M$, 
$I$ being the total number of iterations until convergence.
Evidently, POD becomes computationally advantageous if $I K > (M+1)/2$.

We emphasise that the POD coefficient vector 
is only used for logistic convenience
and not necessary for the clustering algorithm.
Thus, we keep the notation of snapshots for the methodological part 
and only refer to mode coefficients when necessary.

\subsubsection{Clustering algorithm}
\label{SubSubSec:CROM-kmeans}
CROM assumes a sequence of fluctuations $\{\boldsymbol{v}^m\}_{m=1}^M$ 
 at discrete times $t_m$.
In the machine learning context \citep[see][for instance]{Hastie2009book}, 
these snapshots are also called \textit{observations} or \textit{objects}.
We wish to partition these snapshots into $K$ clusters $\mathcal{C}_k$, $k=1,\,\ldots,\,K$, 
characterised in terms of similarity of their elements. 
Here, similarity of objects is defined based on the measure $\slsD_{mn}^{\scriptscriptstyle\bullet}$ introduced in~\eqref{Eqn:Dmn}.

In clustering, each snapshot is connected to the nearest cluster by using 
a characteristic function defined as
\begin{equation}
 \slsT_{k}^{m} := \begin{cases}
                1, &  \mbox{if }\boldsymbol{v}^m\in \mathcal{C}_k,\\
                0, &  \mbox{otherwise}.
               \end{cases}
 \label{Eqn:CharacteristicFunction}
\end{equation}
The number of observations $n_k$ in cluster $k$ is given by
\begin{equation}
 n_k := \sum\limits_{m=1}^M\,\slsT_{k}^{m}.
\end{equation}
The total number of snapshots can be derived from
\begin{equation}
 M = \sum\limits_{k=1}^{K}\,n_k = \sum\limits_{k=1}^{K}\,\sum\limits_{m=1}^M\,\slsT_{k}^{m}.
\end{equation}

The cluster mean or cluster centroid $\boldsymbol{c}_k$ of $\mathcal{C}_k$ is here defined as 
the average of the snapshots belonging to the cluster.
This centroid can be conveniently expressed by the characteristic function:
\begin{equation}
\label{Eqn:DefCentroids}
\boldsymbol{c}_k
= \frac{1}{n_k} \sum\limits_{\boldsymbol{v}^m\in\mathcal{C}_k}\,\boldsymbol{v}^m
= \frac{1}{n_k} \sum\limits_{m=1}^M  \> \slsT_{k}^{m} \boldsymbol{v}^m.
\end{equation}
Thus, all snapshots have the same weight like in 
\citet{Du_Faber_Gunzburger_SIAM_Review_1999} 
and \citet{Burkardt2006siam} 
for the centroidal Voronoi tessellation approach.

The k-means algorithm partitions the data space into $K$ centroidal Voronoi cells 
which are defined as particular Voronoi cells for which the 
generating points of the Voronoi tessellation are equal to the mass centroids of 
the Voronoi regions \citep{Du_Faber_Gunzburger_SIAM_Review_1999}.
The quality of the algorithm is monitored by the total cluster variance,
which sums up the variances of all the clusters $\mathcal{C}_k$:
\begin{equation}
\label{Eqn:TotalClusterVariance}
 J \left(\boldsymbol{c}_1,\dots,\boldsymbol{c}_K \right)
= \sum\limits_{k=1}^K \,\sum\limits_{\boldsymbol{v}^m\in\mathcal{C}_k}\,
\left\Vert\boldsymbol{v}^m - \boldsymbol{c}_k \right\Vert_{\Omega}^2
= \sum\limits_{k=1}^K \,\sum\limits_{m=1}^M \, 
\slsT_{k}^{m}\,\left\Vert\boldsymbol{v}^m - 
 \boldsymbol{c}_k \right\Vert_{\Omega}^2.
\end{equation}
The optimal locations of the centroids are determined as solutions of the minimisation problem 
based on $J$, i.e.
\begin{equation}
\label{Eqn:OptimalTotalClusterVariance}
\boldsymbol{c}_1^{\rm{opt}}, \dots,\boldsymbol{c}_K^{\rm{opt}} = 
{\hbox{argmin}}_{\boldsymbol{c}_1,\dots,\boldsymbol{c}_K}  \> J \left(\boldsymbol{c}_1,\dots,\boldsymbol{c}_K \right),
\end{equation}
where $\boldsymbol{c}_k^{\rm{opt}}$ are given by \eqref{Eqn:DefCentroids}.
In this sense, the ensemble of $M$ snapshots is
optimally represented by 
the $K$ clusters and their centroids.
In the following, the centroids are assumed
to be optimal according to \eqref{Eqn:OptimalTotalClusterVariance}
and the superscript '$\rm opt$' will be dropped.

The k-means algorithm contains three steps:
First, $K$ centroids are initialised. 
Then, each observation is assigned to its closest centroid using the chosen metric. 
This is called the assignment step. 
In the update step, the cluster centres are re-calculated as the mean of 
all observations belonging to this cluster. 
The assignment step and the update step are repeated until convergence is reached. 
Convergence means that either the centroids do not move or do not change below a threshold. 
Since k-means has a sensitivity to initial conditions, 
a cluster centre initialisation using k-means++ \citep{Arthur2007proc} is here applied.

From machine learning perspective,
the clustering algorithm k-means is an unsupervised classification algorithm. 
This means that 
the observations are classified based on their similarity 
without any prior assumption or categorisation of the data. 
For a strictly periodic flow, 
this results in a generalisation of phase averaging. 
For a sufficient number of clusters $K$, each cluster centroid 
represents the phase average of the snapshots in each cluster.

Each of steps 2 and 3 of the k-means algorithm reduces the value of the criterion $J$, 
so that convergence is assured when the number of clusters $K$ is increased. 
In terms of numerical accuracy, the choice $K=M$ is then optimal since in 
this case the centroids become the snapshots and the total cluster variance vanishes. 
On the other hand, in terms of compression of the data, we would 
like to employ as few clusters as possible. 
The choice of the number of clusters 
is then a matter of compromise between numerical accuracy 
and compression of the data. 
Generally, $K$ is fixed at a value $K_c \ll M$ 
corresponding to the number of clusters for which the variation of the 
variance $J$ with $K$ becomes small.
This method is known in machine learning as the ``elbow criterion'' \citep{Hartigan1975,Tibshirani2001jrsb}. 
Recently, \cite{Chiang2010joc} compared in an extensive study different measures 
to determine a suitable number of clusters that represent hidden groups in the data.
In our study, 
we consider $K_c$ small 
to reveal the main physical transition mechanisms with a sufficient numerical accuracy.
In sections \ref{Sec:LorenzModel} and \ref{Sec:MixingLayerModel}, we adopt $K_c$ equal to $10$. 

From a system point of view, 
the cluster analysis can be seen as a method of dimension reduction.
A high-dimensional state space is discretised into $K$ centroidal Voronoi cells.
Thus, the probability distribution over the state space can be discretised as a 
probability distribution over the clusters.
Each cluster $\mathcal{C}_k$ is represented by its centroid $\boldsymbol{c}_k$,
which approximates the state of the set of snapshots in this cluster. 
For this reason, in the following sections,
we decide by convention to refer to the cluster centroids $\boldsymbol{c}_k$ as states.
\subsection{Dynamical model}
\label{SubSec:CROM-dynamics}
In this section, the cluster analysis 
is augmented with a dynamical model for the cluster states.
In \S~\ref{SubSubSec:CROM-Tident}, the cluster probability distribution 
is obtained directly from the snapshot data. 
The major step is the determination
of the cluster transition matrix (CTM) 
which serves as the propagator in terms of probability. 
This matrix is the constitutive element of a dynamical model 
elaborated in \S~\ref{SubSubSec:CROM-Tmodel}.
The derived dynamical model has the Markov property 
which assumes that the system is memoryless.
A discussion of the applicability of the Markov property
is provided in \S~\ref{SubSec:Comparison-MarkovProperty}.

\subsubsection{Identification of the cluster transition matrix}
\label{SubSubSec:CROM-Tident}

In the kinematic part of CROM (\S~\ref{SubSec:CROM-kinematics}),
the snapshots are assumed to fill the state space sufficiently densely 
for some representative statistics.
For a statistical analysis such as clustering, 
the advantage of having uncorrelated snapshots is to minimise redundancy.
For the dynamical component of CROM, the cluster transition matrix, 
we have to request a small time step.
In other words, the time step
$\Delta t = t^{m+1}-t^m$, $m=1,\ldots,M-1$
is constant and small as compared to the characteristic time scale of the flow.

The clustering algorithm yields state space
coarse-grained in centroidal Voronoi cells.
Different points of a cluster have different trajectories.
Thus, on the level of cluster resolution,
we can only make statements about 'the most probable' trajectory.
Hence, we search for a propagator 
of a probability distribution discretised by $K$ clusters.
More precisely, we will describe the dynamics 
of clusters as a Markov chain \citep{Norris1998book}. 
In this paper, the convention is made to use a left stochastic matrix (each column sum is 
equal to $1$) for the dynamical propagator and column vectors of probabilities.

Firstly, a probabilistic representation of the system, equivalent to the 
weighted average realisation of the ensemble \cite[see, \eg][]{Fowler1929book, Niven2009epjb},
can be defined by assigning a probability $q_k$ to each cluster $\mathcal{C}_k$.
Each component $q_k$ is approximated by the number of snapshots in cluster $k$
normalised with the total number of snapshots:
\begin{equation}
\label{Eqn:qk}
 q_k := \frac{n_k}{M} = \frac{1}{M} \sum\limits_{m=1}^M\, \slsT_{k}^{m}.
\end{equation}
Since every cluster contains at minimum one element ($n_k>0$), the probabilities $q_k$ 
are strictly positive ($q_k > 0$).
In addition, \eqref{Eqn:qk} fulfils the normalisation condition $\sum_{k=1}^K q_k=1$.
These probabilities are summarised
in the cluster probability vector $\boldsymbol{q} = \left[q_1,\,\ldots,\,q_K \right]^T$.
This vector shall approximate the fixed point of the dynamical model developed below.

In the following, 
the temporal evolution of a general cluster probability vector $\boldsymbol{p}$ is pursued.
For computational convenience, 
we describe the evolution with the same time step $\Delta t$ as the original data,
i.e.\ pursue a time-discrete formulation. 
The transition to a continuous-time formalism is described below.

The cluster transition matrix (CTM) $\slsP$ is identified as the 
one-step transition probability matrix.
Let $\slsP_{jk}$ be the probability of moving from cluster $\mathcal{C}_k$ to cluster 
$\mathcal{C}_j$ in one forward time-step, the elements of the resulting CTM can be inferred by
\begin{equation}
 \slsP_{jk} = \frac{n_{jk}}{n_k},
 \label{Eqn:TPM-definition}
\end{equation}
where $n_{jk}$, the number of snapshots 
that move from $\mathcal{C}_k$ to $\mathcal{C}_j$, are given by
\begin{equation}
  n_{jk} := \sum\limits_{m=1}^{M-1}\,\slsT_{k}^{m}\,\slsT_{j}^{m+1}\,.
\end{equation}
Note that $M$ consecutive snapshots define only $M-1$ transitions.

All elements of this matrix are non-negative,
\begin{equation}
\label{Eqn:PjkNonNegative}
\slsP_{jk} \ge 0, \quad j,k = 1,\ldots,K.
\end{equation}
The elements of each column sum up to unity 
\begin{equation}
\label{Eqn:PjkColumnSum}
\sum_{j=1}^K\,\slsP_{jk} = 1
\end{equation}
as the total probability for all transitions is one.
This property is necessary to preserve the normalisation condition of the probability vector.

The diagonal element $\slsP_{jj}$ gives 
probability to stay in the cluster $\mathcal{C}_j$ during one time step.  
In order to extract the most probable path, 
the clusters are ordered so that the first cluster has the highest state probability $q_1$. 
The second cluster is chosen that the highest transition 
probability is to move from state $\boldsymbol{c}_1$ to state $\boldsymbol{c}_2$, \etc 
If one cluster has already been utilised, 
the cluster with the second highest transition probability is used, \etc
The eigenvalues of the CTM are sorted by descending moduli, 
$\left|\lambda_1\right|\geq\left|\lambda_2\right|\geq\ldots\geq\left|\lambda_K\right|$.

\subsubsection{Cluster transition matrix model}
\label{SubSubSec:CROM-Tmodel}

The above  cluster transition analysis leads naturally 
to Markov chains.
A Markov chain is characterised by the Markov property
that the probability to move 
from the current state $\boldsymbol{c}_k$ at time step $l$ 
to another state $\boldsymbol{c}_j$ in the next time step $l+1$ 
depends only on the current state and not on past values.
The autonomous nature of the Navier-Stokes equation 
suggests a similar time-independent property 
for the propagator $\slsP$ of the cluster probability vector.
For sufficiently smooth initial conditions and steady boundary conditions, 
the uniqueness of the solution can be proven for two-dimensional flows \citep{Ladyzhenskaya1963book},
and, in general, can be assumed for most three-dimensional numerical simulations.
However, the solution of the Navier-Stokes equation may depend on the history 
if, e.g., the boundary conditions are discontinuous as shown by \cite{Rempfer2006amr}.
We refer to \S~\ref{SubSec:Comparison-MarkovProperty} for a more detailed discussion
of the Markov asumption.

The cluster probability vector with probabilities $p_k^l$ 
to be in state $\boldsymbol{c}_k$ at time $l$ is denoted by
\begin{equation}
\label{Eqn:ProbL}
 \boldsymbol{p}^{l} = \left[ p_1^l, \ldots, p_K^l\right]^T.
\end{equation}
This vector respects that only non-negative probabilities 
\begin{equation}
\label{Eqn:ProbLNonNegative} 
p_k^l\geq 0
\end{equation} 
have a meaning and the normalisation condition 
\begin{equation}
\label{Eqn:ProbLSum}
\sum_{k=1}^K \,p_k^l = 1
\end{equation} 
holds for each time step $l$.
The conditional probability to be in state $\boldsymbol{c}_j$ 
if the previous state has been $\boldsymbol{c}_k$ is 
given by the CTM defined in \eqref{Eqn:TPM-definition}.

The evolution of the cluster probability vector can be described as follows.
Let $\boldsymbol{p}^0$ be the initial probability distribution.
Then, consecutive distributions are described by the iteration formula
\begin{equation}
\label{Eqn:ProbIteration}
 \boldsymbol{p}^{l+1} = \slsP\,\boldsymbol{p}^l\,, \quad  l=0,1,2, \ldots
\end{equation}
The CTM properties \eqref{Eqn:PjkNonNegative} and \eqref{Eqn:PjkColumnSum}
guarantee properties \eqref{Eqn:ProbLNonNegative} and \eqref{Eqn:ProbLSum}
of the cluster probability vector.
The cluster probability vector at time $l$ is compactly given by
\begin{equation}
\label{Eqn:ProbL0}
 \boldsymbol{p}^l = \slsP^l\,\boldsymbol{p}^0.
\end{equation}

The propagator $\slsP$ defines a time-homogeneous Markov chain
with well-known properties \citep{Meyer2000book}: 
\begin{enumerate}
\item[(1)]$\slsP$ is a stochastic matrix with the properties \eqref{Eqn:PjkNonNegative} and \eqref{Eqn:PjkColumnSum}.
\item[(2)]The sequence of probability vectors $\boldsymbol{p}^{l}$, $l=0,1,2,\ldots$  has no long-term memory.
The state at iteration $l+1$ only depends on the $l$-th state and not on any previous iterations.
\item[(3)]The absolute values of all eigenvalues of this matrix do not exceed unity. 
This excludes a diverging vector sequence.
\item[(4)] 
It exists an eigenvalue $\lambda_1(\slsP) = 1$ with algebraic multiplicity $1$
and all other eigenvalues satisfy $\vert \lambda_k (\slsP)\vert < 1$ for $k=2,\ldots,K$.
This is a consequence of the Perron-Frobenius theory for nonnegative matrices \citep{Meyer2000book}.
The eigenvector $\boldsymbol{p}^{ev}_1$ associated with the dominant eigenvalue $\lambda_1$ 
fulfils $\slsP \> \boldsymbol{p}^{ev}_1 = \boldsymbol{p}^{ev}_1$.
Since $\vert \lambda_k(\slsP)\vert< 1$ for $k=2,\ldots,K$,
the vector $\boldsymbol{p}^{ev}_1$ is the only one that survives infinite iterations.
\end{enumerate}

The long-term behaviour can be studied by powers of the CTM 
as defined in \eqref{Eqn:ProbL0}.
The asymptotic probability distribution is obtained by
\begin{equation}
\label{Eqn:ProbInfinity}
 \boldsymbol{p}^{\infty}:= \lim\limits_{l\rightarrow \infty}\,\slsP^l\,\boldsymbol{p}^0
\end{equation}
and the infinite-time CTM $\slsP^{\infty}$ is defined as
\begin{equation}
\label{Eqn:CTMInfinity}
  \slsP^{\infty}:=\lim\limits_{l\rightarrow\infty}\,\slsP^l\,.
\end{equation}
If $\slsP^l$ converges to a unique, stationary matrix $\slsP^{\infty}$, 
the system can be said to be {\it ergodic}, in the sense 
that it will be probabilistically reproducible: 
regardless of the initial region of state space in which it is sampled, 
the ensemble mean will converge in the infinite-time limit to the time mean.  
Mathematically, the stationary probability vector must be identical 
with 
the eigenvector $\boldsymbol{p}^{ev}_1$ associated with the dominant eigenvalue $\lambda_1=1$, 
and thus is a fixed 
point to \eqref{Eqn:ProbL0} for any $l$ 
and with respect to iteration with \eqref{Eqn:CTMInfinity}. 
If, however, $\slsP^{\infty}$ is oscillatory or nonstationary, 
the system will not be probabilistically reproducible, displaying a more complicated 
connection between the initial sampling region and its convergence properties.
The asymptotic rate of convergence at which $\slsP^l$ approaches $\slsP^{\infty}$ is given by its second largest eigenvalue modulus
$\vert\lambda_2\vert$. This is similar to the power method where the rate of convergence is determined by the ratio 
$\vert\lambda_2/\lambda_1\vert^l$ and how fast it approaches zero \citep{Meyer2000book}.

Iterating $P$ amounts to determining the ``most significant'' eigenvector. This procedure is used in many fields, 
the most prominent example being the PageRank algorithm used by the Google search engine \citep{Meyer2012book}.
This reveals the underlying system dynamics, in the sense that a cluster is not simply important by itself,
but rather by the clusters which transition to it. This provides a powerful approach to determine the
\textit{dynamically} important clusters (or structures).

The accuracy of the CTM can be assessed
when it is benchmarked 
against the continuous-time form 
of the evolution equation \eqref{Eqn:ProbIteration}:
\begin{equation}
\label{Eqn:ProbL0Continuously}
\frac{\mathrm d \boldsymbol{p}(t) }{\mathrm dt} 
= \slsP^{\mathrm cont} \boldsymbol{p}(t).
\end{equation}
The discrete-time Markov model \eqref{Eqn:ProbIteration}
can be derived from \eqref{Eqn:ProbL0Continuously}
by an analytical integration over the time $\Delta t$.
The resulting discrete-time transition matrix reads
\begin{equation}
\label{Eqn:CTMContinuously}
\slsP = \exp \left( \slsP^{\mathrm cont} \Delta t\right)   
\quad
\Rightarrow
\quad 
\slsP^{\mathrm cont}   
= \frac{1}{\Delta t} \log \slsP 
\approx \frac{ \slsP - \slsI}{\Delta t}
+ O(\Delta t),
\end{equation}
where $\slsI$ is the identity matrix.
The approximate term on the right-hand side
coincides with the Euler integration scheme
for \eqref{Eqn:ProbL0Continuously}.
Note that the transition matrix $\slsP$ is 
an approximation of the Perron-Frobenius operator
for discrete states and discrete times \citep{Lasota1994book,Cvitanovic2012book}.
We refer to \S~\ref{Sec:UlamGalerkinMethod} for more details.

The reverse operation, 
i.e.\ the derivaton of the continuous-time Markov model 
\eqref{Eqn:ProbL0Continuously}
from the discrete-time version \eqref{Eqn:ProbIteration}
requires a carefully chosen limiting process $\Delta t \to 0$.
Let us  consider a continuous-time trajectory 
over the time interval $[0,T]$.
Let us further assume that the state space
is partitioned into $K$ fixed clusters.
From this trajectory, 
$M$ snapshots are taken at constant sampling rate, 
i.e.\ with time step $\Delta t = T / M$.
Each of these snapshots 
can be attributed to one cluster $k=1,\ldots,K$
and the discrete-time transition matrix $\slsP_M$
can be computed via \eqref{Eqn:TPM-definition} as function of $M$.
As $M \to \infty$ or, equivalently, $\Delta t \to 0$, $(1/\Delta t) \> \log \slsP_M \to \slsP^{\mathrm cont}$.
From $\slsP_{jk}=n_{jk}/n_k$,
the possible values of a transition matrix element 
are discrete as $n_k$ and $n_{jk}$ are integers.
The maximum resolution is controlled by the denominator $1/n_k \ge 1/M$.
Thus, increasing $M$ improves the maximum accuracy of $\slsP_M$. 
The time horizon $T$ for the trajectory 
also plays  an important role.
The number of non-trivial cluster transitions scales with $T$ 
and are well defined in the limit  $\Delta t \to 0$.
Thus, the accuracy of the continuous-time transition matrix
also increases with the time horizon.

\subsubsection{Validity of the Markov model}
\label{SubSec:Comparison-MarkovProperty}
The assumed Markov property of the model requires a plausibility argument:
The Markov property assumes that a system is memoryless, i.e.\
future states of the system 
are solely determined by the current state 
and do not depend on the previous history of the system.
Counter-examples are differential equations 
with history terms (time integral over past states),
like for some non-Newtonian fluids.
Also, the solution of the incompressible Navier-Stokes equation
may indeed depend on the history, 
particularly for time-discontinuous boundary conditions \citep{Rempfer2006amr}.

In the following, we consider
a steady domain with steady boundary conditions
and additional mathematical smoothness properties of the domain and the initial condition.
Under such premises, 
the uniqueness of the Navier-Stokes solution can be proven for two-dimensional flows \citep{Ladyzhenskaya1963book}.
For three-dimensional flows, 
the uniqueness is generally assumed in an adequate numerical Navier-Stokes discretisation,
i.e.\ the computed numerical solution is expected to be essentially unique 
up to the prediction horizon of the initial boundary value problem.
The resulting grid-based or spectral methods yield
an autonomous dynamical system in a finite-dimensional state space.
Traditional Galerkin methods \citep{Fletcher1984book} --- with POD models as a subset --- yield the same type of dynamical systems.
In all these spatial Navier-Stokes discretisations, 
the corresponding evolution of the probability density 
in the state space is described by a Liouville equation.
As will be shown in \S~\ref{Sec:UlamGalerkinMethod}, 
the state space discretisation of the Liouville equation \eqref{Eqn:LiouvilleEquation}
leads naturally to a continuous-time autonomous Markov model \eqref{eq:ulam_galerkin}
with the Markov property.

Under similar premises, smoothly varying boundary conditions
can be expected to lead to a time-dependent Markov model.
We refer to \cite{Vishik1988book} and \cite{Hopf1952jrma} for a discussion of the Liouville equation
for the (undiscretised) Navier-Stokes equation.

\subsection{Attractor characterisation}
\label{SubSec:CROM-attractor}
In the following sections, the CROM is used to educe
geometric and dynamic properties of the flow dynamics 
for the Lorenz attractor (\S~\ref{Sec:LorenzModel}), 
a mixing layer (\S~\ref{Sec:MixingLayerModel}),
and an Ahmed body (appendix~\ref{Sec:AppA:ExampleBroadbandTurbulenceSimpleStructure}), 
respectively.
The properties are quantified by parameters
introduced in this section. 

First, the size of the attractor is represented by the snapshot data
and the size of the clusters is quantified.
The diameter of the attractor represented by the snapshot ensemble is given by
\begin{equation}
\label{Eqn:DiameterAttractor}
 D_A^{\scriptscriptstyle\bullet} := 
\sup_{m,n}\left\{\slsD_{mn}^{\scriptscriptstyle\bullet}\,:\,\boldsymbol{v}^m,\boldsymbol{v}^n 
\in 
\mathrm{attractor}\right\},
\end{equation}
where $\slsD_{mn}^{\scriptscriptstyle\bullet}$ is defined by \eqref{Eqn:Dmn}.
$D_A^{\scriptscriptstyle\bullet}$ constitutes an upper bound 
and reference level to all considered distances.

Analogously, we introduce the diameter of the cluster $\mathcal{C}_k$ as
\begin{equation}
\label{Eqn:DiameterCluster}
 D_k := \sup_{m,n}\left\{\slsD_{mn}^{\scriptscriptstyle\bullet}\,:\,\boldsymbol{v}^m,\boldsymbol{v}^n \in 
\mathcal{C}_k\right\}\,.
\end{equation}
Similar diameters for all clusters indicate a relatively homogeneous partition of the state space.

The standard deviation $R_k$ of the cluster $\mathcal{C}_k$ is given by
\begin{equation}
 \label{Eqn:StandardDeviationCluster}
 R_k =
\sqrt{
\frac{1}{n_k}
\sum\limits_{\boldsymbol{v}^m\in\mathcal{C}_k}
\left\Vert \boldsymbol{v}^m-\boldsymbol{c}_k\right\Vert_{\Omega}^2}
=
 \sqrt{
 \frac{1}{n_k}
 \sum\limits_{m=1}^M\, \slsT_{k}^{m} 
\left\Vert \boldsymbol{v}^m-\boldsymbol{c}_k\right\Vert_{\Omega}^2}.
\end{equation}
This quantity allows to assess the homogeneity of the clusters.

In analogy to the snapshot distance matrix,
we introduce the cluster distance matrix $\slsD=\left( \slsD_{jk} \right)$, 
which incorporates the distances between the cluster centroids: 
\begin{equation}
\label{Eqn:ClusterDistanceMatrix}
 \slsD_{jk} :=\left\Vert\boldsymbol{c}_j-\boldsymbol{c}_k\right\Vert_{\Omega}\,
\quad j,k=1,\ldots,K.
\end{equation}
The cluster distance matrix allows, for instance, 
to estimate the trajectory length for cluster transitions,
\eg $\slsD_{jk}$ for the direct transition from cluster $k$ to $j$.

In addition, 
a characteristic flow norm variance  
represented by the cluster probability distribution 
$\boldsymbol{p}=\left[p_1, \ldots, p_K \right]^T $
is naturally defined by
\begin{equation}
\label{Eqn:Variance}
 R^2(\boldsymbol{p}) := \sum\limits_{j=1}^K\,\sum\limits_{k=1}^K\,p_j\,p_k\,\slsD_{jk}^2.
\end{equation}

The mixing property of the propagator $\slsP$
is characterised by the finite-time Lyapunov exponent (FTLE)
\begin{equation}
\label{Eqn:FTLE}
 \lambda^l_{FTLE}
 = \frac{1}{l}\,\ln 
\left\Vert\frac{\slsP^l\,\delta\boldsymbol{p}^0}{\delta\boldsymbol{p}^0}\right\Vert .
\end{equation}
This exponent measures the divergence of the distance 
$\delta\boldsymbol{p}^l = \boldsymbol{p}^l_a-\boldsymbol{p}^l_b$ 
of initially close probability distributions
$\boldsymbol{p}^0_a$ and $\boldsymbol{p}^0_b$ via \eqref{Eqn:ProbL0}.
The dependence of the FTLE on the particular initial condition
decreases rapidly for increasing $l$.
Eventually, all trajectories converge to the fixed point \eqref{Eqn:ProbInfinity}
and the Lyapunov exponent vanishes 
\begin{equation}
\label{Eqn:LyapunovExponent}
 \lambda_{FTLE} = \lim\limits_{l\to \infty} \lambda^l_{FTLE} = 0.
\end{equation}

Finally,
the transition matrix of the propagator is assessed.
The infinite-time propagator \eqref{Eqn:CTMInfinity} 
with stable asymptotic probability distribution \eqref{Eqn:ProbInfinity} 
can be compared to a transition matrix $\slsQ=(\slsQ_{jk})$ 
constructed from the cluster analysis, with elements
\begin{equation}
\label{Eqn:Qjk}
 \slsQ_{jk} = q_k,\,\quad j,k=1,\ldots,\,K\,.
\end{equation}
Note that $\slsP^{\infty}$ is obtained from a dynamically asymptotic realisation (mode) and
$\slsQ$ from a weighted average realisation (mean).
Hence, one cannot expect them to be identical.
However, in an ergodic system in the limit of a very large number of samples, the two matrices should approach a common value. 
The difference between the two
can be quantified by the 
Kullback-Leibler entropy \citep{Kullback1951ams, Kullback1959book, Noack2012jfm} 
\begin{equation}
\label{Eqn:KullbackLeiblerEntropy}
  H(\slsP,\slsQ)\,
  = - \mathfrak{D}(\slsP,\slsQ)
  := -\sum\limits_{j=1}^K \sum\limits_{k=1}^K \slsP_{jk} \ln 
\frac{\slsP_{jk}}{\slsQ_{jk}} .
\end{equation}
Alternatively, 
the divergence $ \mathfrak{D}$ is given by the negative Kullback-Leibler entropy.
For $\slsP=\slsQ$, the divergence vanishes.
For $\slsP \not = \slsQ$, the divergence assumes positive values.
This entropy measures the uncertainty of $\slsP_{jk}$ relative to $\slsQ_{jk}$ 
as prior (least-informative) expectation, while the divergence can be considered as the information gained 
from the quantification of the finite-time transitions.
More importantly, the entropy provides a measure of the system ergodicity, reflected by convergence towards $H=0$ in the infinite sampling limit 
$\slsP
\to \slsQ$.


\psfrag{a1}[cc][][1][-90]{$a_1$}
\psfrag{a2}[cc][][1][-90]{$a_2$}
\psfrag{a3}[cc][][1][-90]{$a_3$}
\psfrag{a4}[cc][][1][-90]{$a_4$}
\psfrag{ttime}[cc][][1][0]{$t$}
\psfrag{10}[cc][][1][0]{$10$}
\psfrag{0}[cc][][1][0]{$0$}
\psfrag{-10}[cc][][1][0]{$-10$}
\psfrag{400}[cc][][1][0]{$400$}
\psfrag{800}[cc][][1][0]{$800$}
\psfrag{1200}[cc][][1][0]{$1200$}
\psfrag{1600}[cc][][1][0]{$1600$}
\psfrag{2000}[cc][][1][0]{$2000$}
\psfrag{ck}[cc][][1][-90]{$k$}
\psfrag{2}[cc][][1][0]{$2$}
\psfrag{4}[cc][][1][0]{$4$}
\psfrag{6}[cc][][1][0]{$6$}
\psfrag{8}[cc][][1][0]{$8$}

\psfrag{kidx}[cc][][1][0]{$k$}
\psfrag{1}[cc][][1][0]{$1$}
\psfrag{2}[cc][][1][0]{$2$}
\psfrag{3}[cc][][1][0]{$3$}
\psfrag{4}[cc][][1][0]{$4$}
\psfrag{5}[cc][][1][0]{$5$}
\psfrag{6}[cc][][1][0]{$6$}
\psfrag{7}[cc][][1][0]{$7$}
\psfrag{8}[cc][][1][0]{$8$}
\psfrag{9}[cc][][1][0]{$9$}
\psfrag{10}[cc][][1][0]{$10$}

\psfrag{Djk}[cc][][1][-90]{$\slsD_{jk}$}
\psfrag{Pjk}[cc][][1][-90]{$\slsP_{jk}$}
\psfrag{0.01}[cc][][1][0]{$0.01$}
\psfrag{0.1}[cc][][1][0]{$0.1$}
\psfrag{1}[cc][][1][0]{$1$}
\psfrag{5.73}[cc][][1][0]{$5.73$}
\psfrag{11.45}[cc][][1][0]{$11.45$}
\psfrag{17.18}[cc][][1][0]{$17.18$}

\psfrag{ltime}[cc][][1][0]{$l$}
\psfrag{-e2}[cc][][1][0]{$-10^2$}
\psfrag{-e-2}[cc][][1][0]{$-10^{-2}$}
\psfrag{-e0}[cc][][1][0]{$-10^0$}
\psfrag{-e5}[cc][][1][0]{$-10^5$}
\psfrag{-e-5}[cc][][1][0]{$-10^{-5}$}
\psfrag{e0}[cc][][1][0]{$10^0$}
\psfrag{e1}[cc][][1][0]{$10^1$}
\psfrag{e2}[cc][][1][0]{$10^2$}
\psfrag{e3}[cc][][1][0]{$10^3$}
\psfrag{e4}[cc][][1][0]{$10^4$}

\psfrag{Imlambda}[cc][][1][-90]{$\Im(\lambda)$}
\psfrag{Relambda}[cc][][1][0]{$\Re(\lambda)$}
\psfrag{p1k-pinfk}[cc][][1][0]{$p^{ev}_{1,k}\,p^{\infty}_k$}
\psfrag{-1}[cc][][1][0]{$-1$}
 
\psfrag{Dk}[cc][][1][-90]{$D_k$}
\psfrag{diameter}[cc][][1][-90]{$D_k$}
\psfrag{std}[cc][][1][-90]{$R_k$}
\psfrag{Var}[cc][][1][-90]{$R^2(\boldsymbol{p})$}
\psfrag{HPQ}[cc][][1][-90]{$H(\slsP^l,\slsQ)\qquad\quad$}
\psfrag{lambda1}[cr][][1][-90]{$\,\lambda^l_{FTLE}$}

\section{Lorenz attractor as illustrating example}
\label{Sec:LorenzModel}

We illustrate CROM for a nonlinear system of ODEs, the Lorenz equations, 
\begin{subequations}
  \begin{align}
    \frac{\mathrm d\,x}{\mathrm dt} & = \sigma \, \left( y - x \right)\,,\\
    \frac{\mathrm d\,y}{\mathrm dt} & = x\,\left(r-z\right) - y \,,\\
    \frac{\mathrm d\,z}{\mathrm dt} & = x\,y - b\,z\,,
  \end{align}
\end{subequations}
with the system parameters $\sigma=10$, $b = 8/3$, and $r = 28$. 
These equations represent a model for Rayleigh-B\'enard convection as proposed by \citet{Lorenz1963jas}. 
For these values of the parameters, the attractor possesses three unstable 
fixed points at $(0,0,0)$ and $(\pm \sqrt{72},\,\pm \sqrt{72},\,27)$. 
The trajectory is most of the time in the neighbourhood of
the weakly unstable oscillatory fixed points 
$(\pm \sqrt{72},\,\pm \sqrt{72},\,27)$.
These neighbourhoods are loosely referred to as 'ears' of the Lorenz attractor.
Starting from an initial state close to one of the non-zero fixed points, 
the trajectory oscillates for several periods around this point
before it moves to the other ear.
Here, it oscillates in a similar manner before it transitions back again. 
The Lorenz system is solved using an explicit fourth-order Runge-Kutta formula 
with an initial condition chosen to be on the attractor. 
The sampling time step is $\Delta t = 0.005$. 
The cluster analysis is performed on snapshot data $\left\{\boldsymbol{x}\left(t_m\right)\right\}_{m=1}^M$ with 
$\boldsymbol{x}=[x,y,z]^T$ for $M= 100\,000$ time units. 

\begin{figure}
\psfrag{xL}[cc][][1][0]{$x$}
\psfrag{yL}[cc][][1][0]{$y$}
\psfrag{zL}[cc][][1][-90]{$z$}
\psfrag{-20}[cc][][1][0]{$-20$}
\psfrag{20}[cc][][1][0]{$20$}
\psfrag{40}[cc][][1][0]{$40$}
 \centering
 \includegraphics[scale = 1]{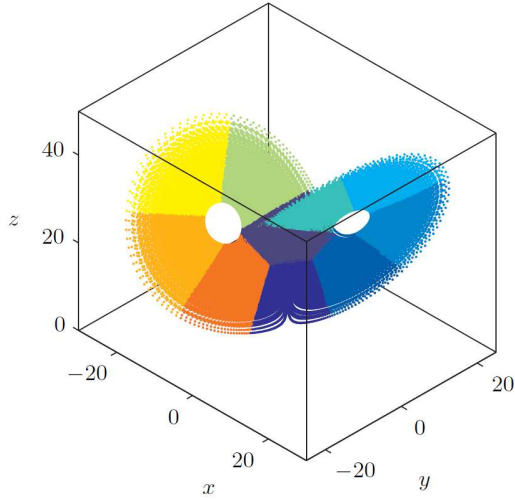}
  \caption{Phase portrait of the clustered Lorenz attractor for $K_c=10$ clusters.
Points with the same colour belong to the same cluster.
Note that the 'ears' of the Lorenz attractor are partitioned into clean sectors. }
  \label{Fig:Figure02}
\end{figure}

Figure~\ref{Fig:Figure02} displays the phase portrait of the Lorenz attractor and the segmentation 
according to the clustering.
We chose $K_c=10$ clusters. 
This value is in good agreement with the ``elbow criterion'' (see \S~\ref{SubSubSec:CROM-kmeans}).
This number is large enough to resolve the main transition mechanism
and small enough to obtain a simple structure for the model.
A qualitative discussion 
on the choice of the number of clusters, 
its lower and upper bounds  
as well as its influence 
on the state and transition resolution
is provided at the end of \S~\ref{Sec:MixingLayerModel}.
Each colour in the figure represents one cluster.
Note that the sawtooth at the outer edge of each cluster is a visualisation artefact.
The cluster analysis divides the attractor into clean sectors around the two repelling fixed points. 
The centroids of the clusters associated with the fixed points
can be considered as phase average with very coarse phase bins.
The distance measure for the clustering algorithm is directly based on the Euclidean norm $\|\cdot\|$ as defined 
in \eqref{Eqn:Dmn} for the POD coefficients.

\begin{figure}
\psfrag{cluster index}[cc][][1][-90]{$k$}
\psfrag{snapshot}[cc][][1][0]{$t$}
\psfrag{12}[cc][][1][0]{$12$}
\psfrag{a1}[cc][][1][-90]{$x$}
\psfrag{a2}[cc][][1][-90]{$y$}
\psfrag{a3}[cc][][1][-90]{$z$}
\psfrag{snapshot}[cc][][1][0]{$t$}
\psfrag{-20}[cc][][1][0]{$-20$}
\psfrag{20}[cc][][1][0]{$20$}
\psfrag{40}[cc][][1][0]{$40$}
\psfrag{15}[cc][][1][0]{$15$}
\psfrag{25}[cc][][1][0]{$25$}
\psfrag{-40}[cc][][1][0]{$-40$}
 \centering
 \centerline{\raisebox{0.14\textheight}{(a)}\hspace{5mm}\includegraphics[scale = 1, trim = 0mm 8mm 1mm 0mm, clip = true]{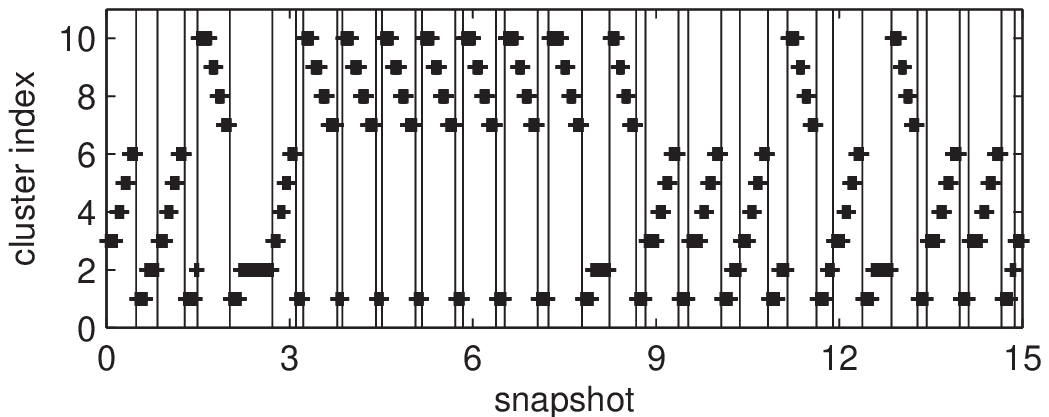}}
 \vfill
 \centerline{\raisebox{0.14\textheight}{(b)}\hspace{3.5mm}\includegraphics[scale = 1, trim = 0mm 8mm 1mm 0mm, clip = true]{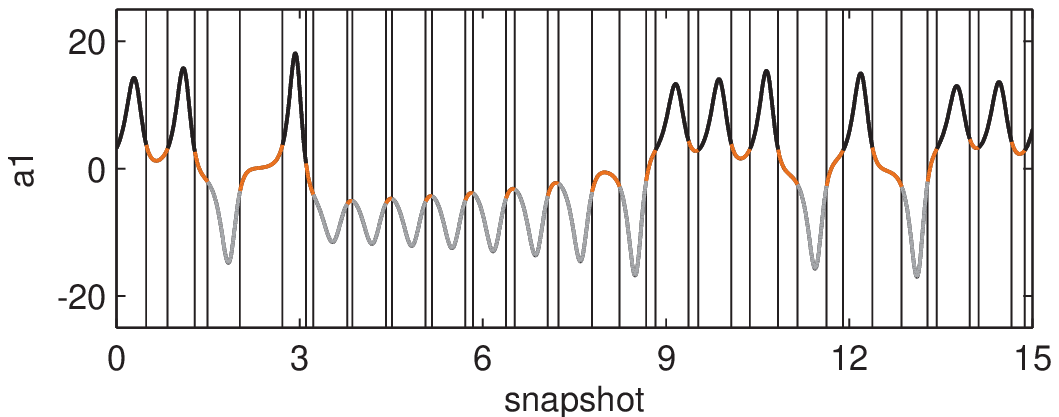}}
 \vfill
 \centerline{\raisebox{0.14\textheight}{(c)}\hspace{3.5mm}\includegraphics[scale = 1, trim = 0mm 8mm 1mm 0mm, clip = true]{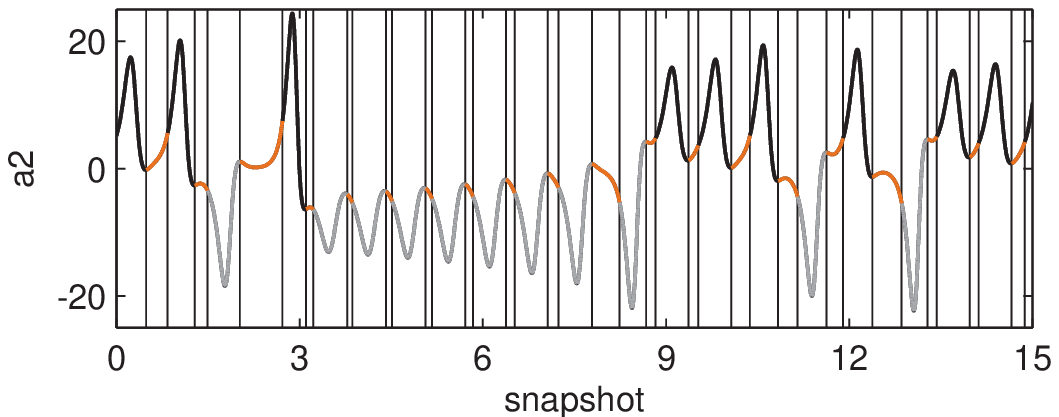}}
 \vfill
 \centerline{\raisebox{0.18\textheight}{(d)}\hspace{5mm}\includegraphics[scale = 1]{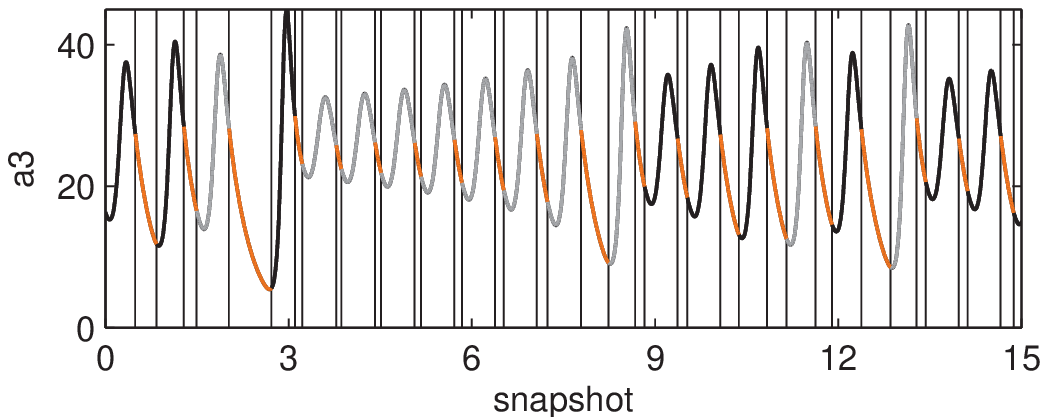}}
 \caption{Temporal cluster analysis of the Lorenz attractor.
The upper figure (a) marks the cluster $k$ to which each snapshot belongs.
Figures\,(b-d) show
the time series for all three variables $x$, $y$, and $z$ 
(top to bottom).
The vertical lines separate between the branching clusters $T$ ($k=1,2$), 
and the two ears (group $E^+$ with clusters $k=3,4,5,6$ and group $E^-$ with $k=7,8,9,10$).
In the time series plots, $T$ is colour marked in orange, $E^+$ in black, and $E^-$ in grey.
}
 \label{Fig:Figure03}
\end{figure}

The evolution of the variables 
is displayed in figure~\ref{Fig:Figure03}\,(b-d). 
The colour marking in the time series plots corresponds to different groups of clusters
($T$ in orange, $E^+$ in black, and $E^-$ in grey) 
which are explained below. Note that these do not represent each cluster as in figure~\ref{Fig:Figure02}.
The first two variables, $x$ and $y$, clearly show the location of the trajectory on the attractor. 
Extrema for positive $x$ and $y$ mark one ear,
and for negative values the other ear.
In figure~\ref{Fig:Figure03}\,(a), the cluster assignment of each snapshot is visualised.
It reveals three groups of clusters, namely clusters $k=1,2$ (group '$T$'), 
$k=3,4,5,6$ (group '$E^+$'), and $k=7,8,9,10$ (group '$E^-$').
Within a group, the transitions are from $k$ to $k+1$ for $T$ and $E^+$, and 
from $k$ to $k-1$ for $E^-$.
The two ears around the fixed points are partitioned 
in two different groups of clusters, $E^+$ and $E^-$.
After each circulation ($E^+$ or $E^-$), the trajectory returns to group $T$. 
Group $T$ serves as a branching region between the two ears.
The colour marking confirms this conclusion.
These visualisations of the clustering  provide revealing insights on the nature of the attractor  
without assuming any prior knowledge of it. 

\begin{figure}
\psfrag{13.39}[cc][][1][0]{$13.4$}
\psfrag{26.79}[cc][][1][0]{$26.8$}
\psfrag{40.18}[cc][][1][0]{$40.2$}
 \centering
\includegraphics[scale = 1, trim = 0cm 0.15cm 0cm 0cm, clip = true]{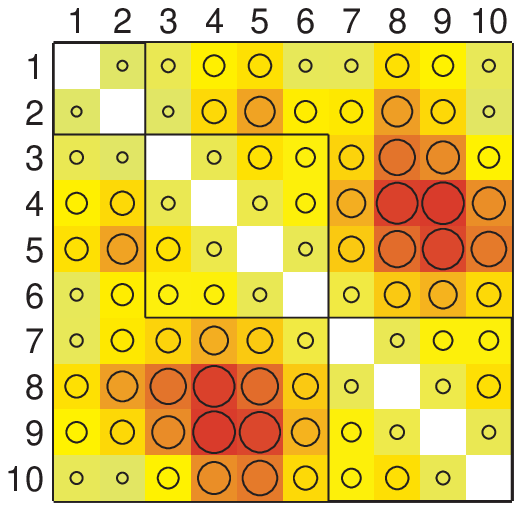}
 \hspace*{20pt}
\includegraphics[scale = 1, trim = 0cm 0.15cm 0cm 0cm, clip = true]{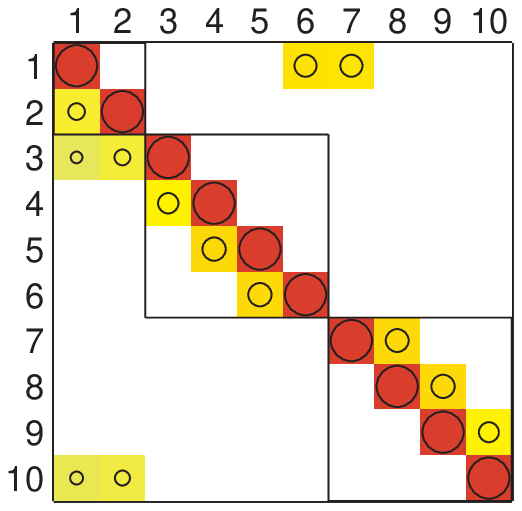}
 \vfill
 {
 \subfigure[]{
\includegraphics[scale=1, trim = 0cm 0cm 0cm 0.29cm, clip = true]{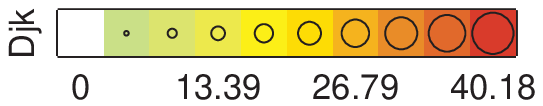}
 \label{SubFig:Figure04a}
} 
 \hspace*{9pt}
 \subfigure[]{
\includegraphics[scale=1, trim = 0cm 0cm 0cm 0.29cm, clip = true]{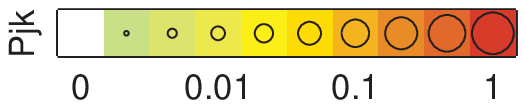}
 \label{SubFig:Figure04b}
}
 }
 \caption{Kinematics and dynamics of the clustered Lorenz attractor.
The geometry of the clusters is depicted with (a) a distance matrix $\slsD$
and the dynamical evolution is quantified by (b) the cluster transition matrix $\slsP$.
All matrix elements are non-negative. 
The value is depicted by background colour 
and increases with the radius of the corresponding circle. 
Note that the scale for the distance matrix is linear, while it is logarithmic for the transition matrix.
The squares mark distances and transitions in the cluster groups
$T$ ($k=1,2$), $E^+$ ($k=3,4,5,6$) and $E^-$ ($k=7,8,9,10$).
}
 \label{Fig:Figure04}
\end{figure}

The distance matrix (see figure~\ref{SubFig:Figure04a}) 
also suggests three groups, $k=1,2$, $k=3,4,5,6$ and $k=7,8,9,10$
in which the centroids within each region are comparably close.
These regions correspond to the groups, $T$, $E^+$, and $E^-$, identified above
and visualised with square boxes.

\begin{figure}
\psfrag{1}[cc][][1][0]{$\colorbox{white}{1}$}
\psfrag{c2}[cc][][1][0]{${\textcolor{white} 2}$}
\psfrag{c3}[cc][][1][0]{${\textcolor{white} 3}$}
\psfrag{c4}[cc][][1][0]{${\textcolor{white} 4}$}
\psfrag{c5}[cc][][1][0]{${\textcolor{white} 5}$}
\psfrag{c6}[cc][][1][0]{${\textcolor{white} 6}$}
\psfrag{c7}[cc][][1][0]{$7$}
\psfrag{c8}[cc][][1][0]{$8$}
\psfrag{c9}[cc][][1][0]{$9$}
\psfrag{c10}[cc][][1][0]{$10$}
\psfrag{AA}[cc][][1][0]{$E^+$}
\psfrag{BB}[cc][][1][0]{$E^-$}
\psfrag{E2}[cc][][1][0]{$E^-$}
\psfrag{E1}[cc][][1][0]{$E^+$}
\psfrag{TT}[cc][][1][0]{\colorbox{white}{T}}
 \centering
\includegraphics[scale = 1]{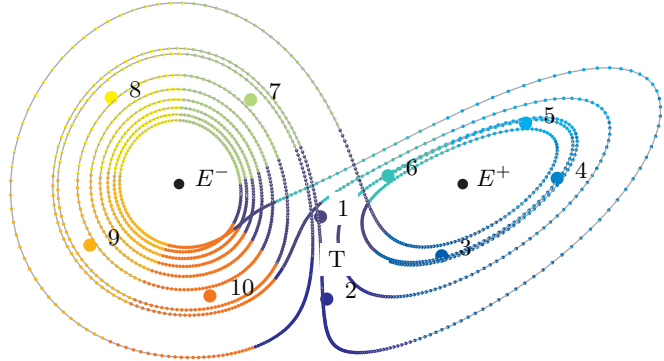}
 \caption{Cluster transitions of the Lorenz attractor.
Cluster centroids are displayed as coloured bullets, 
the black bullets are the two unstable fixed points.
The trajectory is displayed as grey line with coloured dots as discrete states.
For a better visibility of the clusters a non-symmetric perspective has been chosen.
Three qualitatively different groups of clusters can be distinguished:
the cycle groups of the ears $E^+$ and $E^-$ 
as well as the transition clusters $T$.
}
 \label{Fig:Figure05}
\end{figure}

The dynamical behaviour between those groups and each cluster can be studied via 
the cluster transition matrix (CTM) displayed in 
figure~\ref{SubFig:Figure04b}. 
The high probabilities in the principal diagonal indicate a lingering in the cluster. 
This is due to the small time step $\Delta t$ and the large size of the clusters.
For instance, let the initial probability distribution 
be $\boldsymbol{p}^0 = \left[ 1,0,\ldots,0 \right]^T$, i.e.\ only occupies state $k=1$.
Then, the probability $\slsP_{11} \approx 1$ is high that a snapshot will remain in this cluster. 
With lower probability it can move to the states $k=2$, $3$ or $10$. 
Assuming now a unit probability of state $3$ for $\boldsymbol{p}^0$, 
in the next time step, it can stay there or it moves forward in the direction
$3\rightarrow 4$, then $4\rightarrow 5$, and $5\rightarrow 6$. 
In $k=6$ the probability is large for returning to the initial state $1$. 
We have already identified above the states $k=3,4,5,6$ as the oscillation around one ear of the Lorenz attractor. 
The other ear resembled by $k=10,9,8,7$ can be reached by moving from $k=1$ to $k=10$. 
Note that we use the reverse numbering to accentuate the direction of the change of states. 
From the last state $k=7$ it returns to the first state which is always passed after each orbit.
An intuitive picture of the CTM and its state transitions  
is provided in figure~\ref{Fig:Figure05}. 
States are displayed as colour-marked bullets, small colour-marked dots are 
those instants where the trajectory is in the corresponding cluster.

In general, the CTM gives information on the nature of the states and their relation. 
Circulation regions are represented by non-zero main $\slsP_{jj}$ 
and sub-diagonal entries $\slsP_{j+1,j}$ (or $\slsP_{j,j+1}$ depending on the ear). 
An additional non-zero entry that connects the last state of a group with state $k=1$
is necessary for a full orbit.
Distinct states that allow crossing from different groups like $E^+$ and $E^-$ are also identifiable. 
The state $k=1$ serves as the connecting point between the two ears. 
From this state, the trajectories can branch out to the orbits, therefore this state is called branching point.

\begin{figure}
 \centering
 {
 \subfigure[]{
\includegraphics[scale=1, trim = 0cm 0.18cm 0cm 0cm, clip = true]{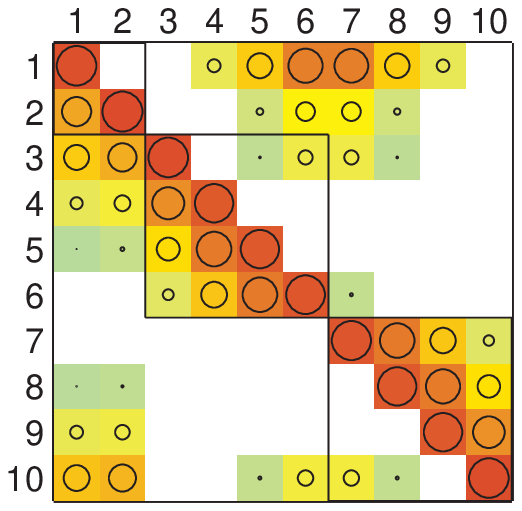}
 \label{SubFig:Figure06a}
} 
 \hspace{20pt} 
 \subfigure[]{
\includegraphics[scale=1, trim = 0cm 0.05cm 0cm 0cm, clip = true]{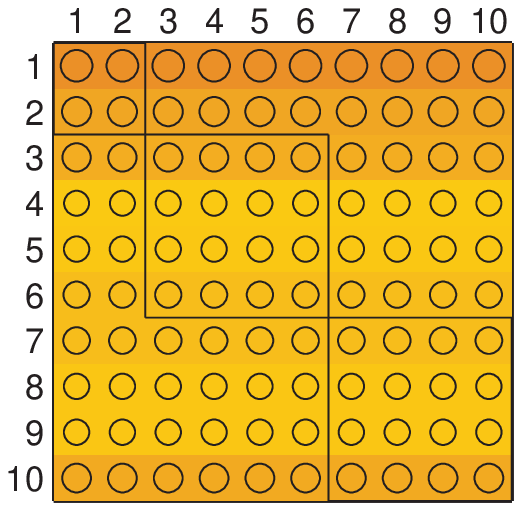}
 \label{SubFig:Figure06b}
}
}
 \caption{Dynamics of the clustered Lorenz attractor.
The figures (a) and (b) illustrate $l=10$ and $l=1000$ 
iterations, respectively,  of the transition matrix $\slsP$.
The matrices are visualised like in figure~\ref{SubFig:Figure04b}.
Note that the increasing iterations make the matrices $\slsP^{l}$ more uniform.}
 \label{Fig:Figure06}
\end{figure}

The dynamical behaviour of the cluster probability vector
can be studied from figure~\ref{Fig:Figure06}
via  powers of the CTM \eqref{Eqn:ProbL0}. 
The cluster dynamics yield information on how the trajectory evolves in time on the attractor. The entries of 
$\slsP_{jk}^l$ give the probability to move from state $k$ to state $j$ in $l$ time steps. 
After $l=10$ iterations of the CTM, the different regions are still visible. 
As an example,
we assume
unit probability in state $k=3$,
i.e. $p_k^0 = \delta_{k3}$, which is
the first cluster of orbit $E^+$.
After $l=10$ time steps the final transitions $3\rightarrow 4$, $3\rightarrow 5$, and $3\rightarrow 6$ 
are possible. 
Since it is not possible to transition to the intermediate state 
$k=1$, $l=10$ is not sufficient to complete a full orbit.
On the other hand, if we assume
$p_k^0 = \delta_{k5}$,
transitions are possible to state $k=1$, to state $k=3$ 
(the begin of another oscillation around the same orbit), 
and to state $k=10$ which marks the starting point of the other orbit $E^-$. 
Thus, we can conclude, a full circulation around one fixed point needs more than $10$ time steps.
In general, the probabilities on the principal diagonal have decreased as a benefit 
to the subdiagonals along the possible paths and scaled 
with the distance. 
With an increasing number of iterations the transition probabilities smooth out, 
and the transition matrix eventually converges 
to $\slsP^{\infty}$ (see figure~\ref{SubFig:Figure06b}).   

\begin{figure}
 \psfrag{cluster index}[cc][][1][0]{$k$}
 \psfrag{qk-pinfk}[cc][][1][0]{$q_k,\, p_k^{\infty}$}
 \psfrag{qkpk}[cc][][1][-90]{$q_k,\, p_k^{\infty}$}
 \psfrag{p1}[cc][][1][-90]{$p_1^l$}
 \psfrag{p9p1}[cc][][1][-90]{$p_9^l, p_1^l$}
 \psfrag{p10k-min-max}[cc][][1][0]{$p_1^l,\,p_9^l$}
 \psfrag{0.05}[cc][][1][0]{$0.05$}
 \psfrag{0.1}[cc][][1][0]{$0.1$}
 \psfrag{0.15}[cc][][1][0]{$0.15$}
 \psfrag{0.2}[cc][][1][0]{$0.2$}
 \psfrag{0.25}[cc][][1][0]{$0.25$}
 \centering
 {
 \subfigure[]{
\includegraphics[scale = 0.95]{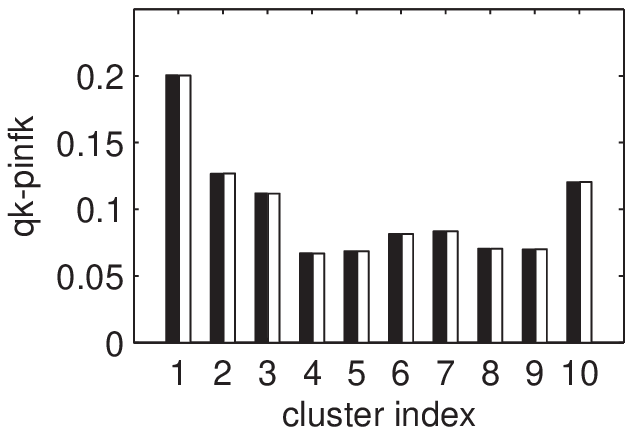}
 \label{SubFig:Figure07a}
} 
 \subfigure[]{
\includegraphics[scale = 0.95]{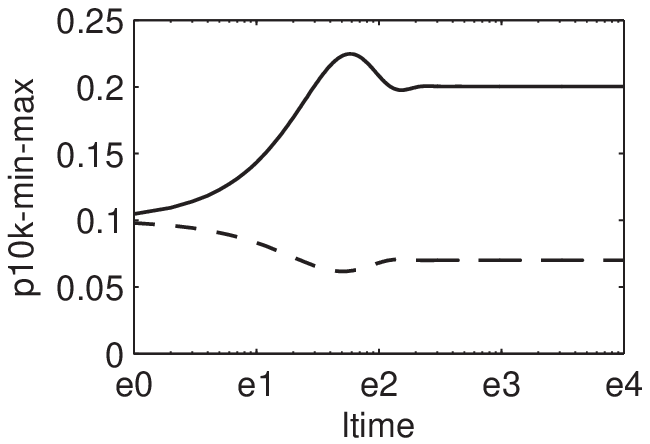}
 \label{SubFig:Figure07b}
}
 }
 \caption{Cluster probability distributions of the Lorenz attractor.
The left figure displays the probability distributions 
from the data ${\bf q}$ \protect\eqref{Eqn:qk} (solid rectangles) 
and from converged iteration of the transition dynamics 
$\boldsymbol{p}^{\infty}$ (open rectangles) \eqref{Eqn:ProbInfinity}.
The right graph shows the convergence of cluster probabilities 
$p_1^l$ (solid line) and $p_9^l$ (dashed line) 
with number of iterations $l$.
The initial condition $\boldsymbol{p}^{0}$ was from equipartition of probabilities.}
 \label{Fig:Figure07}
\end{figure}

In figure~\ref{SubFig:Figure07a} the asymptotic cluster probabilities $\boldsymbol{p}^{\infty}$ 
are compared with the distribution from the data $\boldsymbol{q}$.
The asymptotic probability distribution is inhomogeneous. 
The largest probability $q_1$ results from the distinct 
position this cluster has. As a connection between the two ears this state should 
be passed about twice as often than the others. 
The lowest ones, $q_4$, $q_5$ and $q_8$, $q_9$, correspond to the outer clusters of the two orbits.
Pars pro toto, the evolution of two cluster probabilities, 
the worst defined as $\max|p_k^{10}-q_k|$ and the best defined as $\min|p_k^{10}-q_k|$,
$p_1^l$ and $p_9^l$, respectively, 
are displayed in figure~\ref{SubFig:Figure07b}. 
Although the initial difference $|p_k^{10}-q_k|$ is larger for $p_1$, 
both converge at $l=10^2$. 

\begin{figure}
\psfrag{cluster index}[cc][][1][0]{$k$}
 \psfrag{0.05}[cc][][1][0]{$0.05$}
 \psfrag{0.1}[cc][][1][0]{$0.1$}
 \psfrag{-0.1}[cc][][1][0]{$-0.1$}
 \psfrag{0.15}[cc][][1][0]{$0.15$}
 \psfrag{0.2}[cc][][1][0]{$0.2$}
 \psfrag{-0.2}[cc][][1][0]{$-0.2$}
 \psfrag{0.6}[cc][][1][0]{$0.6$}
 \psfrag{0.8}[cc][][1][0]{$0.8$}
 \centering
 {
 \subfigure[]{
\includegraphics[scale = 1]{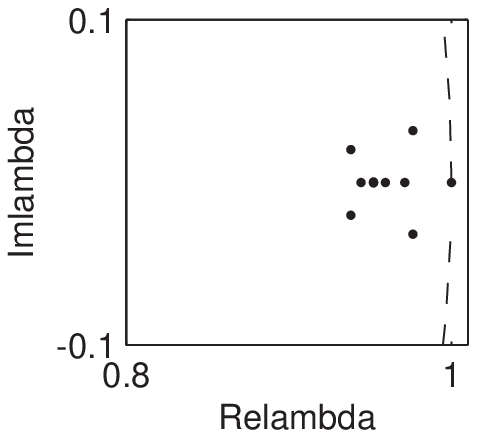}
 \label{SubFig:Figure08a}
} 
 \subfigure[]{
\includegraphics[scale = 1]{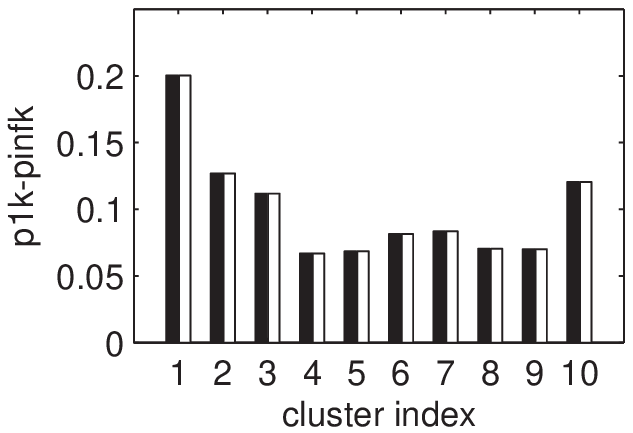}
 \label{SubFig:Figure08b}
}
 }
 \caption{Stability analysis of the transition matrix associated with the clustered Lorenz attractor:
(a) complex spectrum and 
(b) eigenvector $\boldsymbol{p}^{ev}_1$ associated with the dominant eigenvalue $+1$.
Note that all eigenvalues are real or complex conjugate pairs on or within the unit circle (dashed line).
The components of the eigenvector $\boldsymbol{p}^{ev}_1$ are all non-negative, normalised to unit probability
and visualised in a histogram (b) (solid rectangles). 
Note the (expected) agreement between this probability distribution 
and the converged distribution ${\bf p}^{\infty}$ (open rectangles).
}
 \label{Fig:Figure08}
\end{figure}

Another way to access the asymptotic probability distribution is the stability analysis of the CTM 
(see figure~\ref{Fig:Figure08}). 
As expected by the theory (see \S~\ref{SubSubSec:CROM-Tmodel}), 
the complex spectrum yields a dominant eigenvalue $+1$. 
The corresponding eigenvector $\boldsymbol{p}_1^{ev}$
is the only one that survives after sufficiently long iterations and, 
since it is the fixed point of the propagator, 
$\boldsymbol{p}^{\infty}$ has to be equal to $\boldsymbol{p}_1^{ev}$.
Moreover, the rate of convergence 
at which $\boldsymbol{p}^{\infty}$ is approached
is given by the second largest eigenvalue modulus 
as mentioned in \S~\ref{SubSubSec:CROM-Tmodel}
and shown below.

\begin{figure}
\psfrag{cluster index}[cc][][1][0]{$k$}
\psfrag{20}[cc][][1][0]{$20$}
 \centering
 {
 \subfigure[]{
\includegraphics[scale = 1]{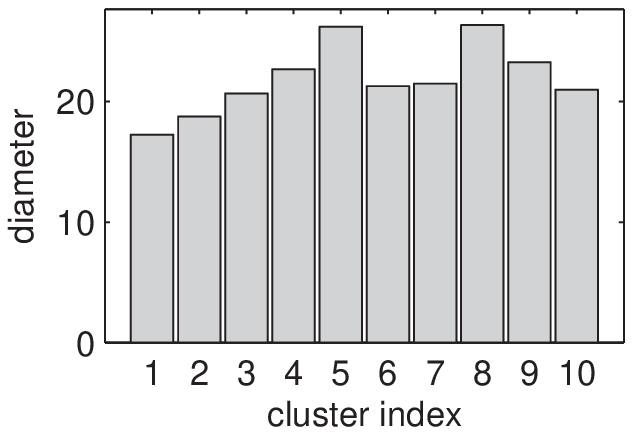}
 \label{SubFig:Figure09a}
}
 \subfigure[]{
\includegraphics[scale = 1]{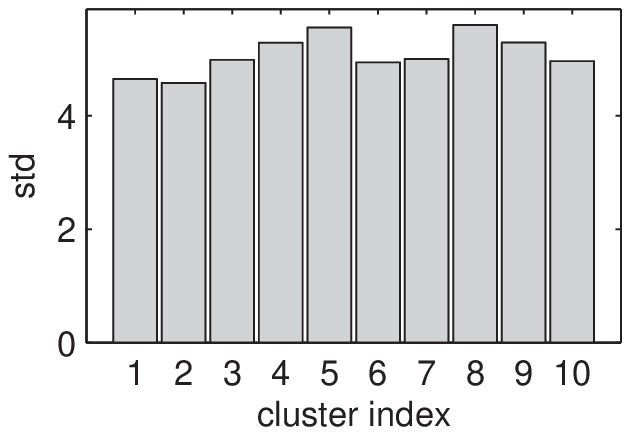}
 \label{SubFig:Figure09b}
}
 }
 \caption{(a) Diameters $D_k$ defined in \protect\eqref{Eqn:DiameterCluster}
and (b) standard deviation $R_k$ as defined in \protect\eqref{Eqn:StandardDeviationCluster} 
of clustered Lorenz attractor. 
Evidently, the clusters have similar sizes.}
 \label{Fig:Figure09}
\end{figure}

The geometric nature of the discretised attractor allows to assess the quality of the discretisation and 
to identify cluster groups and transition regions. 
The similar diameters and standard deviations of the clusters, as 
shown in figure~\ref{Fig:Figure09}, confirm a nearly homogeneous partition of the state space. 

\begin{figure}
\psfrag{eval}[cc][][1][-90]{$\lambda_2^l$}
\psfrag{-20}[cc][][1][0]{$-20$}
\psfrag{-15}[cc][][1][0]{$-15$}
\psfrag{200}[cc][][1][0]{$200$}
\psfrag{600}[cc][][1][0]{$600$}
\psfrag{-5}[cc][][1][0]{$-5$}
\psfrag{-30}[cc][][1][0]{$-30$}
\psfrag{-0.5}[cc][][1][0]{$-0.5$}
\psfrag{0.5}[cc][][1][0]{$0.5$}
\psfrag{1.5}[cc][][1][0]{$1.5$}
 \centering
 \centerline{\raisebox{0.14\textheight}{(a)}\hspace{1cm}\includegraphics[scale = 1, trim = 0 0.77cm 0 0, clip = true]{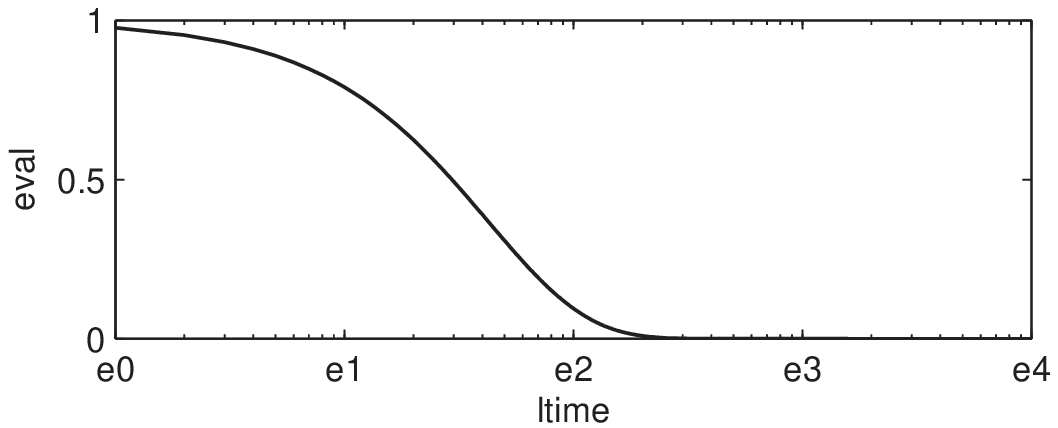}}
 \vfill
 \centerline{\raisebox{0.14\textheight}{(b)}\hspace{0.9cm}\includegraphics[scale = 1, trim = 0 0.77cm 0 0, clip = true]{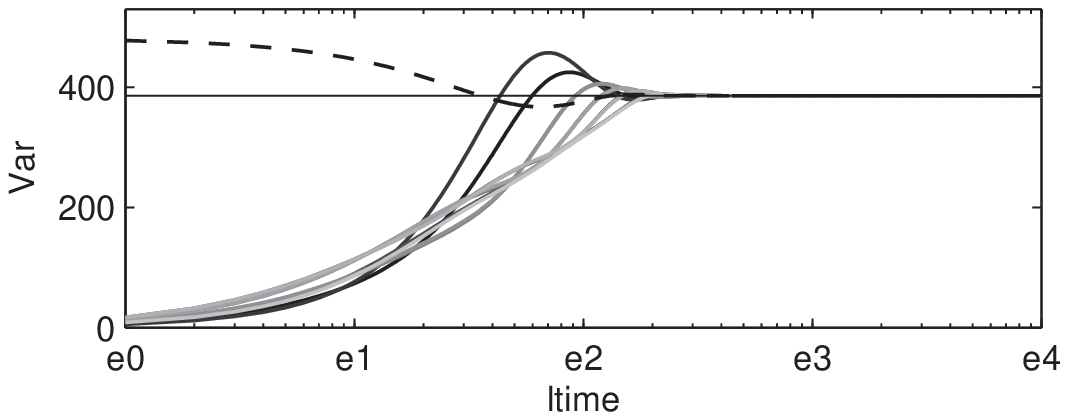}}
 \vfill
 \centerline{\raisebox{0.14\textheight}{(c)}\hspace{1cm}\includegraphics[scale = 1, trim = 0 0.77cm 0 0, clip = true]{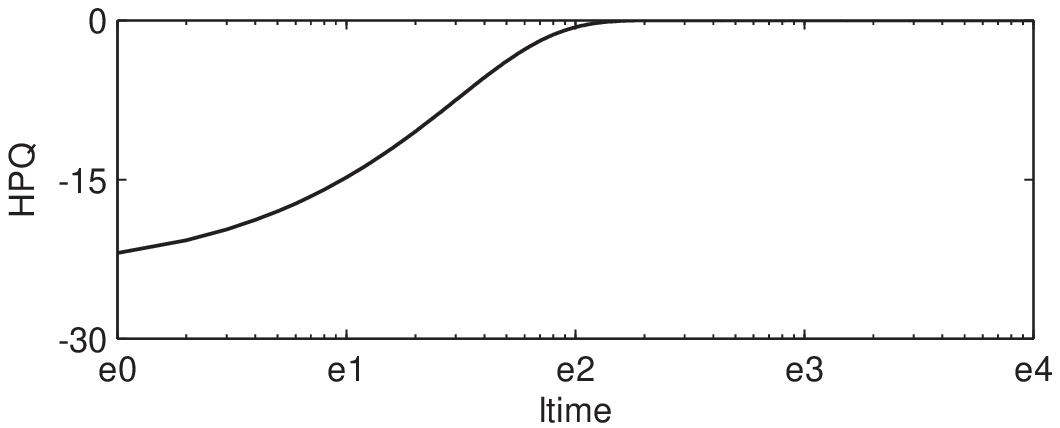}}
 \vfill
 \centerline{\raisebox{0.17\textheight}{(d)}\hspace{1cm}\includegraphics[scale = 1]{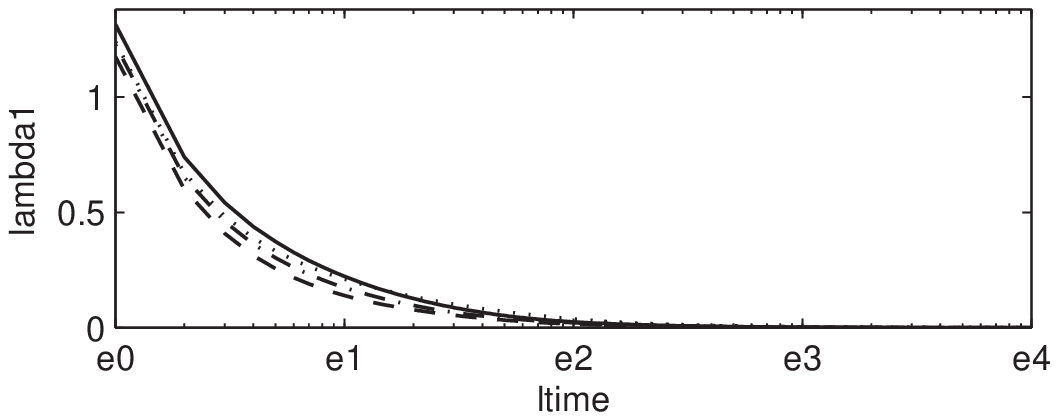}}
 \caption{CROM convergence study for the Lorenz attractor.
Convergence with respect to the number of iterations $l$ is illustrated 
(a) for the second eigenvalue modulus $\vert\lambda\vert_2^l = \vert  0.9762 + 0.0318i\vert^l$ of $\slsP$,
(b) for the variance $R^2(\boldsymbol{p})$ \protect\eqref{Eqn:Variance}
for different initial distributions of (a) the equipartition (dashed line) or
(b) a pure cluster $p_k^0 =\delta_{jk}$
with $j=1$ or $j=2$ is a cluster of the trasnsition region (black) and  
$j=3,\ldots,10$ is a cluster of $E^{-}$ or $E^{+}$ (light grey),
(c) for the Kullback-Leibler entropy $H$ \protect\eqref{Eqn:KullbackLeiblerEntropy}, and
(d) for the FTLE $\lambda_{FTLE}^l$ \protect\eqref{Eqn:FTLE} 
for different initial distances of the Lorenz attractor.
Evidently, all quantities converge at about $l = 
10^2$, i.e.\ each of the measures can be used as convergence indicator.
}
 \label{Fig:Figure10}
\end{figure}

The dynamics on the attractor can be further analysed 
by inspecting the evolution of attractor properties 
(see figure~\ref{Fig:Figure10}). 
As mentioned above, the rate of convergence is given by the second largest eigenvalue modulus.
The power iterations of this quantity are shown in figure~\ref{Fig:Figure10}\,(a),
it converges to zero at about $l=10^2$.

In figure~\ref{Fig:Figure10}\,(b) the variance is displayed for different initial probability conditions. 
Initial distributions $p_k^0 = \delta_{k1}$, $p_k^0 = \delta_{k2}$ \etc are displayed in solid lines,  
and dashed lines indicates equipartition. 
For all of these initial conditions 
except for the initial equipartition,
the spreading increases until it reaches a finite value.
Initial distributions with $p_k^0 = \delta_{k1}$ or $\delta_{k2}$ (dark colour) exhibit a large spreading 
and overshoot which can be linked to 
their special role as a branching region. 
All curves converge after $l=10^2$ iterations.

The evolution of the Kullback-Leibler entropy (see figure~\ref{Fig:Figure10}\,(c)) offers  important insights 
about the Markov model.
As expected, the Kullback-Leibler entropy
\eqref{Eqn:KullbackLeiblerEntropy} of the iterated CTM 
$\slsP^l$
increases with the iteration $l$, 
i.e.\  information is lost by irreversible diffusion.
The entropy converges around $l \approx 10^2$,
when the iterated CTM describes 
the $l$-independent asymptotic probability $\boldsymbol{p}^{\infty}$.
A small negative value of $H$ characterises the small difference
between the inferred cluster probability distribution $\boldsymbol{q}$
of the data and 
the  fixed-point $\boldsymbol{p}^{\infty}$.
Intriguingly, 
there exists an iteration interval  $l<10$
with roughly constant $H$.
In this regime, 
the iterated CTM is well approximated
by a linear Taylor expansion
from  continuous-time Markov model  \eqref{Eqn:ProbL0Continuously},
i.e.\ 
\begin{equation}
 \slsP^l = \exp \left( \slsP^{\mathrm cont} l \Delta t \right)
\approx \slsI + \slsP^{\mathrm cont} l \Delta t + O(\Delta t^2)\, .
\end{equation}
This implies that the information gained 
from $\slsP^l$  are comparable for different $l$. 
In other words, the Kullback-Leibler entropy
does not only indicate the converged state of the iterations
and a difference between Markov model and data.
The entropy also indicates the validity regime 
of the  Euler scheme for the  continuous-time Markov model 
\eqref{Eqn:ProbL0Continuously}.

The evolution of the FTLE is depicted in figure~\ref{Fig:Figure10}\,(d). 
Initially distant trajectories eventually 
converge to the asymptotic distribution $p^{\infty}$. 
All four properties converge at $l\approx 10^2$ and any of them can serve as an adequate measure of convergence.

Summarising, 
the CROM strategy segments the Lorenz attractor 
into nearly homogeneous clusters which are arranged like pieces of a pie 
around the unstable fixed points. 
Without employing any prior knowledge of the nature of the attractor, 
the CTM yields the two characteristic orbits, and identifies the 
states that behave as a branching region.
The convergence behaviour is in good agreement 
with the result of the statistical analysis of the clustered data.
Increasing the number $K$ of clusters does 
not change the main conclusions.
The number of clusters in the two ears (state resolution) does increase
and  fine-structure details of the branching process can be observed.
Note that in the special case $K=2$ 
CROM yields two clusters, each resolving one ear.
Variations of the snapshot time step $\Delta t$  have no strong effect 
on the observed oscillation and transition mechanisms.
The transition probabilities between different clusters depend, 
in the first approximation, linearly on the time step.

\section{Mixing layer model}
\label{Sec:MixingLayerModel}

CROM is applied to a two-dimensional incompressible mixing layer 
with Kelvin-Helmholtz vortices undergoing vortex pairing.
This flow is computed with a direct numerical simulation of the Navier-Stokes equation. 
The velocity ratio is $r=U_1/U_2=3$ where $U_1$ and $U_2$ denote the upper (fast) and lower (slow) stream velocity. 
The Reynolds number
$\Rey=\Delta U\delta_{\omega}/\nu=500$ is based on
the maximum velocity $U_1$, 
the initial vorticity thickness $\delta_{\omega}$, 
and the kinematic viscosity $\nu$. 
A $\tanh$ profile with stochastic perturbation is chosen as inlet profile. 
A snapshot ensemble of $M=2000$ with a sampling time of $\Delta t = 1$,
which is nondimensionalised with respect to $U_1$ and $\delta_{\omega}$,
is provided by a finite difference Navier-Stokes solver.
Details of this solver can be found in \cite{Daviller2010phd} and \cite{Cavalieri2011jsv}.
The computational domain is $140\,\delta_{\omega}$ long, and $56\,\delta_{\omega}$ high.
The resolution of the spatial discretisation increases in the mixing region. 

\begin{figure}
 \centering
 \includegraphics[scale=0.35]{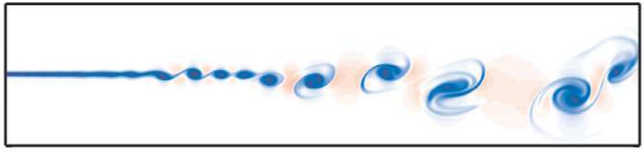}
 \caption{2D DNS simulation of the mixing layer: Colour plot of an instantaneous vorticity distribution.}
 \label{Fig:Figure11}
\end{figure}

The mixing layer exhibits the typical roll-up of initial Kelvin-Helmholtz vortices 
and vortex pairing events further downstream.
A vorticity snapshot is displayed in figure~\ref{Fig:Figure11}. 
In the flow simulations, vortex merging of two and three vortices occur 
which results into a large spreading of the shear layer. 
Moreover, four vortices may merge into two pairs which will form a single structure further downstream.

\begin{figure}
 \centering
\includegraphics[scale = 1]{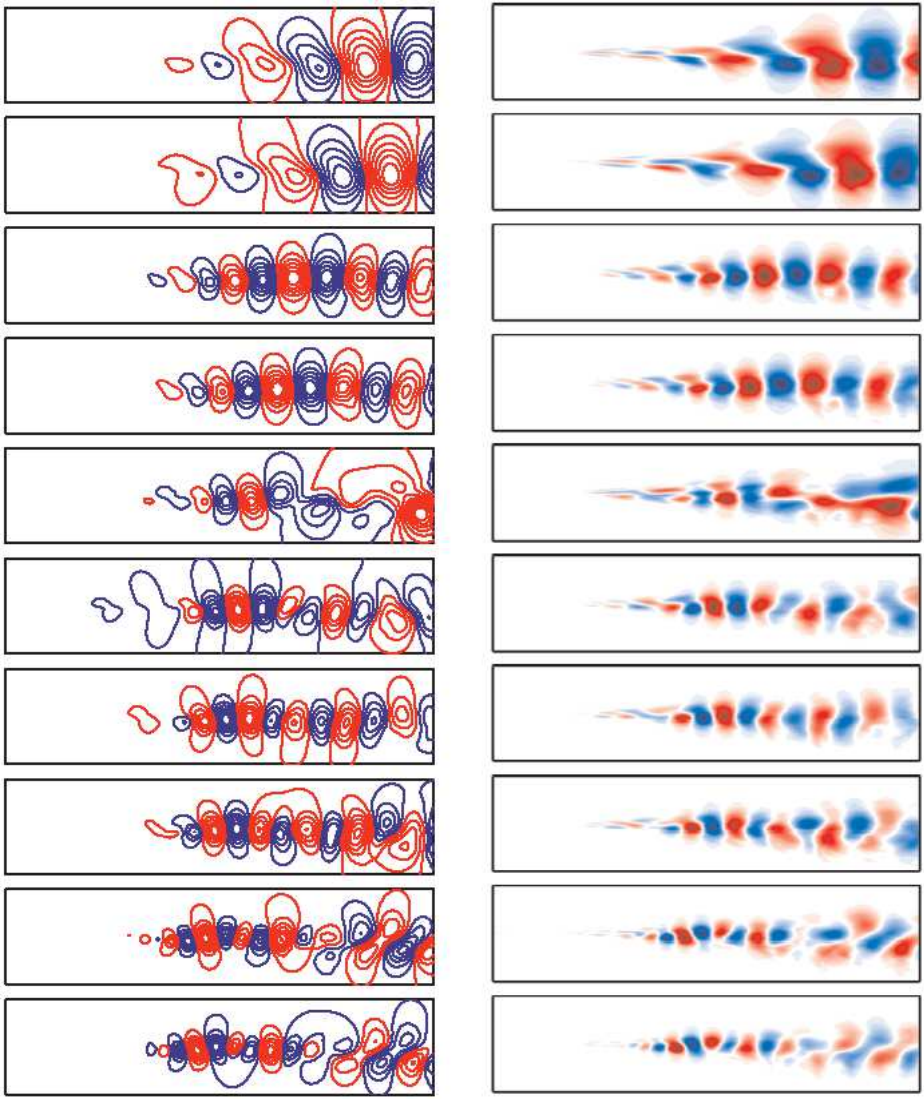}
 \caption{POD modes $\boldsymbol{u}_i$ $i=1,\ldots 10$ (top to bottom) of the mixing layer. 
The flow is visualised by 
iso-contours of transversal velocity component (left)
and by colour plots of the vorticity (right).}
 \label{Fig:Figure12}
\end{figure}

The flow data is compressed with a POD from the snapshot ensemble.
The POD modes are visualised in figure~\ref{Fig:Figure12}
via iso-contours of the transversal velocity component
and colour plots of the vorticity.
The first two mode pairs ($i=1,2,3,4$) represent the convection of 
nearly periodic vortex structures in streamwise direction \citep{Rajaee1994jfm}. 
The first mode pair ($i=1,2$) represents flow structures associated with the subharmonic instability, 
while the second mode pair ($i=3,4$) corresponds to Kelvin-Helmholtz vortices associated 
with a higher frequency and increasing wavenumber in streamwise direction \citep{Noack2004swing, Laizet2010pof}. 
Higher-order modes contain a mixture of different frequencies.

\begin{figure}
\psfrag{cluster index}[cc][][1][-90]{$k$}
\psfrag{snapshot}[cc][][1][0]{$t$}
 \centering
 \centerline{\raisebox{0.14\textheight}{(a)}\hspace{5mm}\includegraphics[scale = 1, trim = 0mm 8mm 1mm 0mm, clip=true]{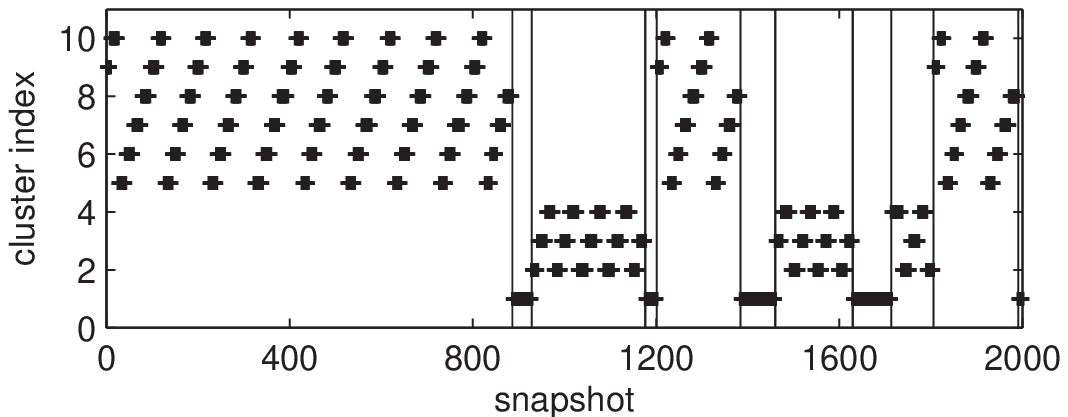}}
 \vfill
 \centerline{\raisebox{0.14\textheight}{(b)}\hspace{4.7mm}\includegraphics[scale = 1, trim = 0mm 8mm 1mm 0mm, clip=true]{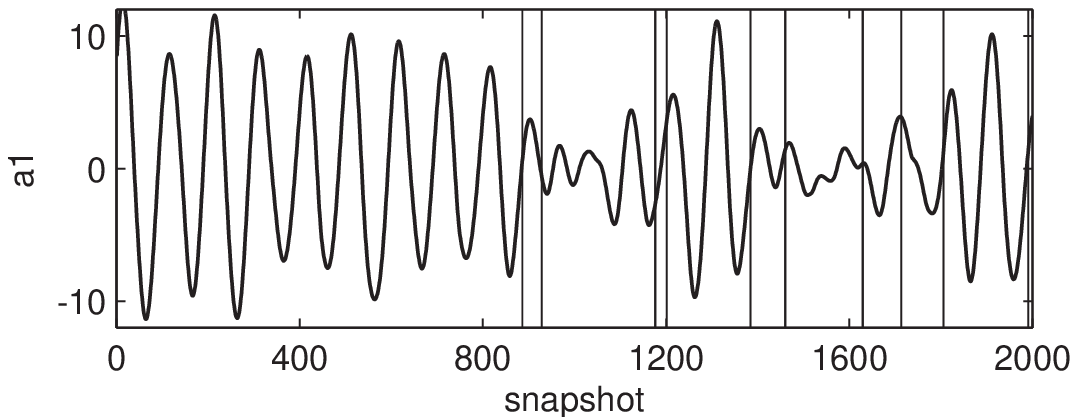}}
 \vfill
 \centerline{\raisebox{0.14\textheight}{(c)}\hspace{4.7mm}\includegraphics[scale = 1, trim = 0mm 8mm 1mm 0mm, clip=true]{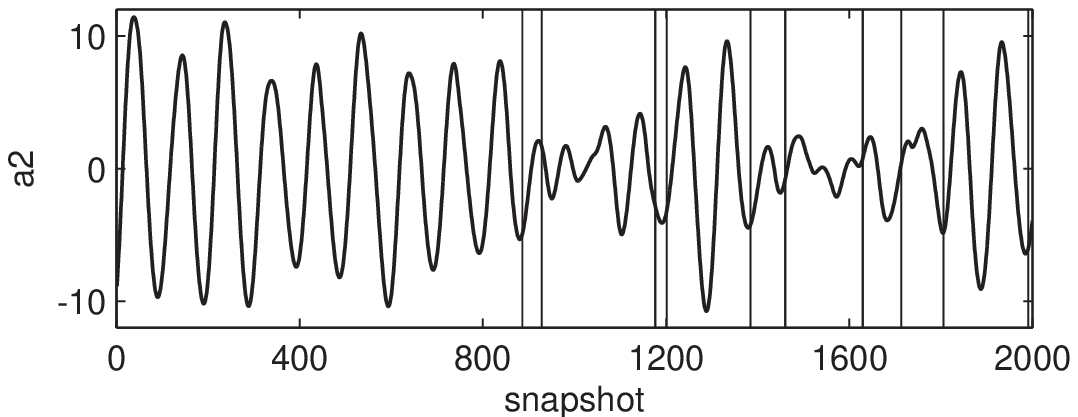}}
 \vfill
 \centerline{\raisebox{0.14\textheight}{(d)}\hspace{4.7mm}\includegraphics[scale = 1, trim = 0mm 8mm 1mm 0mm, clip=true]{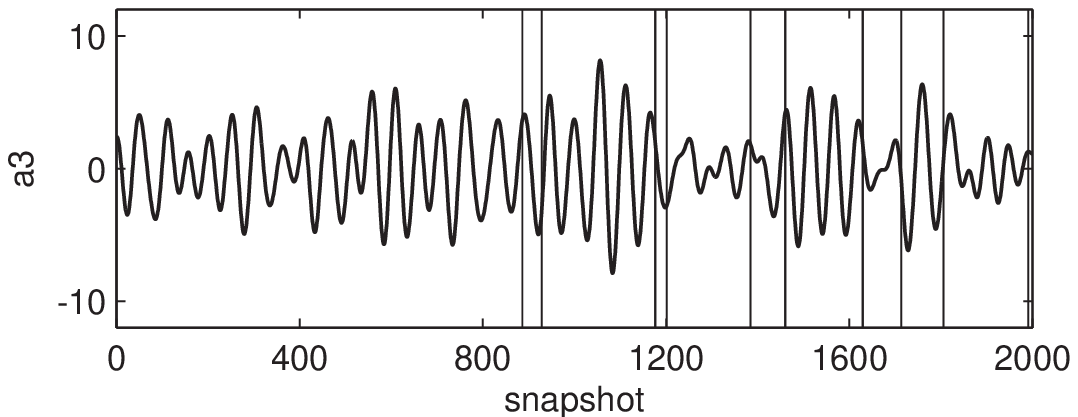}}
 \vfill
 \centerline{\raisebox{0.15\textheight}{e)}\hspace{5mm}\includegraphics[scale = 1]{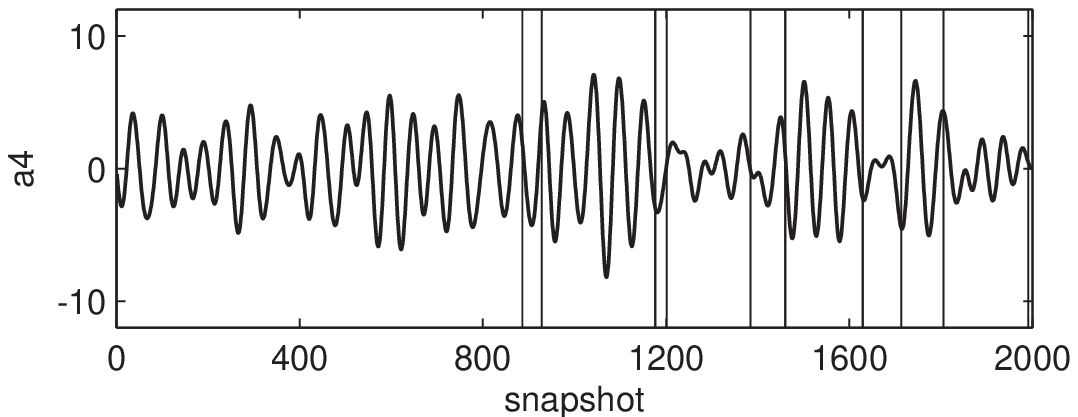}}
 \caption{Temporal cluster analysis of the mixing layer.
Figures (b-d) show
the POD coefficients of first and second POD mode pairs $a_i$, $i=1,2,3,4$ (top to bottom) as function of time $t$.
The upper figure (a) marks the cluster $k$ to which each snapshot belongs.
The vertical lines separate between Kelvin-Helmholtz vortices (cluster $k=2,3,4$)
and vortex pairing dominated flows (remaining clusters).}
 \label{Fig:Figure13}
\end{figure}

The first four temporal POD coefficients are displayed in figure~\ref{Fig:Figure13}\,(b-e).
A simultaneous decrease in the amplitudes of the first mode pair 
and increase in the amplitudes of the second mode pair 
is an indicator for a change of the flow from vortex pairing to predominant Kelvin-Helmholtz vortices. 

For the reasons discussed in \S~\ref{Sec:CROM}, 
the cluster analysis is performed on the POD coefficient vectors 
$\boldsymbol{a}^m$, $m=1,\,\ldots,\,M$, with $\boldsymbol{a} = [a_1,\,\ldots,\,a_{M-1}]^T$.
We chose the number of clusters $K_c=10$, like in the Lorenz attractor.
As for the Lorenz attractor, the choice of $K_c=10$ is in good agreement with the ``elbow criterion''
as explained in \S~\ref{SubSubSec:CROM-kmeans}.
The assignment of each snapshot to the clusters is shown in 
figure~\ref{Fig:Figure13}\,(a). 
Vortex pairing dominated flows and Kelvin-Helmholtz vortices 
are clearly separated into different groups of clusters, 
called $\text{VP}$
with $k=5,6,7,8,9,10$ for the former and
$\text{KH}$ with $k=2,3,4$ for the latter.
The two dynamical regimes are mainly 
connected by cluster $k=1$.
In this case, the flow has to pass this state for the transition from one dynamical regime to the other.

\begin{figure}
 \centering
\includegraphics[scale = 1]{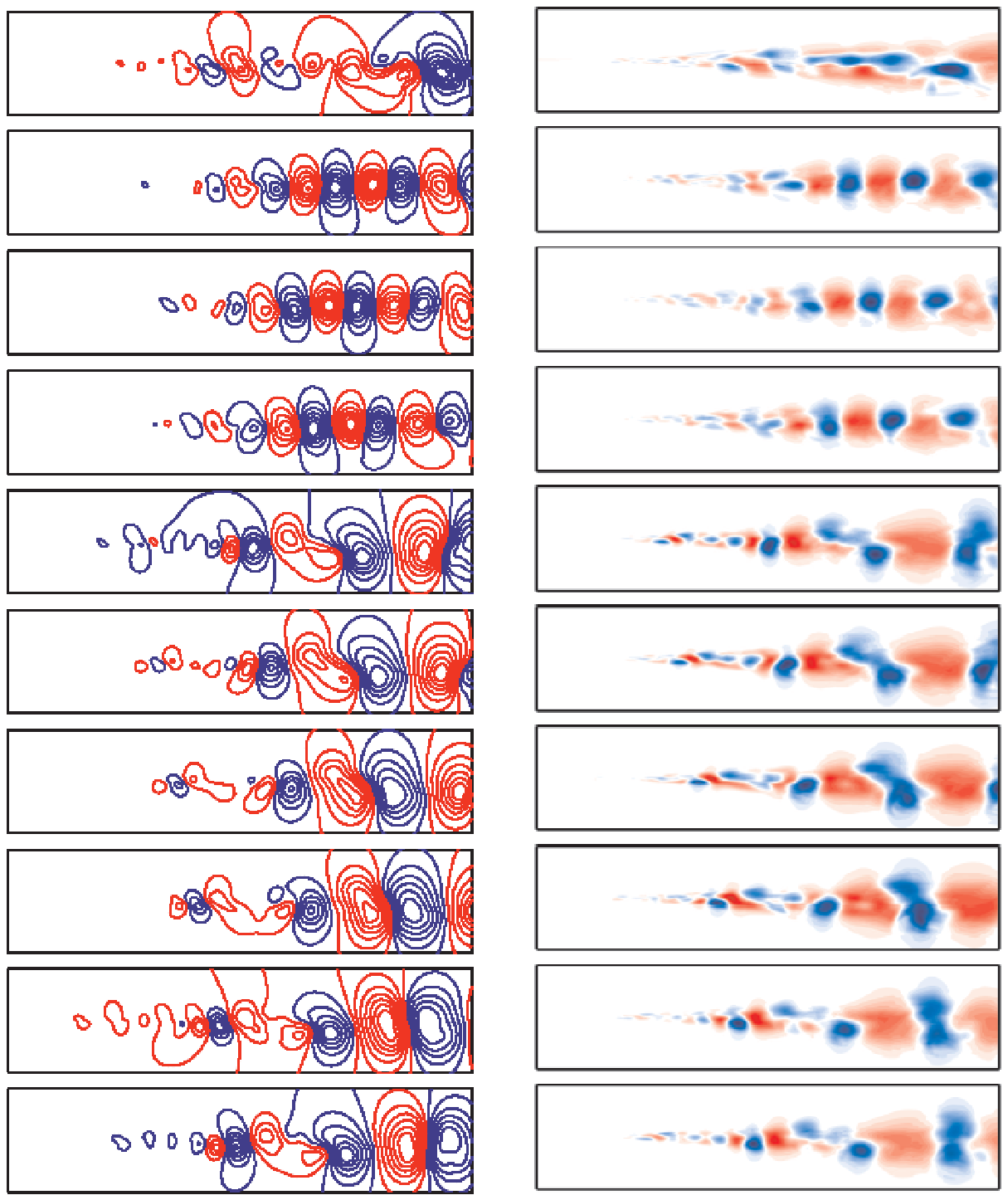}
 \caption{Cluster centroids ${\bf c}_k$, $k=1,\ldots,10$ (top to bottom)
of the mixing layer.
The flow is visualised like in figure~\protect\ref{Fig:Figure12}.}
 \label{Fig:Figure14}
\end{figure}

The centroids associated with each cluster are displayed in figure~\ref{Fig:Figure14}. 
In the subharmonic group $\text{VP}$, 
 the centroids $k+1$ lead the centroids $k$ by a sixth of wavelength of the merged vortices.
In the group $\text{KH}$, 
a similar behaviour can be observed 
but with shifts corresponding to about a third of the wavelength of the Kelvin-Helmholtz structures.
The centroid $k=1$ represents an intermediate state between those two groups.
The wavenumber increases but is somewhat between the wavenumber corresponding to subharmonic and the 
fundamental frequency.
The order of the centroids is clearly aligned with the dynamical evolution of the flow structures.
In contrast, 
the first four 
POD modes represent sine- and cosine-like wave pairs 
associated with the two shedding regimes.
In case of CROM, flow structures are resolved not by energy content, frequency,  and wavelength, but by phase. 
Moreover, transient states can be explicitly captured.

\begin{figure}
\psfrag{11}[cc][][1][0]{$11$}
\psfrag{17}[cc][][1][0]{$17$}
 \centering
\includegraphics[scale=1, trim = 0cm 0.18cm 0cm 0cm, clip = true]{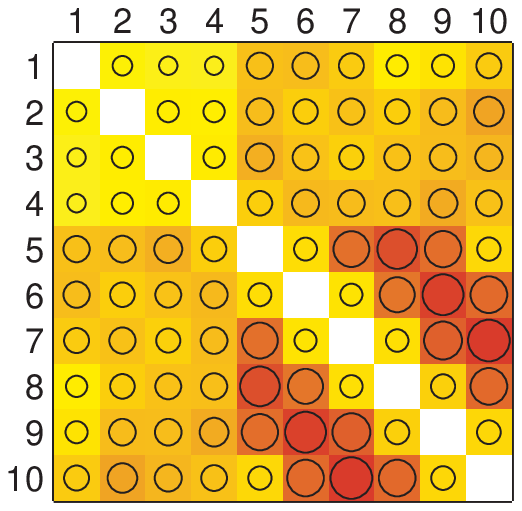}
 \hspace*{20pt}
\includegraphics[scale=1, trim = 0cm 0.18cm 0cm 0cm, clip = true]{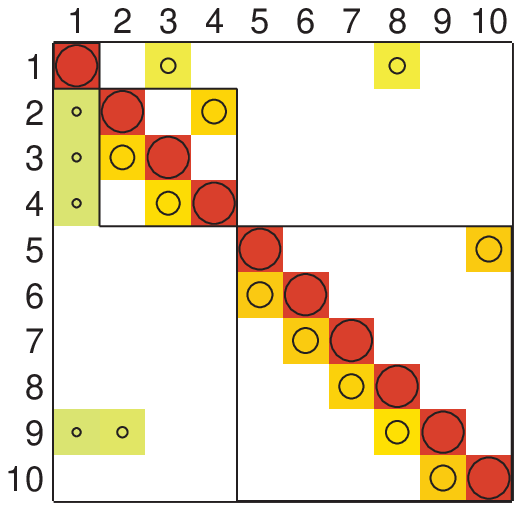}
 \vfill
 {
 \subfigure[]{
\includegraphics[scale=1, trim = 0cm 0cm 0cm 0.29cm, clip = true]{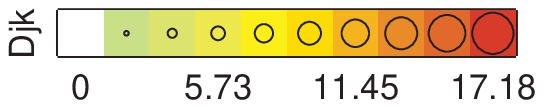}
 \label{SubFig:Figure15a}
}
 \hspace*{10pt}
 \subfigure[]{
\includegraphics[scale=1, trim = 0cm 0cm 0cm 0.29cm, clip = true]{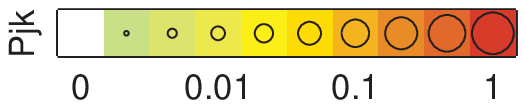}
 \label{SubFig:Figure15b}
}
 }
 \caption{
 Kinematics and dynamics:
 (a) Distance matrix $\slsD$ and (b) transition matrix $\slsP$ 
of the clustered mixing layer data (see figure \ref{Fig:Figure13}).
Values are displayed like in figure~\ref{Fig:Figure04}.}
 \label{Fig:Figure15}
\end{figure}

\begin{figure}
\psfrag{c1}[cc][][1][0]{$\mathbf{{\textcolor{black} 7}}$} 
\psfrag{c2}[cc][][1][0]{$\mathbf{{\textcolor{black} 8}}$}
\psfrag{c3}[cc][][1][0]{$\mathbf{{\textcolor{black} 9}}$}
\psfrag{c4}[cc][][1][0]{$\mathbf{{\textcolor{black}{10}}}$}
\psfrag{c5}[cc][][1][0]{$\mathbf{{\textcolor{black} 5}}$}
\psfrag{c6}[cc][][1][0]{$\mathbf{{\textcolor{black} 6}}$}
\psfrag{c7}[cc][][1][0]{$\mathbf{{\textcolor{black} 1}}$}
\psfrag{c8}[cc][][1][0]{$\mathbf{{\textcolor{white} 2}}$}
\psfrag{c9}[cc][][1][0]{$\mathbf{{\textcolor{white} 3}}$}
\psfrag{c10}[cc][][1][0]{$\mathbf{{\textcolor{white} 4}}$}
\psfrag{AA}[cc][][1][0]{$\text{VP}$}
\psfrag{BB}[cc][][1][0]{$\text{KH}$}
 \centering
\includegraphics[scale = 0.8]{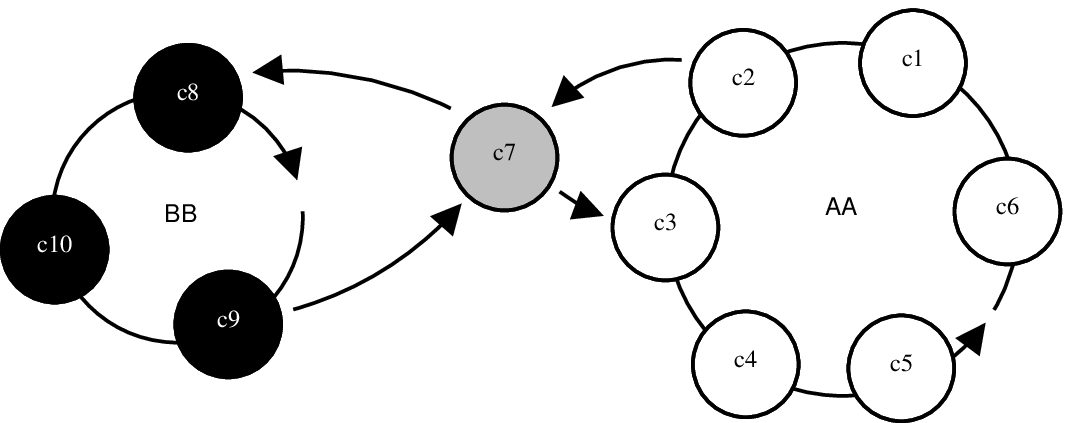}
 \caption{Simplified cluster transitions of the mixing layer. 
The arrows indicate possible transitions above a certain threshold.
This threshold is chosen small enough to guarantee full connectivity.
The graph  highlights two cyclic cluster groups 
associated with Kelvin-Helmholtz shedding (black) 
and with vortex pairing (white).
The grey cluster $k=1$ serves as bifurcation point.  
}
 \label{Fig:Figure16}
\end{figure}

The distance matrix and the CTM (for both see figure~\ref{Fig:Figure15}) 
allow us to analyse the geometric relation of the centroids and their temporal transitions. 
The cluster groups  $\text{VP}$, 
$\text{KH}$ and the transition cluster $k=1$  
are visible in the distance matrix and 
in the CTM as will become clearer below.
Furthermore, the inner cluster distances
between the centroids of group $\text{VP}$ 
are more distant to each other
than the corresponding distances of group $\text{KH}$,
since vortex pairing is associated with larger fluctuation levels
than the more monochromatic Kelvin-Helmholtz shedding.
The CTM shows an inner-group circulation for both, for example
$5\rightarrow 6\rightarrow \dots\rightarrow 10 \rightarrow 5 \rightarrow \dots$ \etc
Cluster $k=1$ has a special significance as the 
'flipper cluster' which acts as a switch between these groups.
As elaborated above, 
this cluster resembles an intermediate state and links both regimes.
All Kelvin-Helmholtz states can be accessed from the flipper cluster.
Moreover, $k=1$ can only be reached by states $k=3$ from $\text{KH}$ 
and $k=8$ of group $\text{VP}$. 
A direct connection between the groups is solely possible by $2\rightarrow 9$ in one direction.
The sketch in figure~\ref{Fig:Figure16} represents a simplified picture of the CTM. 
The most dominant transitions are displayed.
The threshold for selection of $\slsP_{jk}$ is set to a level 
which is just large enough to fulfil the connectivity of all states.
Summarising, the change from vortex pairing to Kelvin-Helmholtz dominated flows
is possible by the crossing of $k=1$ or by the direct transition to $2\rightarrow 9$.
These observations may be important for control.
If one wants to enforce or prevent pure Kelvin-Helmholtz vortices,
the manipulation of the flipper cluster as effective bifurcation point 
appears a particularly effective tool for control.
Its manipulation or use as a precursor
requires further analysis which is beyond the scope of this article.
Nevertheless, a general CROM-based control strategy 
provided in appendix~\ref{Sec:AppB:TowardsCROMbasedFlowControl}
seems promising for the exploitation of nonlinearities.

\begin{figure}
 \centering
 {
 \subfigure[]{
\includegraphics[scale=1, trim = 0cm 0cm 0cm 0.2cm, clip = true]{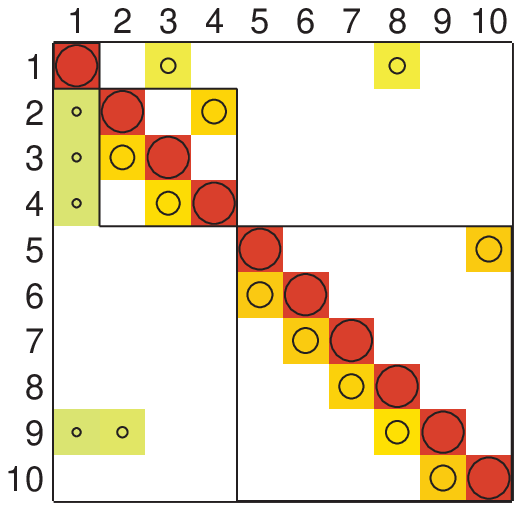}
 \label{SubFig:Figure17a}
}
 \hspace*{20pt}
 \subfigure[]{
\includegraphics[scale=1, trim = 0cm 0cm 0cm 0.2cm, clip = true]{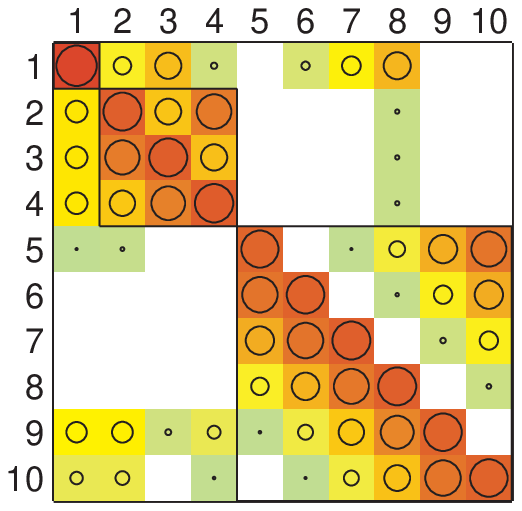}
 \label{SubFig:Figure17b}
}
 \vfill
 \subfigure[]{
\includegraphics[scale=1, trim = 0cm 0cm 0cm 0.2cm, clip = true]{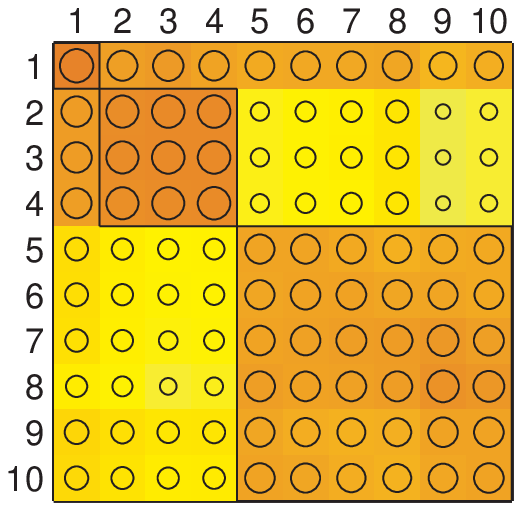}
 \label{SubFig:Figure17c}
}
 \hspace*{20pt}
 \subfigure[]{
\includegraphics[scale=1, trim = 0cm 0cm 0cm 0.2cm, clip = true]{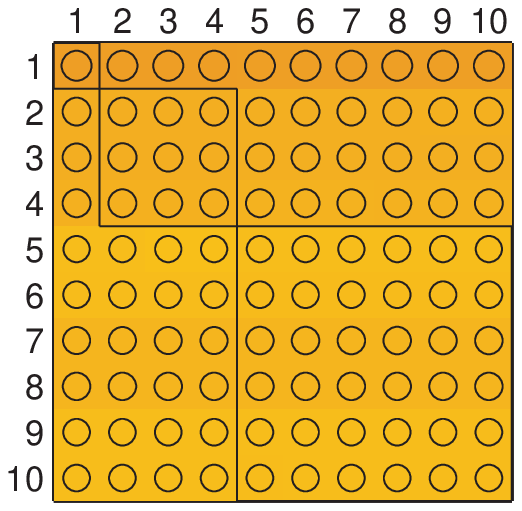}
 \label{SubFig:Figure17d}
}
 }
 \caption{Cluster dynamics of the mixing layer.
 The evolution is quantified by $\slsP^l$ 
 for (a) $l = 10^0$ and (b) $l = 10^1$, 
     (c) $l = 10^2$ and (d) $l = 10^3$.}
 \label{Fig:Figure17}
\end{figure}

From the evolution in time of the transitions (see figure~\ref{Fig:Figure17}), 
we can deduce periodical behaviour in a certain direction similar to the Lorenz attractor. 
With increasing power $l$, the entries of $\slsP$ in each column approach the long-term probability.
Each iteration of the CTM corresponds to the time step $\Delta t$. 

\begin{figure}
 \psfrag{cluster index}[cc][][1][0]{$k$}
  \psfrag{qk-pinfk}[cc][][1][0]{$q_k,\, p_k^{\infty}$}
 \psfrag{p4p1}[cc][][1][-90]{$p_4^l,\,p_1^l$}
 \psfrag{p10k-min-max}[cc][][1][0]{$p_4^l,\,p_6^l$}
 \psfrag{0.05}[cc][][1][0]{$0.05$}
 \psfrag{0.1}[cc][][1][0]{$0.1$}
 \psfrag{0.15}[cc][][1][0]{$0.15$}
 \centering
  {
 \subfigure[]{
\includegraphics[scale = 0.95]{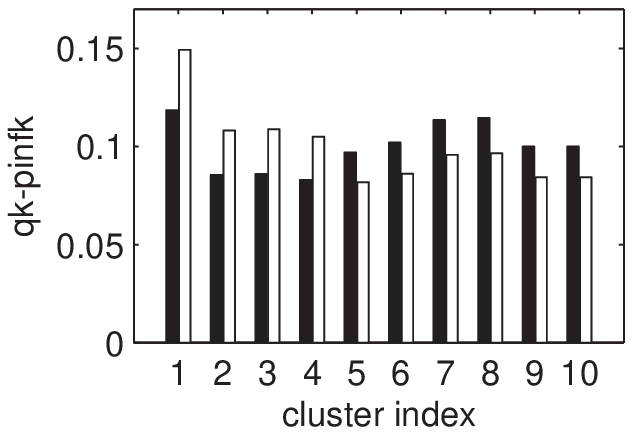}
 \label{SubFig:Figure18a}
}
 \subfigure[]{
\includegraphics[scale = 0.95]{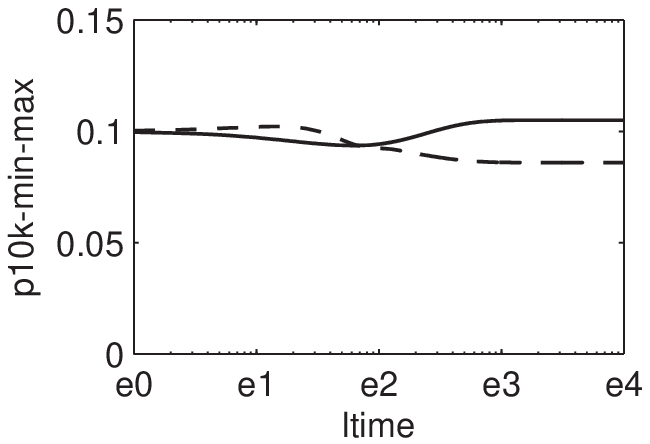}
 \label{SubFig:Figure18b}
}
 }
 \caption{
Cluster probability distributions of the mixing layer 
(like figure~\protect\ref{Fig:Figure07} for the Lorenz attractor).
The left figure displays the probability distributions 
from the data ${\bf q}$ \protect\eqref{Eqn:qk} (solid rectangles) 
and from converged iteration of the transition dynamics 
$\boldsymbol{p}^{\infty}$ (open rectangles) \eqref{Eqn:ProbInfinity}.
The right graph shows the convergence of cluster probabilities 
$p_4^l$ (solid line) and $p_6^l$ (dashed line).
}
 \label{Fig:Figure18}
\end{figure}

For the sake of completeness, 
we give a short description on convergence behaviour and attractor properties.
The cluster probabilities (see figure~\ref{SubFig:Figure18a}) are not strongly biased to any state. 
The converged cluster probabilities 
belonging to the vortex pairing are slightly underestimated,
whereas $p_k^{\infty}$, $k=1,\,2,\,3,\,4$, 
are larger than the cluster probabilities based on the data.
This difference can be attributed 
to discretisation error of the CTM.
Only 1999 snapshot transitions are used 
to estimate 10$\times$10 matrix elements.

In analogy to the CROM of the Lorenz attractor, 
the best cluster $k=4$ is determined as the cluster 
with the minimum value of $|p_k^{10}-q_k|$, 
and the worst cluster $k=6$ is defined as the cluster with the maximum value.
For the mixing layer, these clusters, 
$k=4$ and $k=6$ in figure~\ref{SubFig:Figure18b},
converge both at $l\approx 10^3$.

\begin{figure}
\psfrag{cluster index}[cc][][1][0]{$k$}
\psfrag{0.05}[cc][][1][0]{$0.05$}
\psfrag{0.15}[cc][][1][0]{$0.15$}
\psfrag{0.1}[cc][][1][0]{$0.1$}
\psfrag{-0.1}[cc][][1][0]{$-0.1$}
\psfrag{0.8}[cc][][1][0]{$0.8$}
 \centering
 {
 \subfigure[]{
\includegraphics[scale=1]{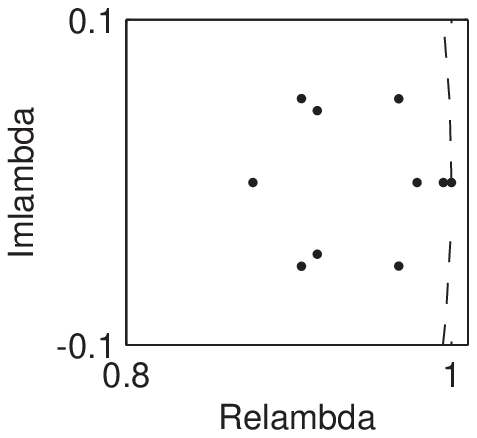}
 \label{SubFig:Figure19a}
}
 \subfigure[]{
\includegraphics[scale=1]{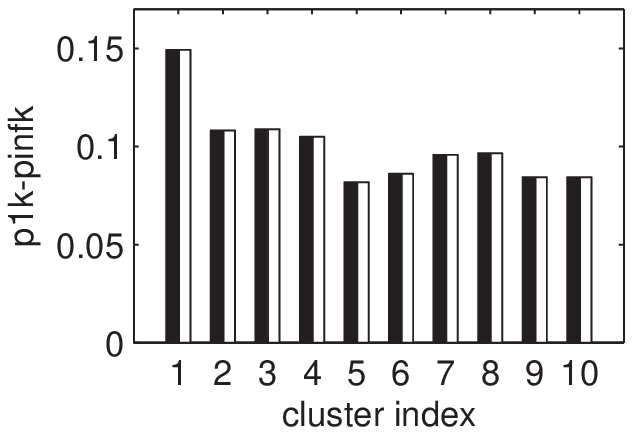}
 \label{SubFig:Figure19b}
}
 }
 \caption{ 
 Stability analysis of the transition matrix associated with the mixing layer:
(a) complex spectrum and 
(b) eigenvector $\boldsymbol{p}^{ev}_1$ associated with the dominant eigenvalue $+1$.
Note that all eigenvalues are real or complex conjugate pairs on or within the unit circle (dashed line).
The components of the eigenvector $\boldsymbol{p}^{ev}_1$ are all non-negative, normalised to unit probability
and visualised in a histogram (b) (solid rectangles). 
Note the (expected) similarity between this probability distribution 
and the converged distribution ${\bf p}^{\infty}$ (open rectangles).
}
 \label{Fig:Figure19}
\end{figure}

As for the Lorenz attractor, 
the eigenvector associated with the dominant eigenvalue
yields the converged cluster probabilities (see figure~\ref{SubFig:Figure19b}). 

\begin{figure}
\psfrag{cluster index}[cc][][1][0]{$k$}
 \psfrag{15}[cc][][1][0]{$15$}
 \psfrag{20}[cc][][1][0]{$20$}
 \centering
 {
 \subfigure[]{
\includegraphics[scale = 1]{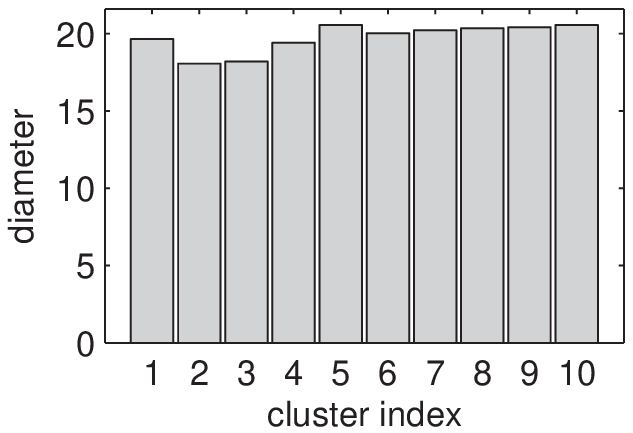}
 \label{SubFig:Figure20a}
}
 \subfigure[]{
\includegraphics[scale = 1]{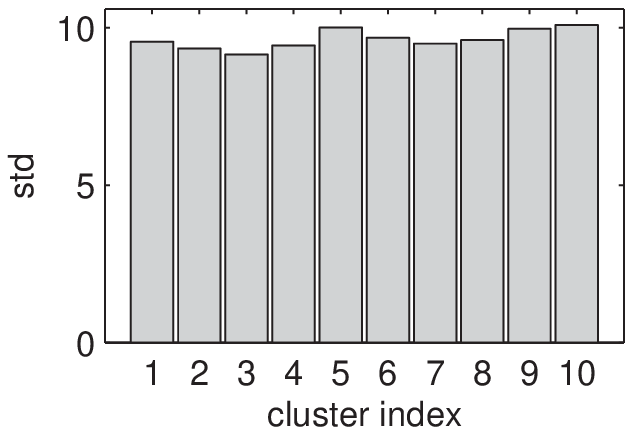}
 \label{SubFig:Figure20b}
}
 }
 \caption{
 (a) Diameters $D_k$ defined in \protect\eqref{Eqn:DiameterCluster}
and (b) standard deviation $R_k$ as defined in \protect\eqref{Eqn:StandardDeviationCluster} 
 like figure \protect\ref{Fig:Figure09}, but for the mixing layer.
Note that the state space partition is relatively homogeneous.
 }
 \label{Fig:Figure20}
\end{figure}

The clustering of the snapshots gives a relatively homogeneous partition of the state space 
as confirmed by the cluster diameters and the cluster standard 
deviations (see figure~\ref{SubFig:Figure20a} and~\ref{SubFig:Figure20b}). 

\begin{figure}
\psfrag{eval}[cc][][1][-90]{$\lambda_2^l$}
\psfrag{50}[cc][][1][0]{$50$}
\psfrag{100}[cc][][1][0]{$100$}
\psfrag{150}[cc][][1][0]{$150$}
\psfrag{-5}[cc][][1][0]{$-5$}
\psfrag{-15}[cc][][1][0]{$-15$}
\psfrag{-30}[cc][][1][0]{$-30$}
\psfrag{-0.5}[cc][][1][0]{$-0.5$}
\psfrag{0.5}[cc][][1][0]{$0.5$}
\psfrag{1.5}[cc][][1][0]{$1.5$}
 \centering
 \centerline{\raisebox{0.15\textheight}{a)}\hspace{1cm}\includegraphics[scale = 1, trim = 0 0.75cm 0 0, clip = true]{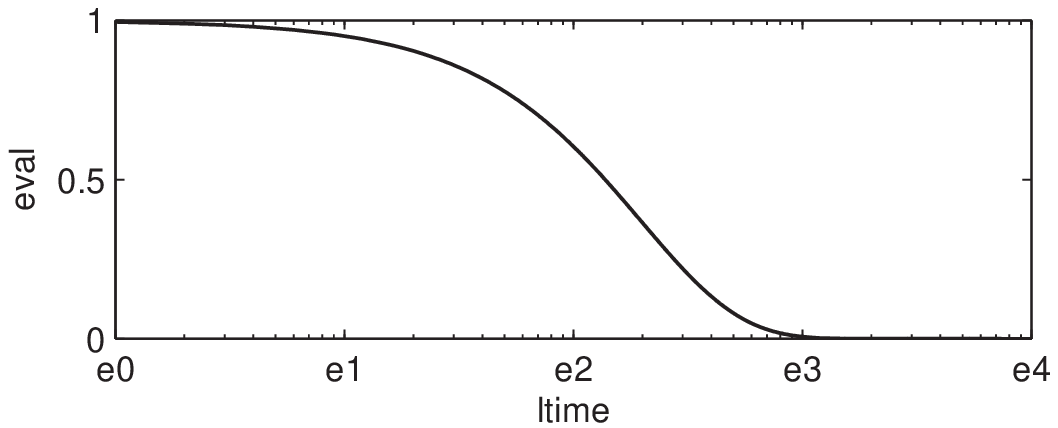}}
 \vfill
 \centerline{\raisebox{0.15\textheight}{b)}\hspace{0.9cm}\includegraphics[scale = 1, trim = 0 0.8cm 0 0]{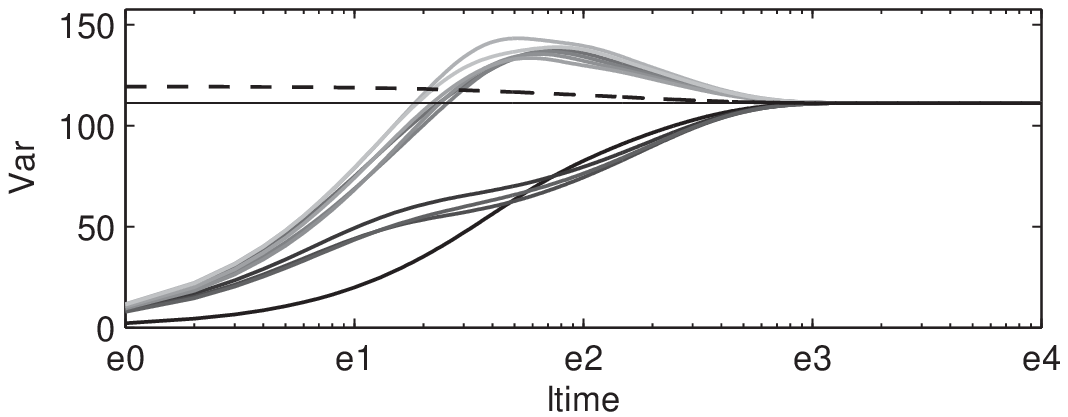}}
 \vfill
 \centerline{\raisebox{0.15\textheight}{c)}\hspace{1cm}\includegraphics[scale = 1, trim = 0 0.75cm 0 0]{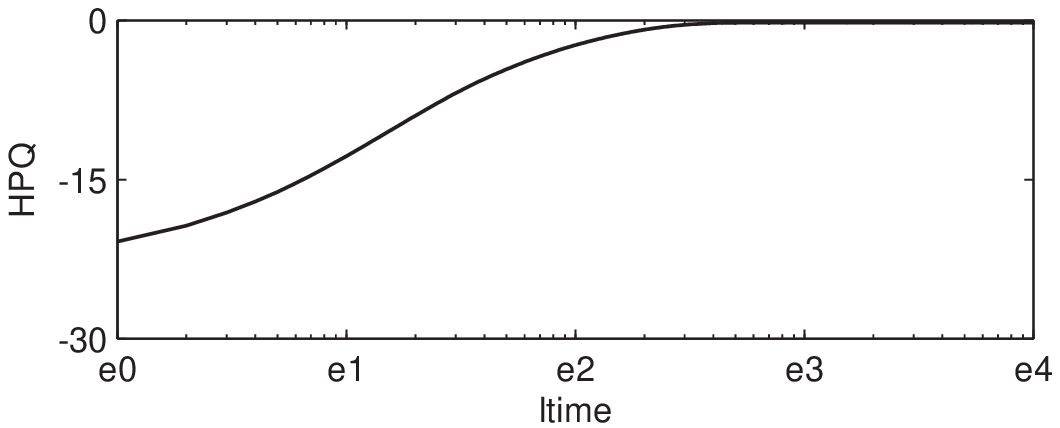}}
 \vfill
 \centerline{\raisebox{0.18\textheight}{d)}\hspace{1cm}\includegraphics[scale = 1]{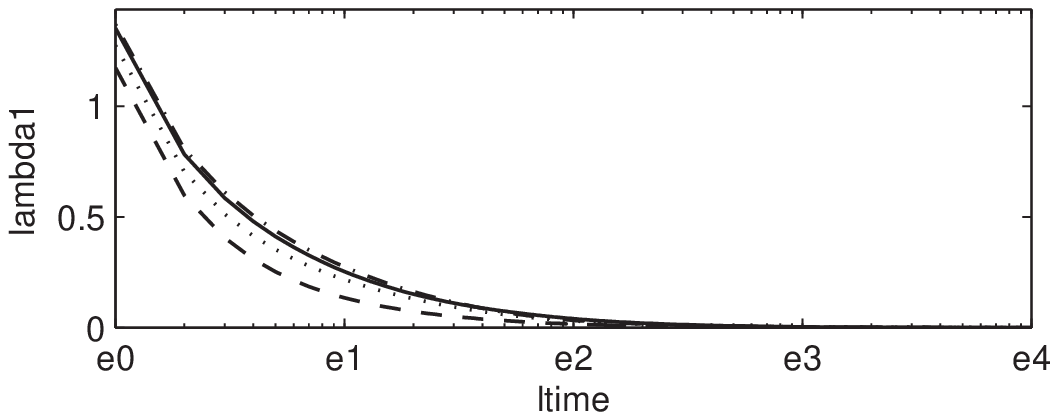}}
 \caption{
 Analogously to figure~\protect\ref{Fig:Figure10},
 CROM convergence study for the mixing layer.
Convergence with respect to the number of iterations $l$ is illustrated 
(a) for the second eigenvalue modulus $\vert\lambda_2\vert^l = \vert 0.995\vert^l$ of $\slsP$,
(b) for the variance $R^2$ \protect\eqref{Eqn:Variance} 
for different initial distributions of a pure cluster $p_k^0 =\delta_{jk}$
with $j=1$ (black), $j$ is a cluster of 'VP' (light grey), and $j$ is a cluster of 'KH' (dark grey),
(c) for the Kullback-Leibler entropy $H$ \protect\eqref{Eqn:KullbackLeiblerEntropy}, and
(d) for the FTLE $\lambda^l_{FTLE}$ \protect\eqref{Eqn:FTLE} 
for different initial distances of the mixing layer.
Note that the converged value for $R^2$ is indicated with the horizontal thin line.
Evidently all quantities converge at about $l = 10^3$, i.e.\ each of the measures can be used as convergence indicator.
}
 \label{Fig:Figure21}
\end{figure}

Figure~\ref{Fig:Figure21} displays the convergence of the second eigenvalue modulus, 
the variance, the Kullback-Leibler entropy, and the FTLE. 
All converge at about $l=10^3$.
The grey lines in figure~\ref{Fig:Figure21}\,(b) correspond to initial distributions with unit probability, 
$p_k^0=\delta_{kj}$ for $j=5,6,7,8,9,10$, 
in one of the vortex pairing states.
These distributions spread out faster and have a larger variance at an intermediate stage than 
those belonging to Kelvin-Helmholtz states which are displayed in black lines. 
The variance of an initial equipartition (dashed lines) is already close to the converged value. 
Nevertheless, all distributions converge after the same number of iterations.
The above results confirm a good state space partition 
and convergence behaviour of the transition matrix.

In general, the performance of a model can be evaluated by
how well and how long it is able to predict future states.
Here, we are interested in how long the model is capable of 
predicting future states from the initial conditions.
We can conclude from the convergence study that
the prediction horizon must be related to the number of iterations
until the asymptotic state is reached.
In the following, 
we define the model's prediction horizon
in terms of the Kullback-Leibler entropy as we have shown it is 
a suitable measure for the convergence of the model for the Lorenz attractor 
as well as for the mixing layer
(compare figures~\ref{Fig:Figure10} and \ref{Fig:Figure21}).

\begin{figure}
\centering
\setlength{\unitlength}{1mm}
\begin{picture}(120,80)
 \put(60,2){$\alpha_1$}
 \put(18,39){$\alpha_2$}
 \psfrag{15}[cc][][1][0]{\raisebox{0pt}{$15$}}
 \psfrag{10}[cc][][1][0]{\raisebox{0pt}{$10$}}
 \psfrag{5}[cc][][1][0]{\raisebox{0pt}{$5$}}
 \psfrag{0}[cc][][1][0]{\raisebox{0pt}{$0$}}
 \psfrag{-5}[cc][cc][1][0]{\raisebox{0pt}{$-5$}\hspace{0.25cm}}
 \psfrag{-10}[cc][cc][1][0]{\raisebox{0pt}{$-10$}\hspace{0.25cm}}
 \psfrag{-15}[cc][cc][1][0]{\raisebox{0pt}{$-15$}\hspace{0.25cm}}
 \put(25,5){\includegraphics[scale = 1]{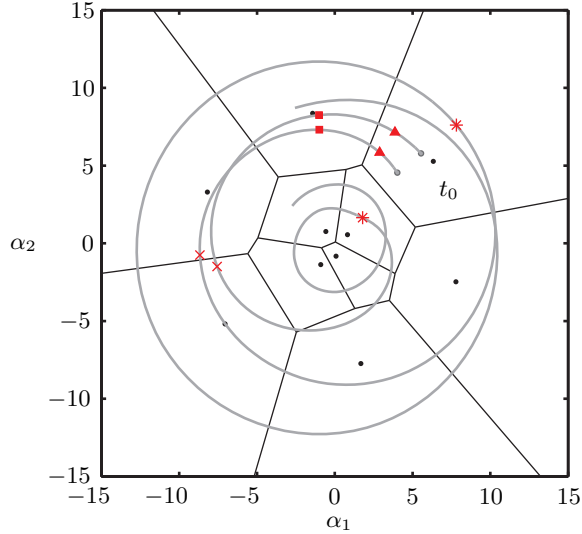}}
  \put(75,46){$t_0$}
\end{picture}
 \caption{
 Prediction horizon of the mixing layer:
 Two different initially close
 trajectories (grey lines) obtained from the integration of the Galerkin system 
 with a non linear eddy-viscosity model as in \cite{Cordier2013ef} for $N=20$
 are displayed. The initial condition are visualised as $\color{gray}{\bullet}$.
 The prediction horizon $t_H$ defined in \eqref{Eqn:PredictionHorizon}
 is  displayed as ${\color{red}\boldsymbol{\ast}}$. 
 Other values correspond to 
 $\frac{1}{4}\, t_H$ (${\color{red}\blacktriangle}$),
 $\frac{1}{2} \, t_H$ (${\color{red}\blacksquare}$), and
 $\frac{3}{4} \, t_H$ (${\color{red}\boldsymbol{\times}}$).
 The clusters are displayed as a Voronoi diagram (black lines)
 where ${\bullet}$ corresponds to a centroid.
 The two-dimensional visualisation is obtained as explained in appendix~\ref{Sec:AppD:Visualisation}.
 The diffusion of the probability distribution is related to the divergence of trajectories
 of the dynamical system.}
 \label{Fig:Figure22}
\end{figure}

In figure~\ref{Fig:Figure22}, exemplarly, 
two trajectories (grey lines) are displayed for two different but closely located initial conditions
(grey bullets) obtained from the integration 
of the Galerkin system for the mixing layer \citep{Cordier2013ef}. 
The thin black lines display the borders of the Voronoi cells as obtained from the clustering of the data
and the black bullets represent the corresponding centroids.
For the two-dimensional visualisation, 
the centroids and the data points are projected onto the first two POD modes obtained from the 
centroids' covariance matrix as described in appendix~\ref{Sec:AppD:Visualisation}.

Let the prediction horizon $t_H:=l_{0.9}$ be defined by
\begin{equation}
\label{Eqn:PredictionHorizon}
 H(\slsP^{t_H}, \slsP^{\infty}) - H(\slsP^1, \slsP^{\infty}) 
 \approx 0.9\; \left(H(\slsP^{\infty}, \slsP^{\infty}) - H(\slsP^1, \slsP^{\infty})\right)
\end{equation}
in terms of the Kullback-Leibler entropy $H(\slsP^{l}, \slsP^{\infty})$ with the transition matrix $\slsP$ as determined  
for the mixing layer (see figure~\ref{SubFig:Figure15b}). 
Then, the red markers 
correspond to $\gamma\,t_H$ with $\gamma=\frac{1}{4},\frac{1}{2},\frac{3}{4},1$.

The initial conditions of the two trajectories 
are in the same cluster which is associated with one shedding regime.
The trajectories evolve similarly in the state space until about $\frac{3}{4}t_H$
when they start to diverge and one trajectory switches to the other shedding regime.
This divergence of trajectories is associated with a 
diffusion of the probability distribution.
After exceeding the prediction horizon, 
the probability distribution has converged to the model's asymptotic state and
initial conditions, i.e., the initial cluster 
can not be recovered anymore from backwards iteration. 
In the asymptotic state, 
the invariant distribution corresponds to the fraction of time 
the trajectory spends in each cluster.

\begin{figure}
\begin{center}
\setlength{\unitlength}{1mm}
\subfigure[]{
\begin{picture}(60,55)
 \put(0,8){\rotatebox{90}{\figurefont{Total number of clusters $K$}}}
 \put(5,7){\vector(0,1){43}}
 \put(7,6){$K=1$}
 \put(20,6){\parbox{3.5cm}{\figurefont{No resolution}}}
 \put(20,14){\parbox{3.5cm}{\figurefont{Main transition mechanism}}}
 \put(7,26){$K = K_c$}
 \put(20,26){\parbox{3.5cm}{\figurefont{Good compromise between resolution and simple structure of $\slsP$}}}
 \put(20,40){\parbox{3.5cm}{\figurefont{Refined state resolution}}}
 \put(7,50){$K=M$}
 \put(20,50){\parbox{3.5cm}{\figurefont{Maximum resolution}}}
\end{picture}
\label{SubFig:Figure23a}}
\hfill
\subfigure[]{
\begin{picture}(60,55)
 \put(0,8){\rotatebox{90}{\figurefont{Time step $\Delta t$ of data}}}
 \put(5,7){\vector(0,1){43}}
 \put(7,6){$\Delta t\rightarrow 0$}
 \put(25,6){\figurefont{noise}}
 \put(25,14){\parbox{3.5cm}{\figurefont{Small scale resolution}}}
 \put(7,26){$\Delta t= \Delta t_{opt}$}
 \put(25,26){\parbox{3.5cm}{\figurefont{Good compromise between resolution and length of prediction horizon}}}
 \put(25,40){\parbox{3.5cm}{\figurefont{Stroboscopic view of clusters}}}
 \put(7,50){$\Delta t=T$}
 \put(25,50){\parbox{3.5cm}{\figurefont{Sampling time}}}
\end{picture}
\label{SubFig:Figure23b}}
\end{center}
\caption{
Impact of parameter variation:
  The influence of (a) the number of clusters $K$ 
  and (b) the time step $\Delta t$ of the data on the resulting CROM.
  The time length $T$ (physical time length of measurement) of the data is assumed to be constant which
  allows a variation of the number of observations $M$ as a result of 
  a variation of $\Delta t$.
}
\label{Fig:Figure23}
\end{figure}
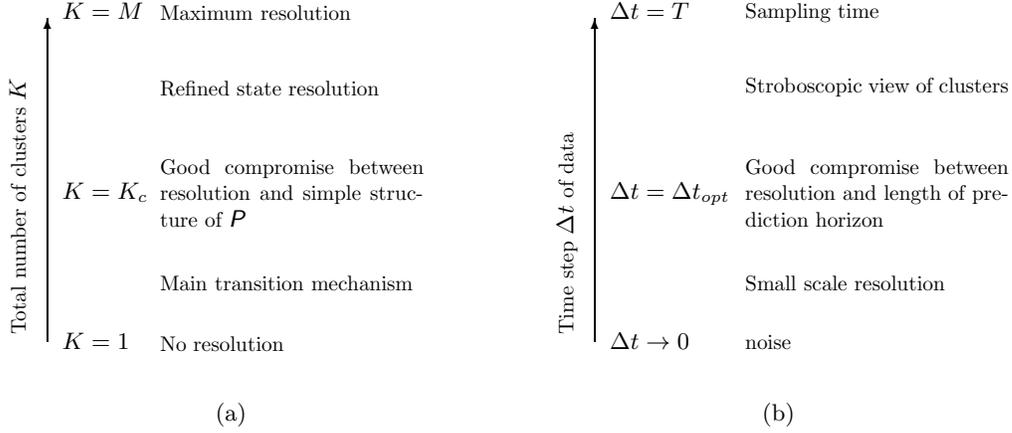
The diffusion of probability, 
which is directly linked to the prediction horizon,
is induced by the cluster size and the time step.
Figure~\ref{Fig:Figure23} serves as a guide for the
following discussion.

For a fixed number of clusters $K$ the prediction horizon can be
optimised by varying the time step $\Delta t$ of the data.
The trajectory spends a long time in each cluster and 
only few transitions to other clusters occur.
The large ratio of inter-cluster transitions
to intra-cluster transitions introduces uncertainty in the model.
This ratio can be decreased by
a suitable choice for the time step
such that, ideally, a stroboscopic view of the clusters is obtained.
We consider as example a limit cycle 
which is discretised into $K$ phase bins 
of equal size with centroids $\boldsymbol{c}_k$ 
corresponding to the phase $2\pi k/K$, $k=1,\ldots,K$. 
Then, the ideal time step would be $2\pi/K$ for which
CROM yields a transition matrix 
with minimised diffusion ($P_{jj}\rightarrow0$ while $P_{jj+1}\rightarrow 1$ for $j=1,\ldots,K-1$)
and maximised prediction horizon.
A necessary upper bound for $\Delta t$ is determined by the required
time-resolution of the data for the estimation of the CTM.
The situation is more complex if several frequencies occur in the data
for which the optimal time step $\Delta t$ is a good compromise
between a stroboscopic view for one frequency and 
a sufficient resolution of the higher frequency.

The prediction horizon can also be optimised with respect to the 
chosen number of clusters $K$.
A larger number of clusters reduces the ratio of inter-cluster transitions
to intra-cluster transitions resulting in a longer prediction horizon.
The lower bound is given by $K=1$ for which
all snapshots $\boldsymbol{u}^m$ 
are in one cluster and its centroid represents the mean flow $\boldsymbol{u}_0$.
It should be noted that $K$ is limited by the number of snapshots $M$,
as $K\times K$ CTM elements are estimated from $M-1$ snapshot transitions.
The resulting CTM for $K=M$ is completely deterministic having entries $P_{jj+1}=1$ 
for $j=1,\ldots,K-1$. 
Its last state $\boldsymbol{c}_K = \boldsymbol{c}_M$ is absorbing, 
i.e. it is not possible to leave this state, since 
there is no transition to the first snapshot of the data.
A good compromise between the resolution of the main transition mechanism
and a refined state resolution can be found by the elbow criterion.

In summary, CROM resolves the two prominent dynamical regimes of the mixing layer, 
Kelvin-Helmholtz vortices and a flow governed by vortex pairing.
The centroids give a revealing picture of the evolution of the flow structures and 
the intermediate role of a
'flipper cluster'
is discovered.
The CROM naturally filters out the dominant wavenumbers 
associated with the characteristic features of the flow.  
The data partitioning is relatively homogeneous 
and the convergence behaviour is in good agreement 
with the result of the statistical analysis of the clustered data.
Increasing the cluster number $K$ 
does not change the principal picture 
of two dynamical regimes with a transition region.
However, one may obtain a refined resolution 
of the metamorphosis from one state to the other.
CROM has a built-in uncertainty due to the cluster size 
which results in a diffusion of probability.
As a consequence, 
the model converges to an asymptotic state
which defines its prediction horizon.
The snapshot time step $\Delta t$ 
has no strong effect on the observed oscillation and transition mechanisms 
as long as it is small compared to the characteristic period.
However, adjusting the time step may improve the prediction horizon.
In contrast,
POD models can be integrated 
infinitely long after exceeding their prediction horizon
leading to erroneous model predictions.
CROM's asymptotic state resembles the discretised
ergodic measure on the attractor.
As such it has a physical meaning 
and provides additional information on the flow data
such as the identification of transients.

\section{Interpretation of CROM as a generalised Ulam-Galerkin method}
\label{Sec:UlamGalerkinMethod}
For a complex dynamical system, 
a probabilistic description corresponding to the evolution of 
a swarm of points in the state space is often more 
insightful than the track of an individual trajectory \citep{Cvitanovic2012book}.
One of the reasons is that 
the time evolution of probability densities 
is governed by a linear
evolution operator called Perron-Frobenius \citep{Lasota1994book}. 
It can be shown \citep{PhysRevE.51.74} that 
this operator is directly related to a Liouville equation 
which describes the evolution of a probability density function (PDF) 
of the state variables. 
In this section, 
we demonstrate that, from a theoretical point of view, 
CROM can be considered as a generalisation of 
the Ulam-Galerkin method \citep{Li1976jat}. 
This method is classically utilised for the approximation 
of the Perron-Frobenius operator 
associated with the Liouville equation \citep{Bollt2013book}. 
For more information on the relationship 
between the Liouville equation and the Perron-Frobenius operator
as well as its adjoint, the Koopman operator, 
the reader is referred to appendix~\ref{Sec:AppC:Comparison-KoopmanModes}.

The evolution equation for the mode coefficient vector $\boldsymbol{a}$
can be derived from a Galerkin projection onto the Navier-Stokes equation \citep{Noack2011book}.
This projection leads to the system
\begin{equation}
\label{Eqn:GalerkinSystem}
\frac{d\boldsymbol{a}}{dt}=\boldsymbol{f}(\boldsymbol{a})
\end{equation} 
with a constant-linear-quadratic propagator
$\boldsymbol{f}:= [f_1,f_2,\ldots,f_N]^T$ with $f_i = c_i + \sum_{j=1}^N l_{ij}a_j + \sum_{j,k=1}^N q_{ijk} a_j a_k$. As it is well known \citep{PhysRevE.51.74}, \eqref{Eqn:GalerkinSystem} induces a Liouville equation for the evolution of the probability density of the POD coefficients
\begin{equation}
\frac{\partial}{\partial t} p(\boldsymbol{a},t)
=
- \nabla_{\boldsymbol{a}} \cdot 
\left[ \boldsymbol{f}(\boldsymbol{a})p(\boldsymbol{a},t) \right]
= \hat{\slsL}p(\boldsymbol{a},t)
\label{Eqn:LiouvilleEquation}
\end{equation}
where $\hat{\slsL}$ is the Liouville operator.
Here and in appendix~\ref{Sec:AppC:Comparison-KoopmanModes}, 
we adopt the symbol $\hat{}$ to denote operators.

The aim of the Ulam-Galerkin method \citep{Bollt2013book} is to 
determine a finite-rank approximation of the Perron-Frobenius operator 
via a Galerkin projection of the Liouville equation 
\eqref{Eqn:LiouvilleEquation} onto a set of basis functions $\{\Psi_k\}_{k=1}^N$. 
Let $\Omega_k$,  $k=1,\ldots,K$ represent a 
Voronoi tessellation of the state space 
such that $\Omega=\cup_{k=1}^K\Omega_k$ with $\Omega_k\cap\Omega_j=\emptyset$  for $k\neq j$. 
The basis functions for the Ulam-Galerkin method 
are a family of characteristic functions
\begin{equation}
\Psi_k(\boldsymbol{a}) = 
\begin{cases}
1/|\Omega_k|^{1/2} & \text{if $\boldsymbol{a} \in \Omega_k$},\\
0 & \text{otherwise},
\end{cases}
\label{Eqn:PDF_TestFunction}
\end{equation}
where $|\Omega_k|:=\int_{\Omega_k}\!\mathrm{d}\boldsymbol{a}$ denotes the volume of the $k-$th cell.

The PDF $p(\boldsymbol{a},t)$ 
shall be approximated by a modal expansion
\begin{equation}
 p (\boldsymbol{a},t) = \sum\limits_{k=1}^K \rho_k(t) \> \Psi_k(\boldsymbol{a}),
\label{Eqn:PDF_expansion}
\end{equation} 
where $\rho_k(t)=\int_{\Omega_k} d \boldsymbol{a} \; p(\boldsymbol{a},t)\Psi_k(\boldsymbol{a})$ 
so that $\sum_{k=1}^K\, |\Omega_k|^{1/2}\,\rho_k = 1$.

Substituting \eqref{Eqn:PDF_expansion} in \eqref{Eqn:LiouvilleEquation}, 
multiplicating with the basis function 
$\Psi_j(\boldsymbol{a})$
and integrating over $\Omega$
yields
\begin{equation}
\frac{d}{dt} \rho_j  
= \sum\limits_{k=1}^K \mathcal{P}_{jk}^{cont} \> \rho_k, \quad  j=1,\ldots,K,
\label{Eqn:LiouvilleVFM}
\end{equation}
where
\begin{equation*}
\mathcal{P}_{jk}^{cont} 
=
\int\limits_{\Omega} \! d\boldsymbol{a} \> \Psi_j(\boldsymbol{a}) \hat{\slsL}\Psi_k(\boldsymbol{a}).
\end{equation*}

$\mathcal{P}^{cont}=(\mathcal{P}_{jk}^{cont})$ corresponds to the 
finite-rank approximation of the Perron-Frobenius operator
which is normalised such that $\sum_{j=1}^K \mathcal{P}_{jk}^{cont} =1$.
If only observations $\{\boldsymbol{a}^m\}_{m=1}^M$ 
of the discrete map 
\begin{equation}
 \boldsymbol{a}^{m+1} = \slsF(\boldsymbol{a}^m)
\end{equation}
are available, 
it is commonly accepted in the Ulam-Galerkin framework that $\mathcal{P}_{jk}^{cont}$ 
can be approximated by
\begin{equation}
\mathcal{P}_{jk} 
= 
\frac{\text{card}\left\{\boldsymbol{a}^m\,\vert\,\boldsymbol{a}^m\in\Omega_j\,
\text{and}\,\boldsymbol{F}(\boldsymbol{a}^m)\in\Omega_k\right\}}{
\text{card}\left\{\boldsymbol{a}^m\in\Omega_j\right\}}
\label{eq:ulam_galerkin}
\end{equation}
where $\text{card}\left\{\cdot\right\}$ denotes 
the cardinality of a set.
Then,
the $jk-$th entry of the matrix $\mathcal{P}^{cont}$ 
can be interpreted as the ratio of the fraction of the cell volume $\Omega_j$ 
that will be mapped inside the cell $\Omega_k$
after one action of the map $\slsF$ to the cell volume of $\Omega_j$
\citep{Santitissadeekorn2007pd,Froyland2013jna}.
In the appendix A of \citet{Bollt2013book}, 
an implementation of the Ulam-Galerkin method is described where \eqref{eq:ulam_galerkin} 
is used in conjunction with a tessellation of the state space 
to approximate the Perron-Frobenius operator. 

The cluster transition matrix $\slsP$ as 
determined by \eqref{Eqn:TPM-definition} is
another finite-rank approximation of the Perron-Frobenius operator. 
From the above description, 
CROM can be considered as an extension of the 
classical Ulam-Galerkin method for which 
the tessellation of the state space is 
not obtained from a 
Delaunay triangulation relying only on geometrical principles but 
from a cluster analysis based on the k-means algorithm. 
Thus,
CROM is more flexible since 
the cluster analysis can be easily modified 
by changing the definition of the distance metric used in the analysis.


\section{Discussion}
\label{Sec:Discussion}
\begin{figure}
\figurefont
\psfrag{CROM}[cc][][1][0]{\figurefont{\textbf{CROM}}}
\psfrag{POD GM}[cc][][1][0]{\figurefont{\textbf{POD GM}}}
\psfrag{u1}[cc][][1][0]{\raisebox{0pt}{$\boldsymbol{u}_1$}}
\psfrag{u3}[cc][][1][0]{\raisebox{0pt}{$\boldsymbol{u}_3$}}
\psfrag{c5}[cc][][1][0]{\raisebox{0pt}{$\boldsymbol{c}_5$}}
\psfrag{c7}[cc][][1][0]{\raisebox{0pt}{$\boldsymbol{c}_7$}}
\psfrag{c9}[cc][][1][0]{\raisebox{0pt}{$\boldsymbol{c}_9$}}
\psfrag{a1}[cc][][1][0]{\raisebox{0pt}{$a_1$}}
\psfrag{aN}[cc][][1][0]{\raisebox{0pt}{$a_N$}}
\psfrag{...}[cc][][1][0]{\raisebox{0pt}{$\dots$}}
\psfrag{Snapshot ensemble}[cc][][1][0]{Snapshot ensemble}
\psfrag{State space}[cc][][1][0]{State space}
\psfrag{compression}[cc][][1][0]{compression}
\psfrag{Cluster analysis}[cc][][1][0]{Cluster analysis}
\psfrag{POD expansion}[cc][][1][0]{POD expansion}
\psfrag{POD}[cc][][1][0]{\figurefont POD}
\psfrag{Kinematics}[cc][][1][0]{Kinematics}
\psfrag{DEFck}[cc][][1][0]{$\boldsymbol{c}_k \!= \!\!\frac{1}{n_k}\!\!\!\sum\limits_{m=1}^M  \!\!\! \slsT_{km} \boldsymbol{v}^m$}
\psfrag{DEFui}[cc][][1][0]{$\boldsymbol{u}' = \sum\limits_{i=1}^N  \> a_{i} \boldsymbol{u}_i$}
\psfrag{Statistical analysis}[cc][][1][0]{Statistical analysis}
\psfrag{Markov model}[cc][][1][0]{Markov model}
\psfrag{Galerkin}[cc][][1][0]{Galerkin}
\psfrag{projection}[cc][][1][0]{projection}
\psfrag{Dynamics}[cc][][1][0]{Dynamics}
\psfrag{Transition}[cc][][1][0]{Transition}
\psfrag{dynamics}[cc][][1][0]{dynamics}
\psfrag{Temp. nonlinear}[cc][][1][0]{Nonlinear}
\psfrag{interaction}[cc][][1][0]{interaction}
\psfrag{plPlp0}[cc][][1][0]{$\frac{\mathrm d}{\mathrm dt}\boldsymbol{p} = \slsP_F \boldsymbol{p}$}
\psfrag{ddtaFa}[cc][][1][0]{$\frac{\mathrm d}{\mathrm dt}\boldsymbol{a} = \boldsymbol{f}\left(\boldsymbol{a}\right)$}
\psfrag{Physical}[cc][][1][0]{Physical}
\psfrag{mechanisms}[cc][][1][0]{mechanisms}
\psfrag{Probabilistic}[cc][][1][0]{Probabilistic}
\psfrag{Deterministic}[cc][][1][0]{Deterministic}
\psfrag{Linear evolution}[cc][][1][0]{Linear evolution}
\psfrag{equation for}[cc][][1][0]{equation for}
\psfrag{probability}[cc][][1][0]{probability}
\psfrag{distribution}[cc][][1][0]{distribution}
\psfrag{Preserves nonlinear}[cc][][1][0]{Preserves nonlinear}
\psfrag{dynamics of PDE}[cc][][1][0]{dynamics of PDE}
\psfrag{Control}[cc][][1][0]{Control}
\psfrag{Control design}[cc][][1][0]{Control design}
\psfrag{for ensemble of}[cc][][1][0]{for ensemble of}
\psfrag{trajectories}[cc][][1][0]{trajectories}
\psfrag{Standard control}[cc][][1][0]{Standard control}
\psfrag{and observer}[cc][][1][0]{and observer}
\psfrag{design}[cc][][1][0]{design}
\psfrag{dpdt=Pf+Pgp}[cc][][1][0]{$\frac{\mathrm d}{\mathrm dt}\boldsymbol{p} = [ \slsP_F + \slsP_G ] \boldsymbol{p}$}
\psfrag{dadt=fa+ga}[cc][][1][0]{$\frac{\mathrm d}{\mathrm dt}\boldsymbol{a} = \boldsymbol{f}\left(\boldsymbol{a}\right) + 
\slsB\boldsymbol{b}$}
 \centering
 \includegraphics[scale=0.7]{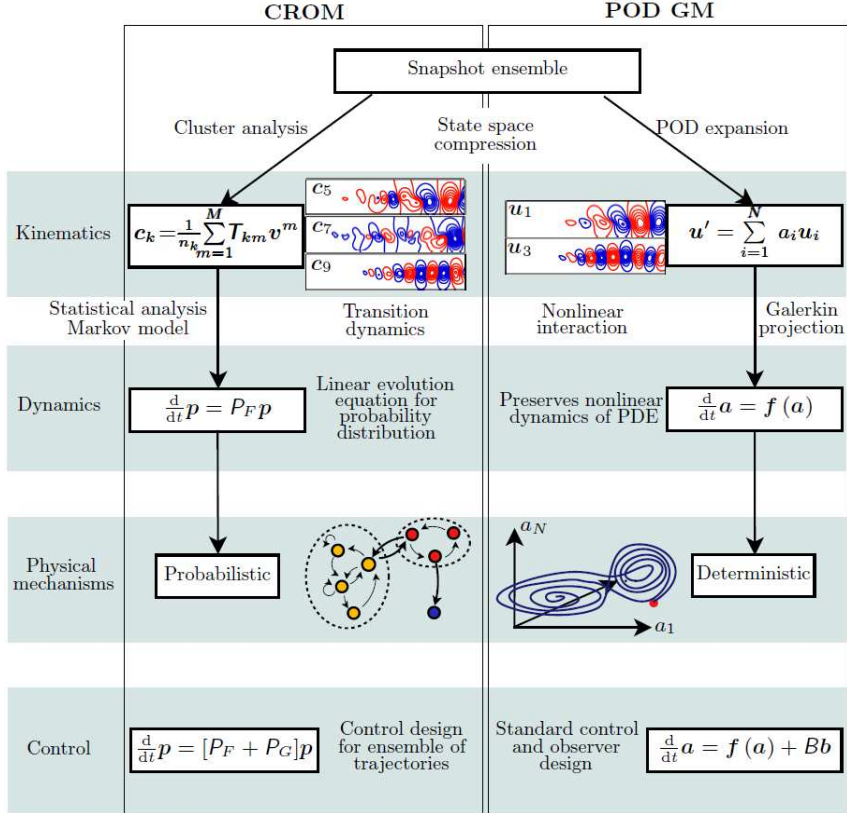}
 \caption{CROM versus POD models. For details see text.}
 \label{Fig:Figure24}
\end{figure}

The derivation of low-dimensional models in fluid dynamics
is driven by 
the need to understand nonlinear mechanisms in turbulent flows
and their promising potential for control purposes.
Different perspectives, 
e.g. a deterministic and a probabilistic viewpoint,
can complement each other 
to give a broader picture
on the governing dynamical system. 
CROM yields a probabilistic dynamical model 
for the evolution of the state variables.
By benchmarking it against a deterministic POD model,
we can identify their respective
capabilities and challenges as a consequence 
from their methodological difference.
The POD model serves --- pars pro toto ---
as an example for many other
data-driven models \citep{Hyvaerinen2012ptrsa,Kutz2013book}.
Figure~\ref{Fig:Figure24} and table~\ref{Tab:Table1} 
serve to guide the following discussion.

POD models and CROM share
many methodological similarities
suggesting a critical assessment 
of their relative strengths and weaknesses.
The similarities include the following:
(1) Both reduced-order models (ROM)
are based on velocity snapshots.
(2) The ROMs yield a small number of modes
from which the full state is approximated.
(3) These modes (POD or centroids) are linear combinations 
of the original snapshots.
(4) The temporal evolution of the coarse-grained state 
is described by an autonomous system.
For a better comparability, 
the continuous-time Markov process 
\eqref{Eqn:ProbL0Continuously} is considered for CROM.

Yet, there are also important differences between POD models and CROM.
POD models minimise the averaged residual
of the corresponding Galerkin expansion.
This expansion defines a linear subspace
parametrised by mode coefficients $\boldsymbol{a}=\left[a_1,\ldots,a_N \right]^T$.
The individual mode $\boldsymbol{u}_i$ is just a basis vector.
As such it has \textit{per se} no physical meaning
and may not be close to any flow realisation.
The uncertainty, i.e.\ the residual of a POD expansion, 
is in the orthogonal complement of the linear POD space.
The notion of a modal expansion and orthogonality implies
that POD models are rooted in Hilbert spaces with an inner product.

In contrast, 
CROM centroids $\boldsymbol{c}_k$ minimise the total cluster variance, 
i.e.\ each centroid shall approximate 
all snapshots of the corresponding cluster.
The minimal uncertainty of the centroidal flow representation 
is quantified by the 'radius' of the cluster,
i.e.\ the maximum distance between the centroid 
and all snapshots of the corresponding cluster.
Evidently, CROM requires a metric space with distance definition 
and no Hilbert space with an inner product.

The difference between a POD model and CROM 
becomes very apparent for periodic flow.
For such data, 
POD generally yields cosine- and sine-like wave pairs 
for the first and higher frequency harmonics 
\citep{Deane1991pfa,Rempfer1994jfm2,Noack2005jfm}.
In contrast, 
CROM discretises the limit cycle into sectors,
i.e.\ phase bins.

POD and CROM have a different weighting of rare short-term events.
POD performs a statistical weighting of events, 
i.e.\  rare events tend to be ignored if their contribution
to the fluctuation energy is small.
In contrast, CROM is better conceptualised as a geometric
partitioning covering all snapshots 
and less dependent on the frequency of events.

Another important difference lies in the state spaces 
spanned by POD coefficients or by cluster probabilities.
A POD model yields a deterministic description
in the corresponding subspace.
In principle, 
the mode coefficient vector  $\boldsymbol{a}$ 
can assume arbitrary values by construction.
In contrast, 
CROM state descriptions range from the specification
of a pure cluster, \eg
${\boldsymbol{p}}= \left [\delta_{k1}, \ldots, \delta_{kK} \right]^T$
for cluster $k$,
to the ergodic measure for all clusters,
i.e.\ a statistically stationary state.
By construction, the cluster probabilities are bounded: 
by $0 \le p_k $ and $\sum_{k=1}^K p_k = 1$.

The Galerkin system
\eqref{Eqn:GalerkinSystem}
can be derived from a projection onto the Navier-Stokes equation.
This projection leads
to a constant-linear-quadratic propagator
$\boldsymbol{f}$
in the dynamical system \citep{Holmes2012book}.
The propagator may also be identified 
directly from data \citep{Cordier2010ijnmf}, like in CROM.
In contrast, 
the Markov model of CROM is a linear evolution equation.

\begin{table}
  \begin{center}
\def~{\hphantom{0}}
  \begin{tabular}{l|p{5.75cm}|p{5.75cm}}
       & \textbf{CROM}  & \textbf{POD Galerkin model}  \\    
      \rule{0pt}{15pt}
      Dynamics & Statistical closure (ergodic measure) with transients and built-in uncertainty (cluster size) &       
      Nonlinear phase dynamics (fixed points, limit cycle, etc.; stability analysis) \\[3pt]
      Transients & Suitable for transient phenomenology &
      Challenges for transient dynamics \\[3pt] 
      Control & Control design for ensemble of trajectories (ensemble control) &
      Standard control design (controller \& observer) \\[3pt]
      Resolution & Complete probabilistic resolution of the whole fluctuation energy &
      Complete deterministic resolution in low-energy subspace \\[3pt] 
      Sensitivity & Sensitive to cluster size and time step\newline
      Computationally robust for long-term integration &
      Sensitive to subscale closures \\
  \end{tabular}
  \caption{
  Capabilities and challenges of CROM and POD models.
  }
  \label{Tab:Table1}
  \end{center}
\end{table}

A linear evolution equation 
describing an unsteady nonlinear Navier-Stokes equation
may appear initially surprising.
The key to the explanation is that a Markov model
describes the evolution of a probability distribution
(ensemble of trajectories),
not the evolution of a single trajectory 
(see appendix~\ref{Sec:AppC:Comparison-KoopmanModes}).
Such evolution equations 
can be derived from the Navier-Stokes equation,
starting with the linear Liouville equation 
of a suitable probability space.
The \citet{Hopf1952jrma} formalism
for the  Navier-Stokes equation is a prominent example.
A simpler version constitutes the Liouville equation 
for a Galerkin system
as in \eqref{Eqn:LiouvilleEquation}.
The reader is referred to \citet{Noack2012jfm} for a detailed discussion.
Thus, CROM is closely aligned with closure schemes,
in which a stable fixed point 
represents the ergodic measure 
for the unsteady attractor in velocity space.

As a final note, we wish to highlight that 
CROM can also be utilised for flow control purposes.
A CROM-based control strategy with the goal to bring the probability distribution
as close as possible to a predefined distribution
is outlined in appendix~\ref{Sec:AppB:TowardsCROMbasedFlowControl}.

\section{Conclusions}
\label{Sec:Conclusions}
In the present study,
a novel cluster-based reduced-order modelling (CROM) strategy is proposed
for identifying physical mechanisms in an unsupervised manner.
CROM is a generalisation of the Ulam-Galerkin method
for the approximation of the Perron-Frobenius operator
and yields a dynamical model for the evolution of an ensemble of trajectories
in terms of a probability distribution.
In contrast to the Ulam-Galerkin method, 
which is based on a Voronoi tessellation of the state space, 
CROM rests on a cluster analysis of the data, e.g. via the k-means algorithm.
The definition of the distance metric enters the approach as a new parameter.
One example is the
observable-inferred decomposition (OID) 
generalising POD for aerodynamic 
or aeroacoustic observables \citep{Schlegel2012jfm}.
CROM can be formulated with the 
distance metric of OID.
Then, the distance definition is based on the observable 
and not the flow itself. 
In principle, one can also think of an entropic distance metric 
taking into account information flux \citep{Simovici2008book} between clusters.

CROM is applied to the Lorenz attractor (in \S3), 
the two-dimensional incompressible mixing layer undergoing vortex pairing (in \S4), 
and the three-dimensional incompressible turbulent wake of an Ahmed body (in appendix A).
For these examples, CROM has been shown
 (1) to identify quasi-attractors like the two shedding regimes of the mixing layer
or the bi-modal states of the Ahmed body, 
and their intrinsic oscillatory or periodical behaviour.
 (2) For periodic flows, the cluster analysis of CROM is a generalisation of phase averaging
 for which the average phase is represented by a centroid 
 and the phase bins are of similar size.
 (3) Main transition processes between those quasi-attractors 
 are characterised by branching regions (Lorenz attractor), 
 periodic transition regions (Ahmed body) or by a flipper cluster like for the mixing layer.
Note that for the mixing layer 
the transition between vortex pairing and Kelvin-Helmholtz shedding 
occurs only during one particular shedding phase.
 (4) An analysis of the spectrum provides additional information 
 on characteristic frequencies of the data and
 the convergence of the model to its asymptotic state, 
 which determines the model's prediction horizon 
 (see appendix~\ref{Sec:AppA:ExampleBroadbandTurbulenceSimpleStructure}).
 (5) The prediction horizon can be defined in terms of the Kullback-Leibler entropy
 measuring the uncertainty of the model compared to its least-informative state. 
 The convergence of the Kullback-Leibler entropy is validated with the 
 rate of convergence obtained from the spectral decomposition.

CROM has a built-in uncertainty due to the cluster size
which results in a diffusion of probability.
Thus, the dynamical model converges to an asymptotic state
which resembles the discretised ergodic measure on the attractor
and yields additional statistical information on the data.
Care has to be taken in the choice of data.
Evidently, the model can only reveal the dynamics that are covered by the data ensemble.
If the data ensemble contains transients, 
they can be easily detected as their asymptotic probability is zero.

A brief discussion shall give an overview of 
CROM's connection to other efforts with regard
to nonlinear flow analysis and state space partitioning.
CROM as an approximation of the Perron-Frobenius operator
is strongly connected to its adjoint, the Koopman operator.
Recently, \citet{Bagheri_JFM2013} 
studied the conditions under which the 
Dynamic Mode Decomposition \citep{{Schmid2010jfm}}
approximates the Koopman modes.
This analytical approach is clearly unthinkable 
for a complex dynamical system, 
and new approaches must be 
considered where CROM can play an important role.

Also, there exists an intriguing analogy 
between the cluster transition model
and the $\epsilon$-entropy for symbolic dynamics \citep{Abel2000prl,Abel2000physd}. 
The symbol $S_k$ in these references corresponds 
to the cluster $\boldsymbol{c}_k$,
 $\epsilon$ to the average distance between the centroids,
and $\Delta t$ to the sampling time $\tau_r$ 
in equation (14) of \cite{Abel2000physd}.
The exit times relate to the transitions between the clusters.

CROM is also connected to ergodic partitioning.
As far back as the 1960s, 
\citet{Kolmogorov1959umn} developed entropic partitioning methods for 
the state space which also include ergodic partitioning.
Ergodic partitioning as used by Mezic's group \citep[see e.g.][]{Mezic1999chaos}
is a concept related to measure-preserving flows 
which can be divided into different sets inside which
the flow is ergodic.
If the attractors are separated in the state space, 
the cluster analysis of CROM is identical to ergodic partitioning.
If not, ergodic partitioning will yield different results.

Summarising,
CROM sufficiently exposes the characteristic 
features of the attractor and identifies quasi-attractors
from the data.
The dynamical connection between 
different flow states 
can be uncovered and 
singular events, meta-stable states, bi-modality, 
hysteresis of several states, multi-attractor behaviour \etc may be detected.
As such, CROM is not restricted to fluid flows. 
CROM is just based on a sequence of observations 
with distance measure. 
The distance metric enters as an additional parameter
allowing, e.g., entropic or OID-based formulations of CROM.
There are no restrictions on the type of data, 
e.g.\ time series, snapshots, \etc, 
or the system from which they originate. 
CROM provides tools to 
identify 
transition regions, which can be crucial for mixing enhancement,
and precursors 
that might lead to undesirable events.
Thus, critical states can be determined 
which have to be manipulated in order to avoid those events.
A promising CROM-based flow control strategy employs linear control laws
taking into account nonlinear actuation dynamics.
In future work, the authors will
test CROM for flow control applications.

\begin{acknowledgments}
\section*{Acknowledgements}
The authors acknowledge the funding and excellent working conditions 
of the  Chair of Excellence
'Closed-loop control of turbulent shear flows 
using reduced-order models' (TUCOROM)
supported by the French Agence Nationale de la Recherche (ANR)
and hosted by Institute PPRIME.
This work is also supported by the NSF PIRE grant OISE-0968313.
EK also thanks for the support by the region Poitou-Charentes, France.
MS would like to acknowledge the support of the LINC
project (no. 289447) funded by EC's Marie-Curie ITN program
(FP7-PEOPLE-2011-ITN). 
We thank the Ambrosys GmbH (Society for Complex Systems Management)
and the Bernd Noack Cybernetics Foundation for additional support.

We appreciate valuable stimulating discussions
with the TUCOROM team and others:
Diogo Barros,
Jacques Bor\'ee, 
Jean-Paul Bonnet,
Steven Brunton,
Jo\"el Delville, 
Thomas Duriez, 
Nicolai Kamenzky, 
Nathan Kutz,
Jacques Lewalle,
Jean-Charles Laurentie,
Marek Morzy\'nski,
Michael Schlegel,
Vladimir Parezanovic,
and Gilles Tissot.
Last but not least, we thank the referees
for many important suggestions.

Special thanks are due to Nadia Maamar 
for a wonderful job in hosting the TUCOROM visitors.
\end{acknowledgments}


\bibliographystyle{jfm}

\bibliography{CROM}


\appendix
\section{Example with broadband turbulence and simple CROM structure}
\label{Sec:AppA:ExampleBroadbandTurbulenceSimpleStructure}
\begin{figure}
\psfrag{x}[cc][][1][0]{\figurefont $x$}
\psfrag{y}[cc][][1][0]{\figurefont $y$}
\psfrag{z}[cc][][1][0]{\figurefont $z$}
 \centering
 \subfigure[]{
 \includegraphics[width = 0.07\textwidth, trim = 23cm 9cm 26cm 11cm, clip = true]{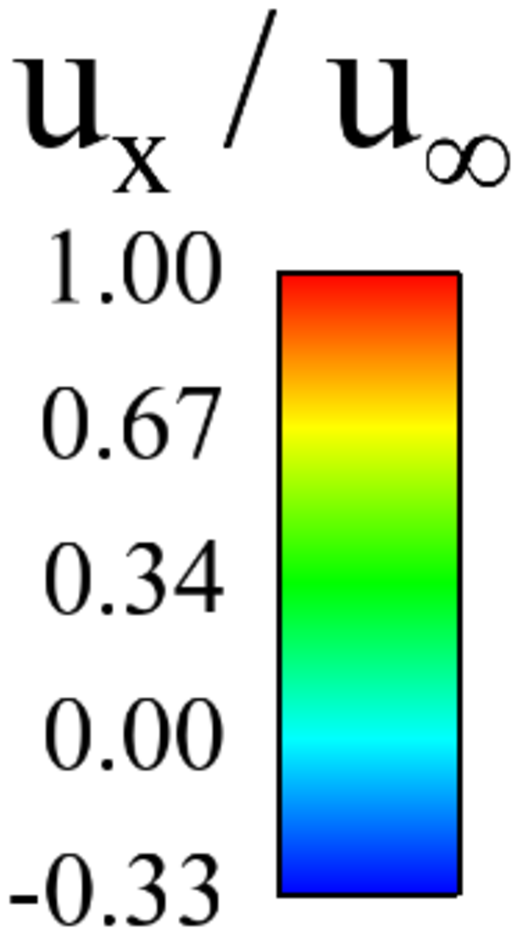}
 \includegraphics[scale = 0.3]{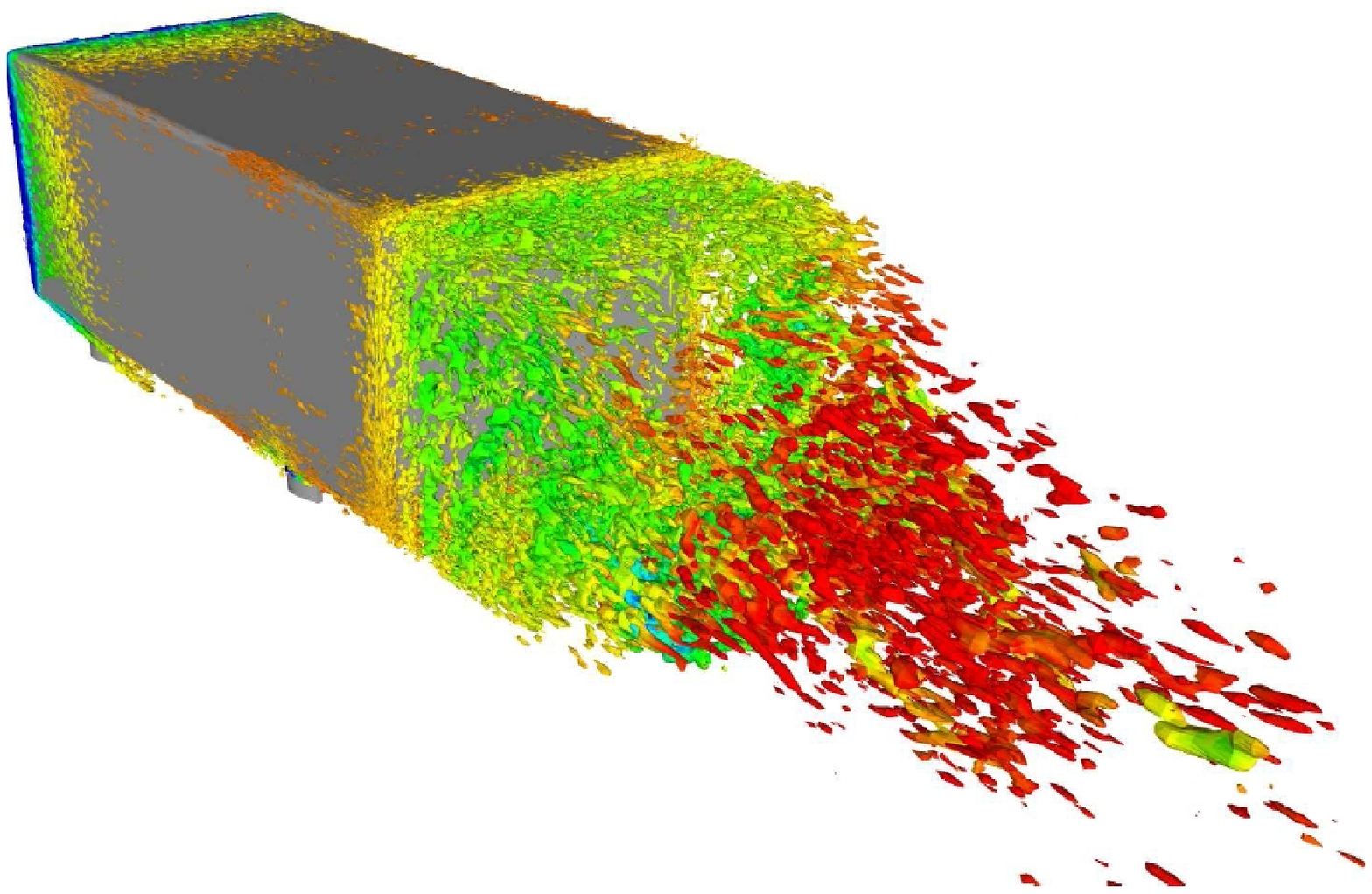}
 \label{SubFig:Figure25a}
}
 \subfigure[]{
 \includegraphics[scale = 0.3, trim = 0cm 0cm 0cm 0cm, clip = true]{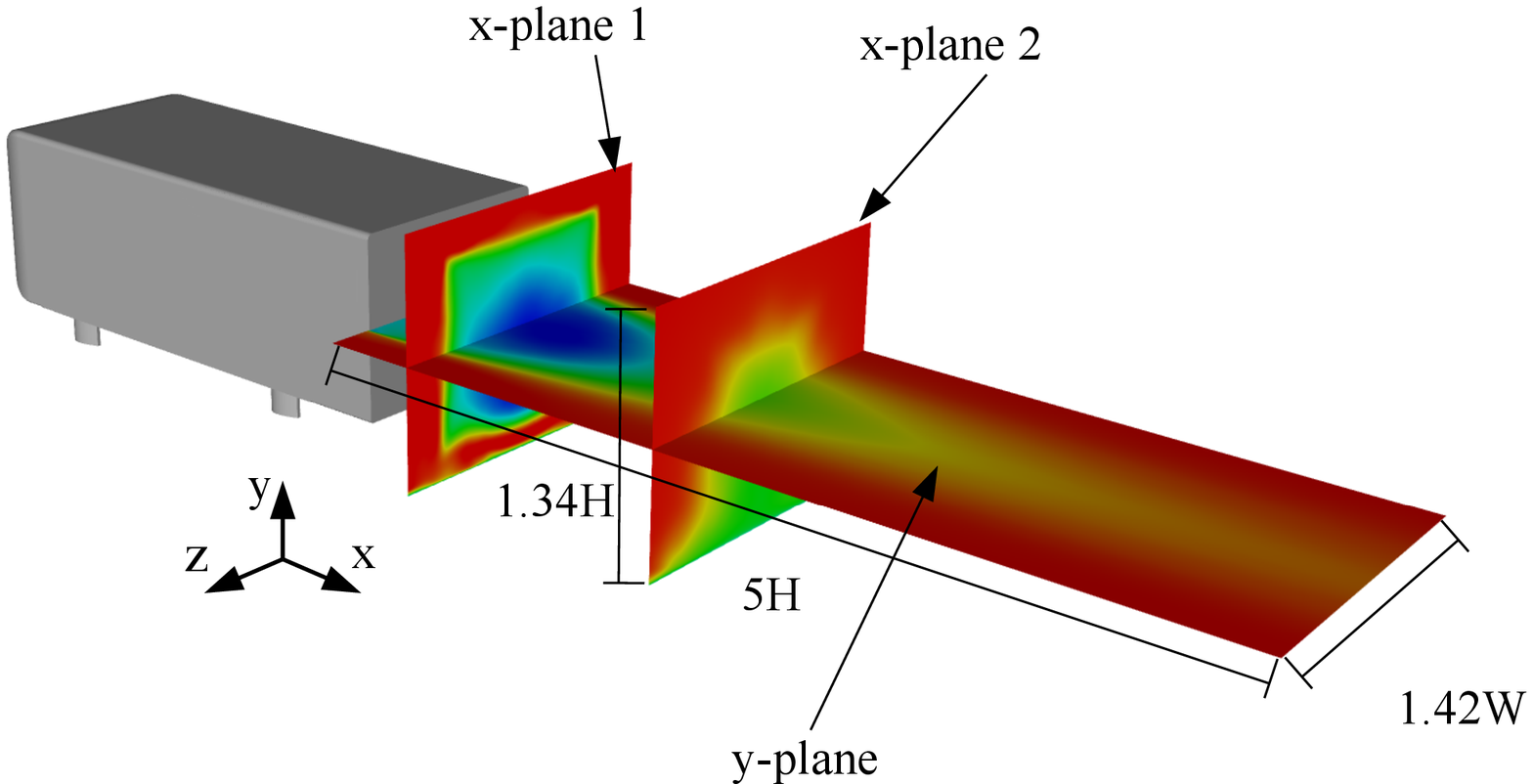}
 \label{SubFig:Figure25b}
}
 \caption{
 Bluff body wake:
 (a) Instantaneous iso-surface of $Q = 150 000$ 
 which is coloured by the pressure coefficient $c_p = \Delta p/\frac{1}{2}\rho U_{\infty}^2$ with 
 $\Delta p = p-p_{\infty}$ where $p_{\infty}$ is the pressure in the freestream.
 The wake is characterised by a broadband spectrum exhibiting no obvious large-scale structures.
 (b) Principal sketch of the bluff body model. Results are visualised in the plane $y=0$ 
 which is coloured here by the streamwise velocity component.}
 \label{Fig:Figure25}
\end{figure}

In this section, 
CROM is applied to an incompressible three-dimensional turbulent wake flow of a vehicle bluff body
with the Reynolds number $Re_H = U_\infty H/\nu=3\cdot 10^5$ based 
on the freestream velocity $U_\infty$, 
the height of the body $H$, 
and the kinematic viscosity $\nu$.
The vehicle bluff body is a generic car model, also known as Ahmed body, 
and widely used to elucidate the relationship between observable flow structures
and drag and lift forces arising typically on passenger cars.
The employed model has a square-back 
and is placed on four cylindrical supports (see figure~\ref{SubFig:Figure25b}).

In this study,
the flow is highly turbulent and exhibits a broadband spectrum in the wake.
The dominant frequeny is $St = fH/U \approx 0.2$ corresponding 
to the global shedding frequency of the wake.
An instantaneous realisation of an iso-surface of the second invariant
of the velocity gradient tensor 
$Q = - \frac{1}{2}\frac{\partial\overline{u_i}\partial\overline{u_j}}{\partial x_i\partial x_j}$ 
colour-marked by the pressure gradient
is shown in figure~\ref{SubFig:Figure25a}.

Recently, 
\cite{Grandemange2013jfm} showed that 
the wake of a square-back body is characterised by a bi-modal behaviour.
Over large time scales of the order of 
$T \sim 10^2 - 10^3 H/U_{\infty}$, 
the flow switches between two asymmetric states 
which possess an asymmetric pressure distribution 
at the base with respect to the $z-$direction.
The presence of this bi-modal behaviour
has been found from Reynolds numbers ranging 
from $Re \sim 10^2$ to $Re \sim 10^6$ \citep{Grandemange2012pre,Grandemange2013jfm}.
The employed database exhibits this low-frequency base flow modulation
as elaborated by \cite{Oesth2013jfm}.
In their study, 
they derived successfully a POD Galerkin model
which is capable of predicting these events.

The database of $M=1000$ velocity snapshots with sampling time $\Delta t = 0.5$,
which is the normalised time step with respect to $U_{\infty}$ and $H$,
is computed with a Large Eddy Simulation (LES).
The computational domain comprises $[L_x\times L_y\times L_y] = [(8H+L+20H)\times 5.33H\times 8.05H$
where $L$ is the length of the car model.
The snapshot data is preprocessed by a statistical symmetrisation with regard to the $z-$plane 
and then compressed with POD. 
A detailed description of the solver and the employed data 
as well as its validation with experimental data can be found in \citet{Oesth2013jfm}.
All results below are visualised with regard to the horizontal plane $y=0$ 
as displayed in figure~\ref{SubFig:Figure25b}.

\begin{figure}
\psfrag{0.32}[cc][][1][0]{$0.32$}
\psfrag{0.63}[cc][][1][0]{$0.63$}
\psfrag{0.95}[cc][][1][0]{$0.95$}
 \centering
\includegraphics[scale=1, trim = 0cm 0.18cm 0cm 0cm, clip = true]{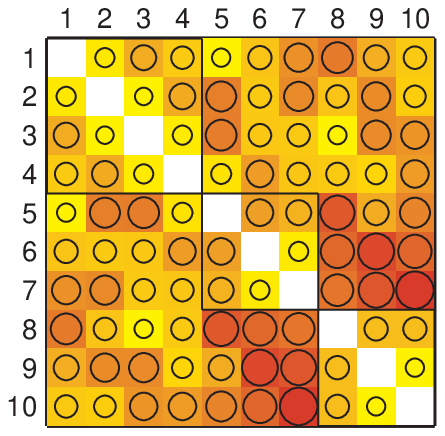}
 \hspace*{20pt}
\includegraphics[scale=1, trim = 0cm 0.18cm 0cm 0cm, clip = true]{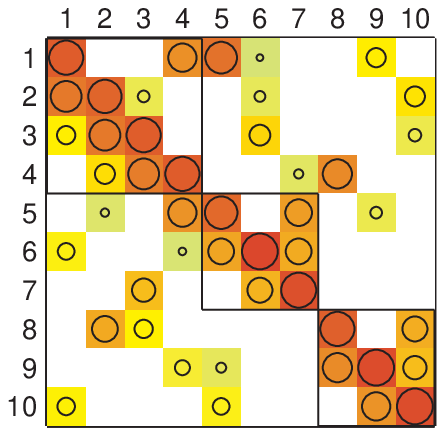}
 \vfill
 {
 \subfigure[]{
\includegraphics[scale=1, trim = 0cm 0cm 0cm 0.29cm, clip = true]{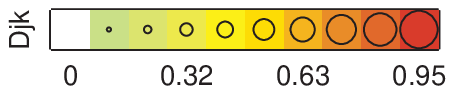}
 \label{SubFig:Figure26a}
}
 \hspace*{10pt}
 \subfigure[]{
\includegraphics[scale=1, trim = 0cm 0cm 0cm 0.29cm, clip = true]{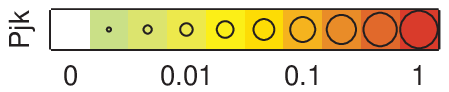}
 \label{SubFig:Figure26b}
}
 }
 \caption{
 Kinematics and dynamics:
 (a) Distance matrix $\slsD$ and 
 (b) transition matrix $\slsP$ 
of the clustered Ahmed body data.
Values are displayed like in figure~\ref{Fig:Figure04}.
Three groups of clusters are distilled, 
namely groups 'T' ($k=1,2,3,4$), 'B1' ($k=5,6,7$), and 'B2' ($k=8,9,10$).}
 \label{Fig:Figure26}
\end{figure}

The cluster analysis is performed in the full POD space for $K_c=10$ clusters.
The distance matrix is visualised in figure~\ref{SubFig:Figure26a}
displaying three groups of clusters.
Two groups, namely clusters $k=5,6,7$ ('B1') and $k=8,9,10$ ('B2'),
are comparably far away from each other,
while they are equally close to the group of clusters $k=1,2,3,4$ ('T').
Moreover, 
these groups are characterised by an intrinsic periodical behaviour
as concluded from the cluster transition matrix (see figure~\ref{SubFig:Figure26b}).
Similar to the mixing layer, 
high probabilities in the first sub diagonal  
and a closing transition
from the last cluster to the first cluster inside the group
indicate a periodical oscillation.
While the groups 'B1' and 'B2' show nearly no connectivity,
the group 'T' serves as a transition region between them.
The most significant transitions are
$3\rightarrow 7$ (T $\rightarrow$ B1) and $2\rightarrow 8$ (T $\rightarrow$ B2) 
as well as 
$5 \rightarrow 1$ (B1 $\rightarrow$ T) and $8 \rightarrow 4$ (B2 $\rightarrow$ T).

\begin{figure}
 \psfrag{0}[cc][][1][0]{}
 \psfrag{u}[cc][][1][0]{u}
 \centering
 \centerline{
 \raisebox{0.06\textheight}{1)}\hspace{0.3cm}
 \includegraphics[scale = 0.25, trim=1.9cm 1.8cm 0cm 0cm,clip=true,keepaspectratio]{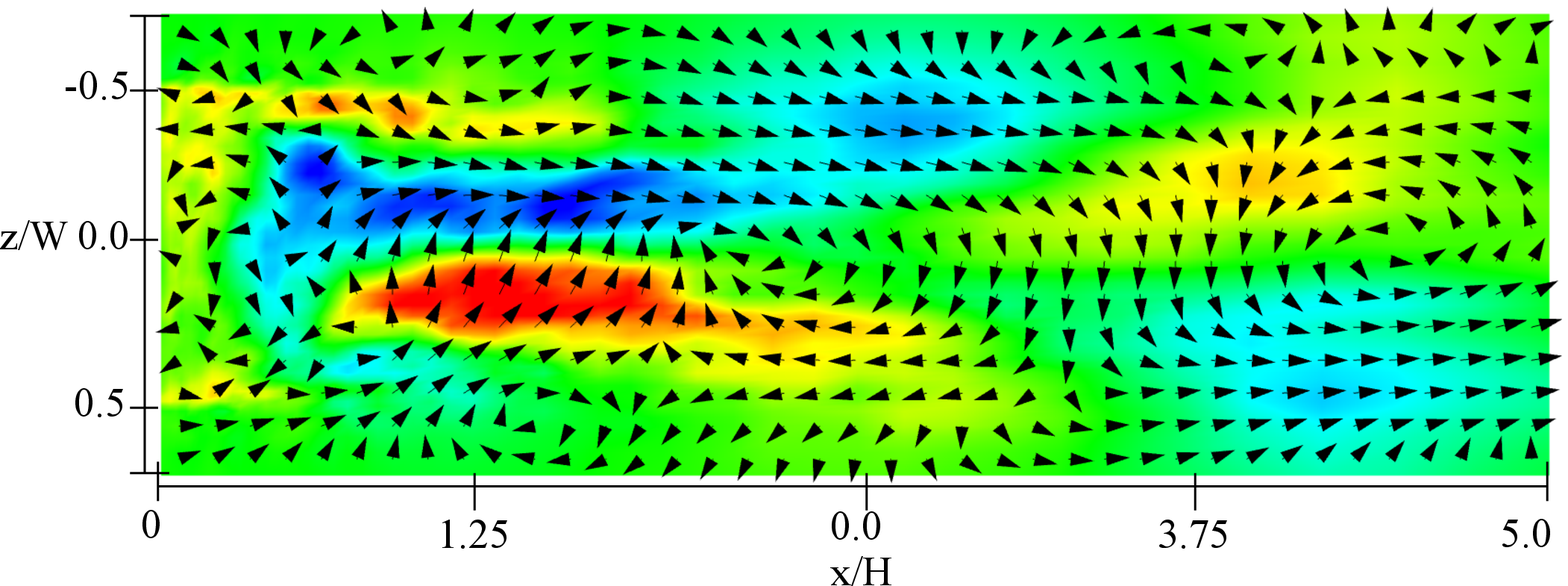}
 \hspace{0.4cm}
 \raisebox{0.06\textheight}{6)}\hspace{0.3cm}
 \includegraphics[scale = 0.25, trim=2cm 1.8cm 0cm 0cm,clip=true,keepaspectratio]{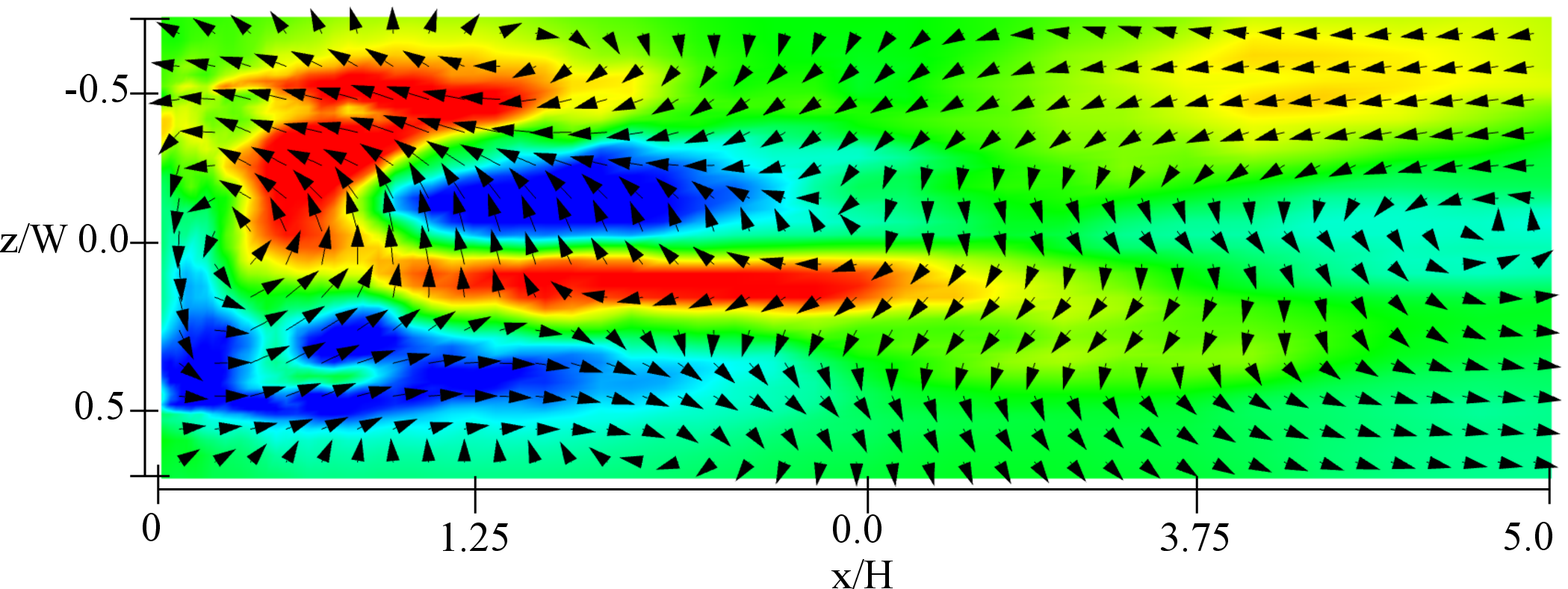}}
 \vfill
 \centerline{
 \raisebox{0.06\textheight}{2)}\hspace{0.3cm}
 \includegraphics[scale = 0.25, trim=1.9cm 1.8cm 0cm 0cm,clip=true,keepaspectratio]{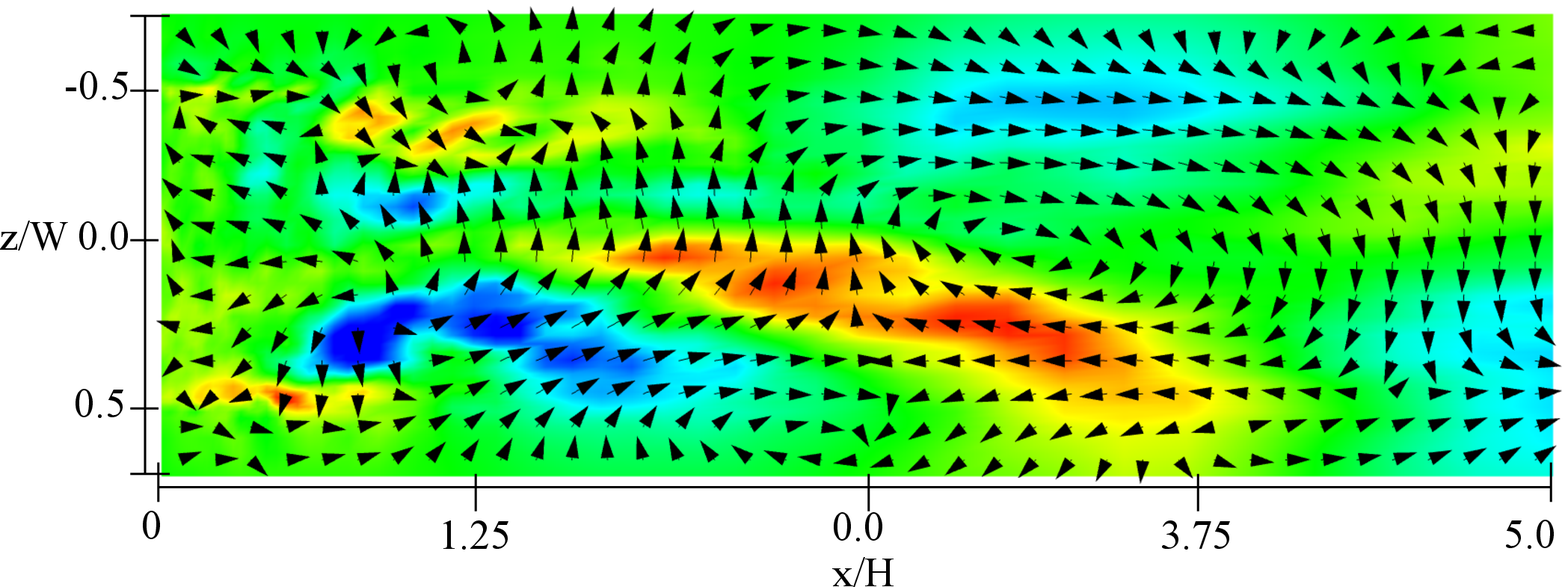}
 \hspace{0.4cm}
 \raisebox{0.06\textheight}{7)}\hspace{0.3cm}
 \includegraphics[scale = 0.25, trim=2cm 1.8cm 0cm 0cm,clip=true,keepaspectratio]{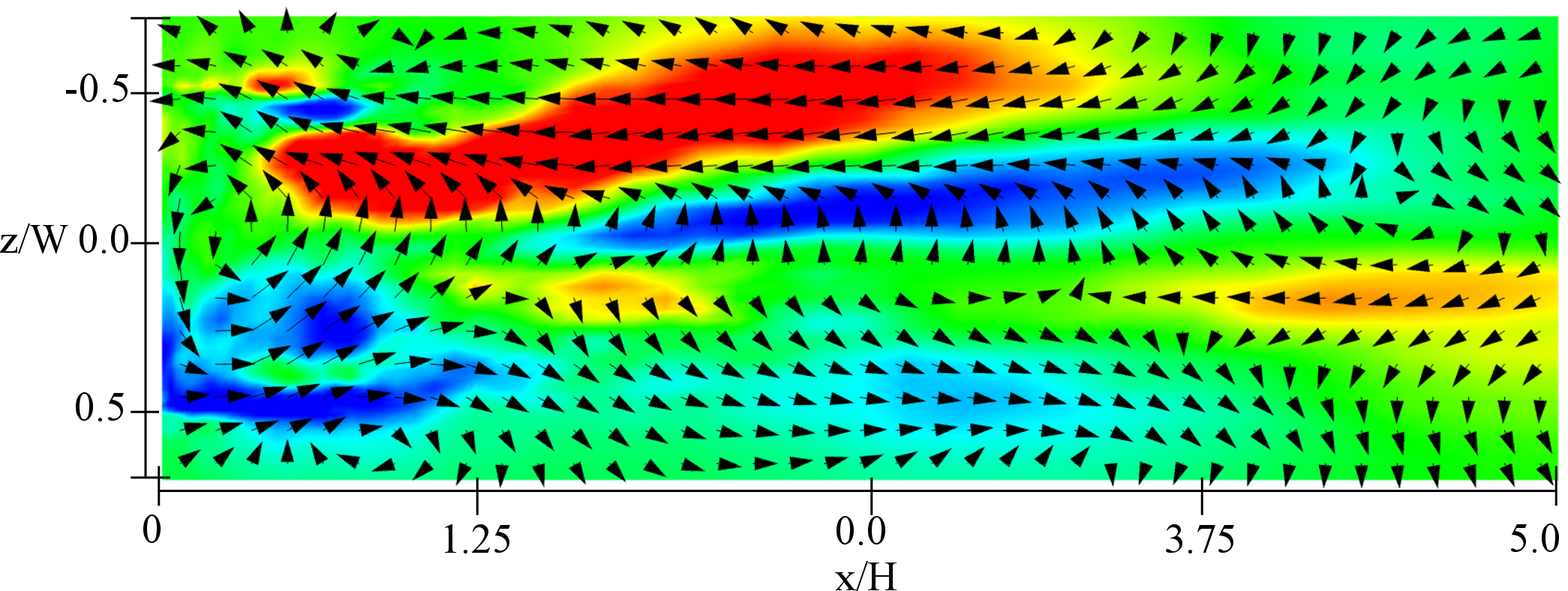}}
 \vfill
 \centerline{
 \raisebox{0.06\textheight}{3)}\hspace{0.3cm}
 \includegraphics[scale = 0.25, trim=1.9cm 1.8cm 0cm 0cm,clip=true,keepaspectratio]{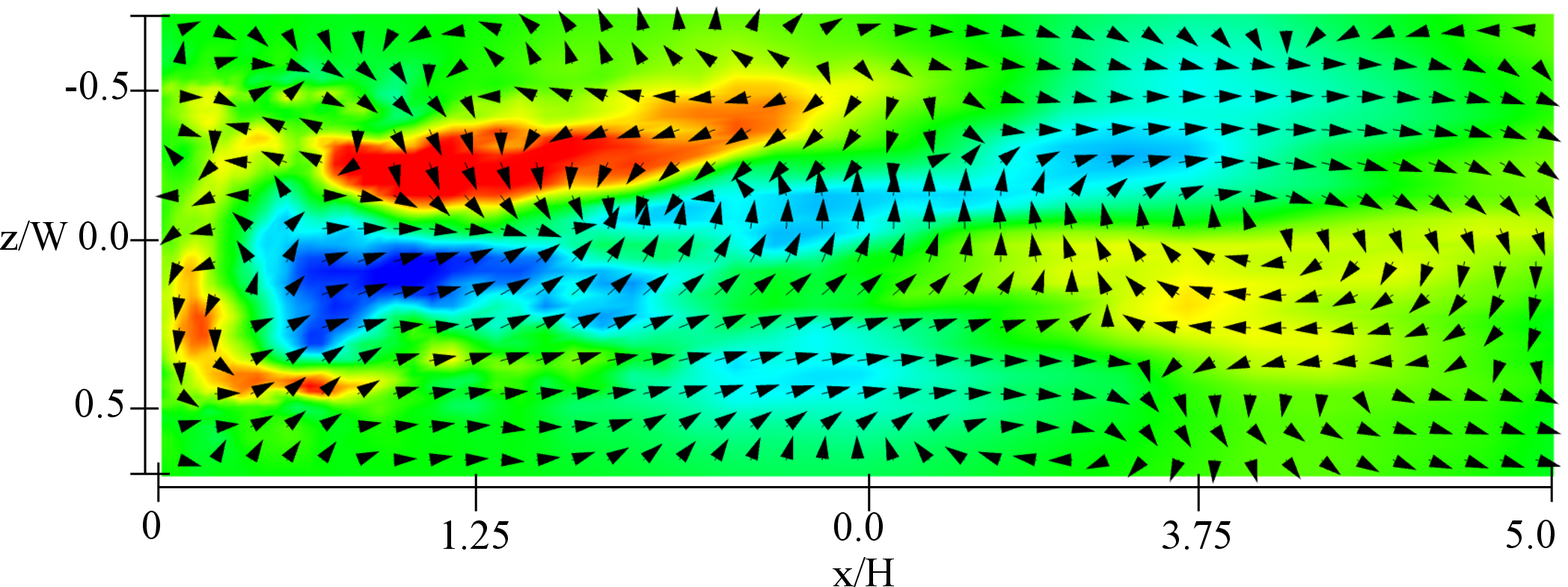}
 \hspace{0.4cm}
 \raisebox{0.06\textheight}{8)}\hspace{0.3cm}
 \includegraphics[scale = 0.25, trim=2cm 1.8cm 0cm 0cm,clip=true,keepaspectratio]{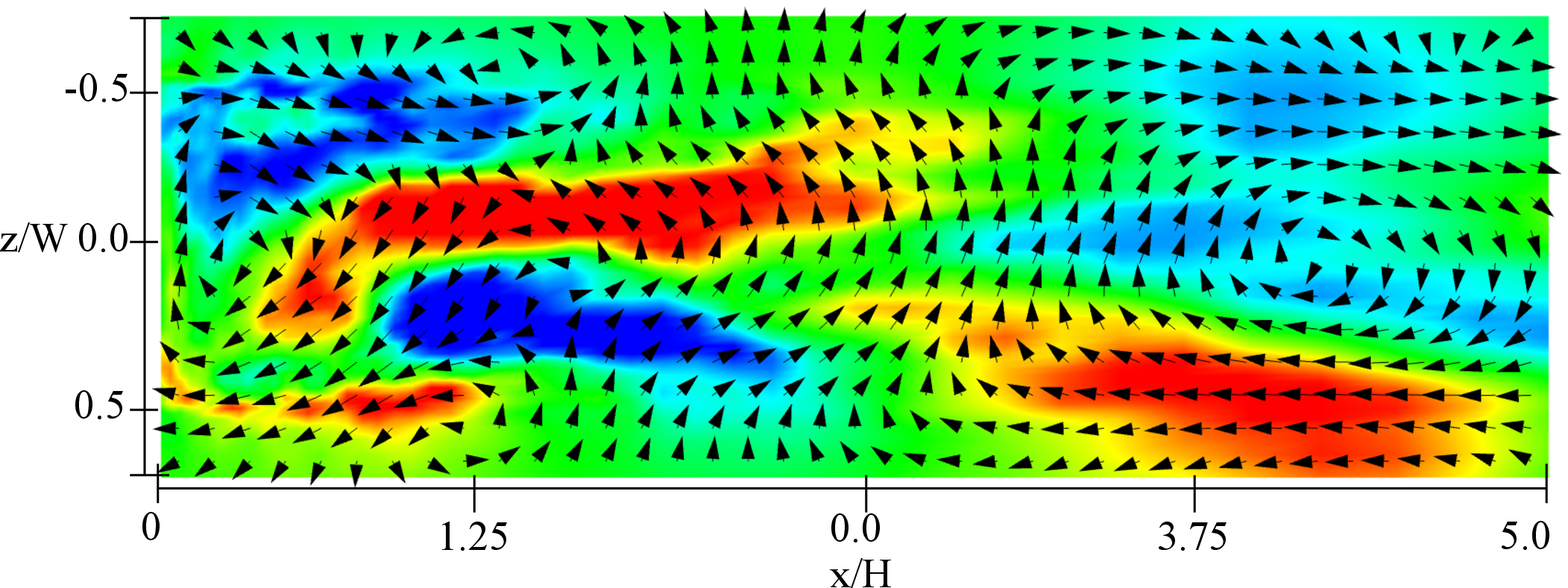}}
 \vfill\centerline{
 \raisebox{0.06\textheight}{4)}\hspace{0.3cm}
 \includegraphics[scale = 0.25, trim=1.9cm 1.8cm 0cm 0cm,clip=true,keepaspectratio]{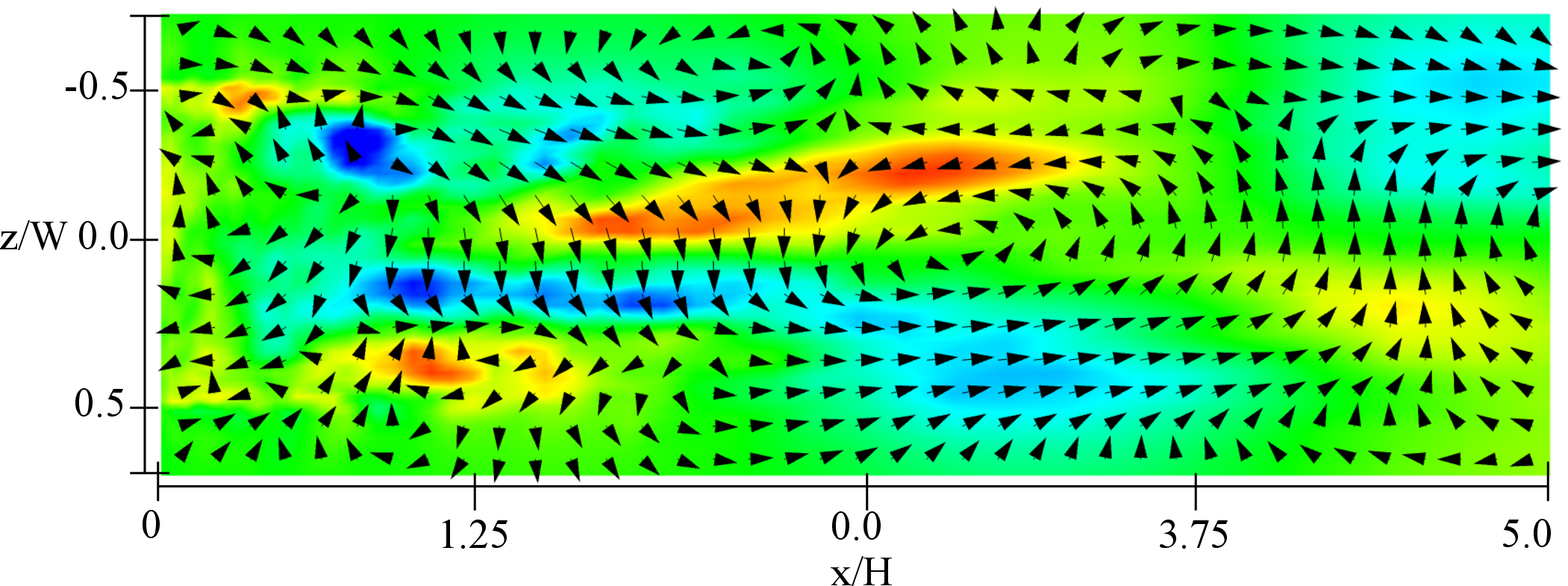}
 \hspace{0.4cm}
 \raisebox{0.06\textheight}{9)}\hspace{0.3cm}
 \includegraphics[scale = 0.25, trim=2cm 1.8cm 0cm 0cm,clip=true,keepaspectratio]{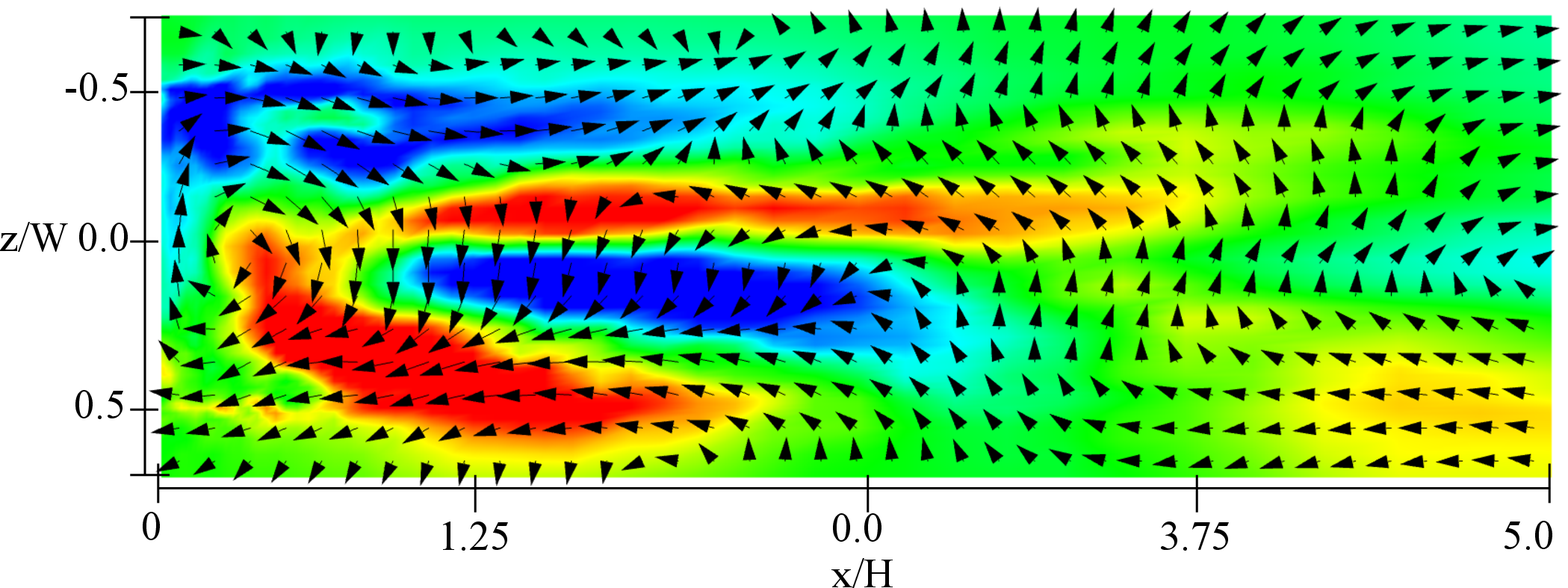}}
 \vfill\centerline{
 \raisebox{0.06\textheight}{5)}\hspace{0.3cm}
 \includegraphics[scale = 0.25, trim=1.9cm 1.8cm 0cm 0cm,clip=true,keepaspectratio]{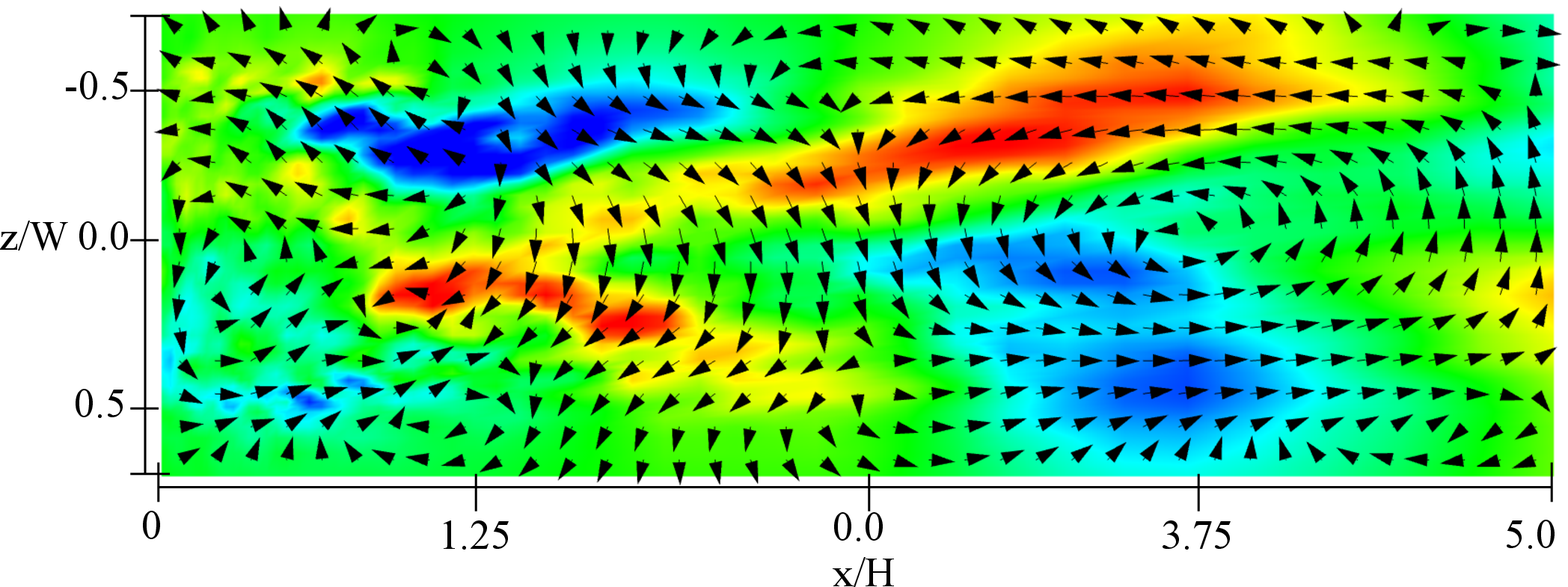}
 \hspace{0.4cm}
 \raisebox{0.06\textheight}{10)}\hspace{0.3cm}
 \includegraphics[scale = 0.25, trim=2cm 1.8cm 0cm 0cm,clip=true,keepaspectratio]{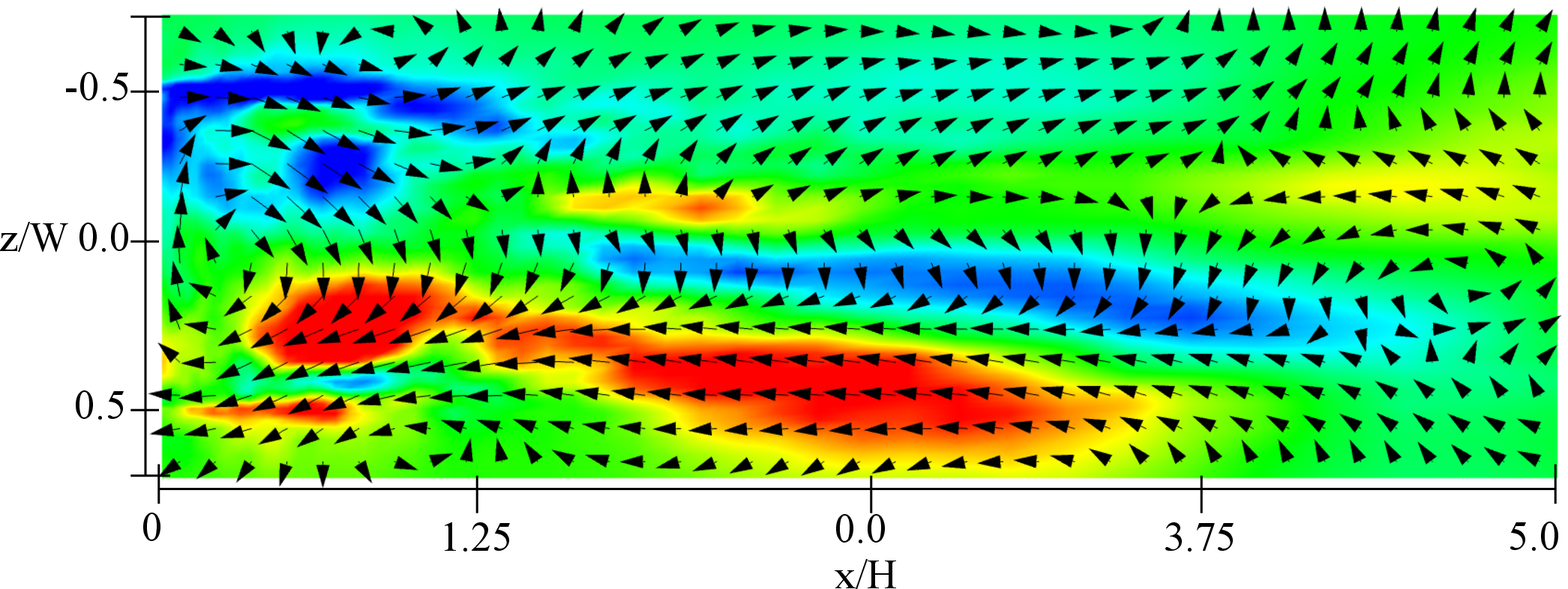}}
 \vfill
 \includegraphics[scale = 0.3, trim=2cm 11cm 1cm 12cm,clip=true,keepaspectratio]{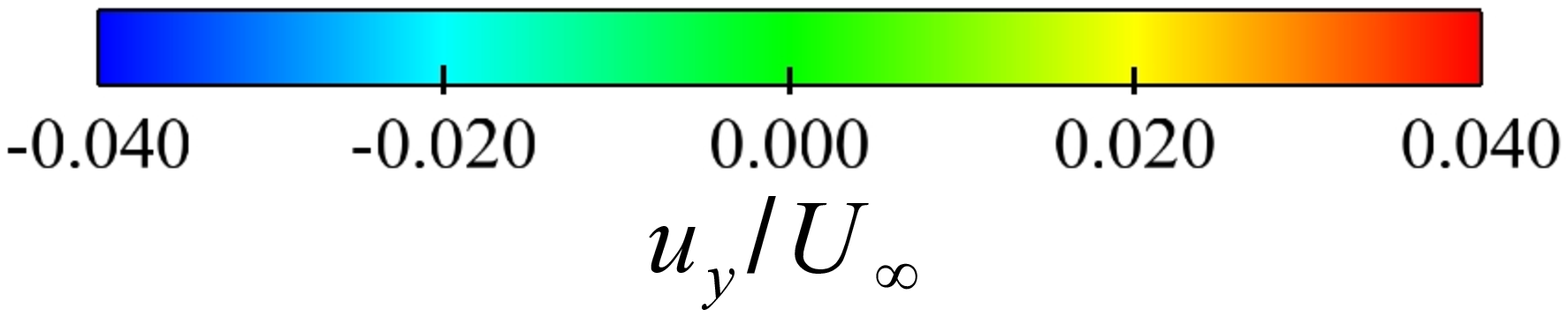}
 \caption{
 Cluster centroids of the Ahmed body. Visualisations of the $y-$planes coloured by $u_y/U_{\infty}$.  
 Cluster group 'T' ($k=1,2,3,4$) resembles the symmetric transition region, 
 cluster groups 'B1' ($k=5,6,7$) and 'B2' ($k=8,9,10$) represent the two semi-modal asymmetric states.}
 \label{Fig:Figure27}
\end{figure}

The cluster centroids computed as 
the mean of the velocity fluctuations in each cluster
are displayed in figure~\ref{Fig:Figure27}.
The clusters in 'B1' and 'B2' are
characterised by strong vortical structures
in opposite direction resembling the two semi-modal flow states.
Cluster $k=5$ is an exception which is addressed below.
The strength of the clusters in 'T' is weaker
and the vector fields seem to be more symmetric
with regard to the centerline $z=0$.
The principal branching clusters, 
similar as the flipper cluster of the mixing layer,
are $k=5$ and $k=8$ connecting group 'T' with 'B1' and 'B2', respectively.
These are intermediate states 
which explains why $k=5$ seems more similar to the states in 'T'
and $k=8$ shares more dominant features with 'B2'.

\begin{figure}
\psfrag{0}[cc][][1][0]{}
\psfrag{c1}[cc][][1][0]{$\mathbf{{\textcolor{black} 1}}$} 
\psfrag{c2}[cc][][1][0]{$\mathbf{{\textcolor{black} 2}}$}
\psfrag{c3}[cc][][1][0]{$\mathbf{{\textcolor{black} 3}}$}
\psfrag{c4}[cc][][1][0]{$\mathbf{{\textcolor{black} 4}}$}
\psfrag{c5}[cc][][1][0]{$\mathbf{{\textcolor{black} 5}}$}
\psfrag{c6}[cc][][1][0]{$\mathbf{{\textcolor{black} 6}}$}
\psfrag{c7}[cc][][1][0]{$\mathbf{{\textcolor{black} 7}}$}
\psfrag{c8}[cc][][1][0]{$\mathbf{{\textcolor{white} 8}}$}
\psfrag{c9}[cc][][1][0]{$\mathbf{{\textcolor{white} 9}}$}
\psfrag{c10}[cc][][1][0]{$\mathbf{{\textcolor{white} {10}}}$}
\psfrag{AA}[cc][][1][0]{$\text{B1}$}
\psfrag{BB}[cc][][1][0]{$\text{B2}$}
\psfrag{TT}[cc][][1][0]{$\text{T}$}
 \begin{minipage}[t]{\textwidth}
 \hspace{0.4cm}\raisebox{0.15\textheight}{a)}\hspace{1.6cm}
 \includegraphics[scale = 0.8]{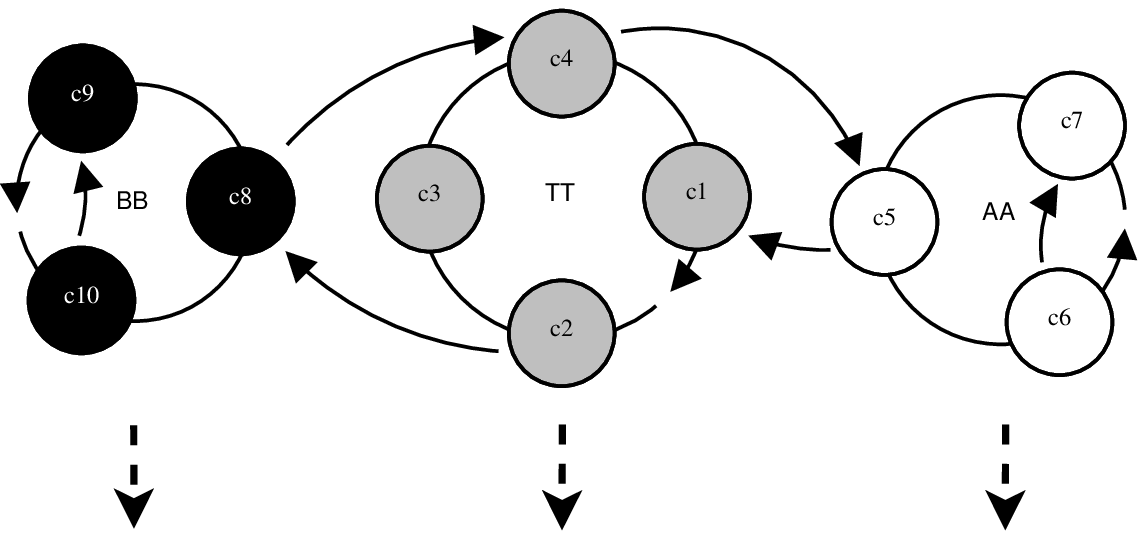}
 \hfill
 \end{minipage}
 \vfill
 \begin{minipage}[t]{\textwidth}
   \centerline{
 \raisebox{0.045\textheight}{b)}
 \includegraphics[scale = 0.2,trim=2cm 1.8cm 0cm 0cm,clip=true,keepaspectratio]{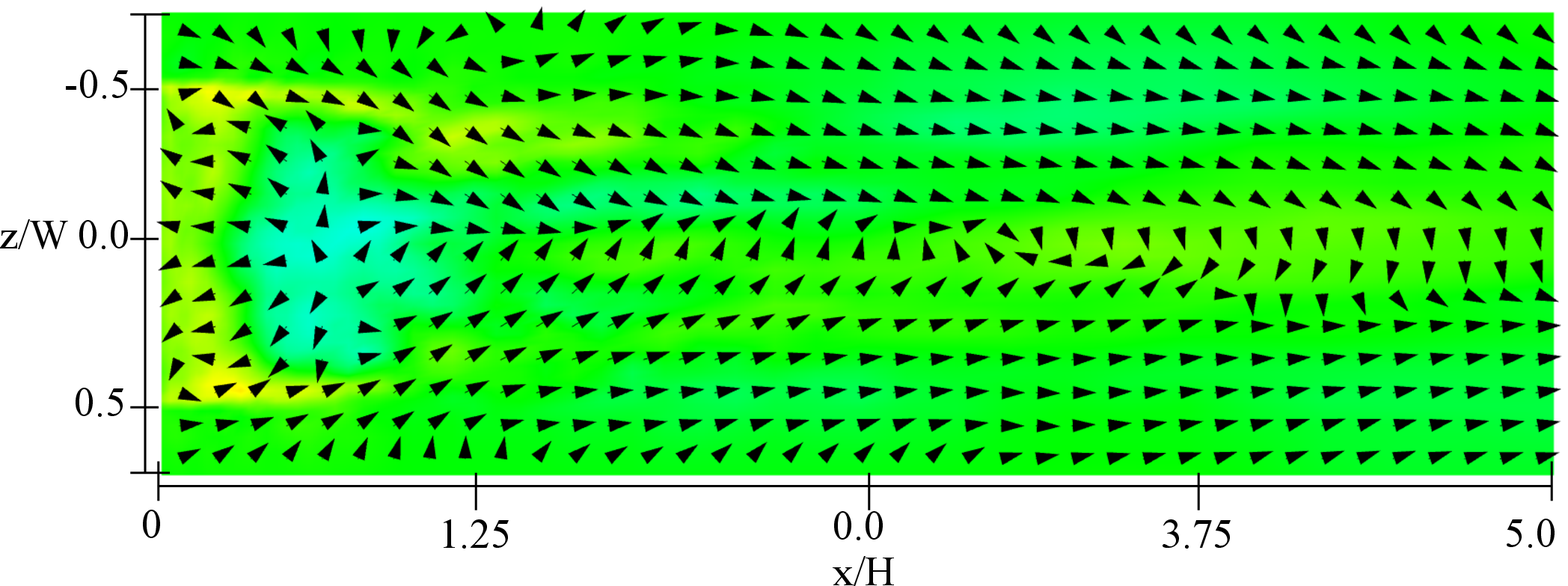}
 \raisebox{0.045\textheight}{c)}
 \includegraphics[scale = 0.2,trim=2cm 1.8cm 0cm 0cm,clip=true,keepaspectratio]{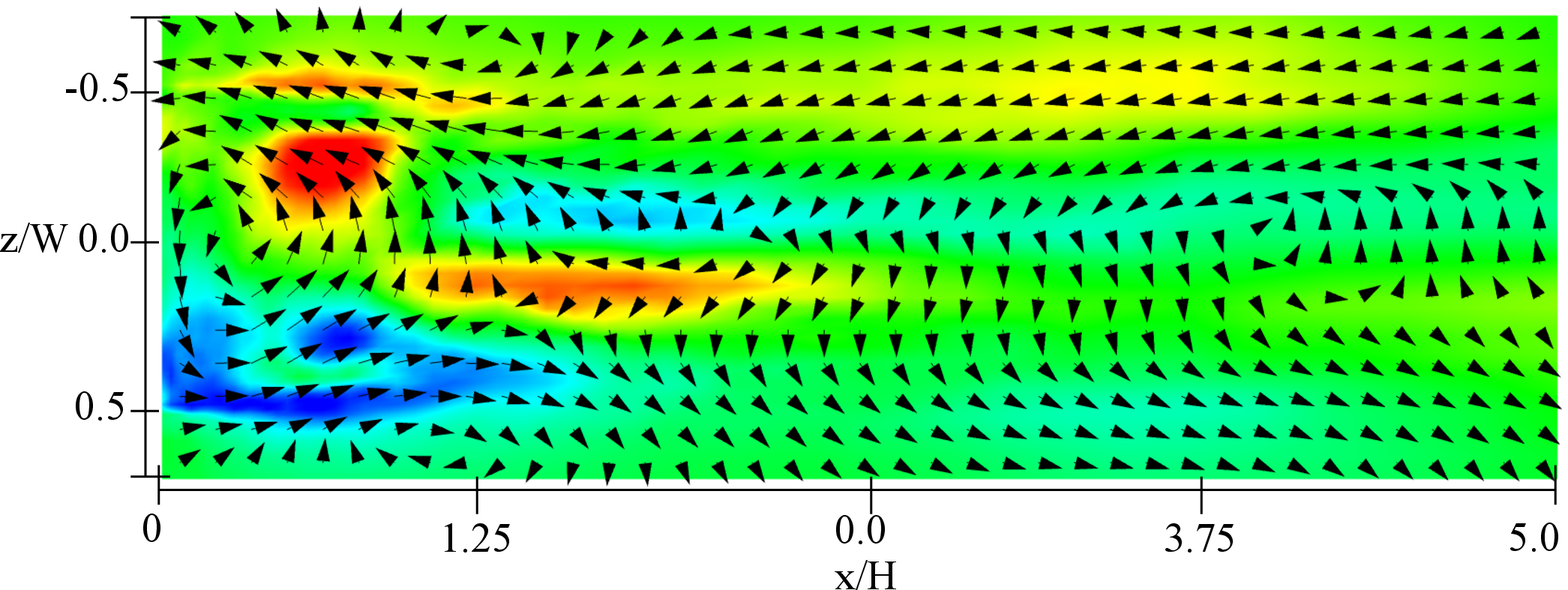}
 \raisebox{0.045\textheight}{d)}
 \includegraphics[scale = 0.2,trim=2cm 1.8cm 0cm 0cm,clip=true,keepaspectratio]{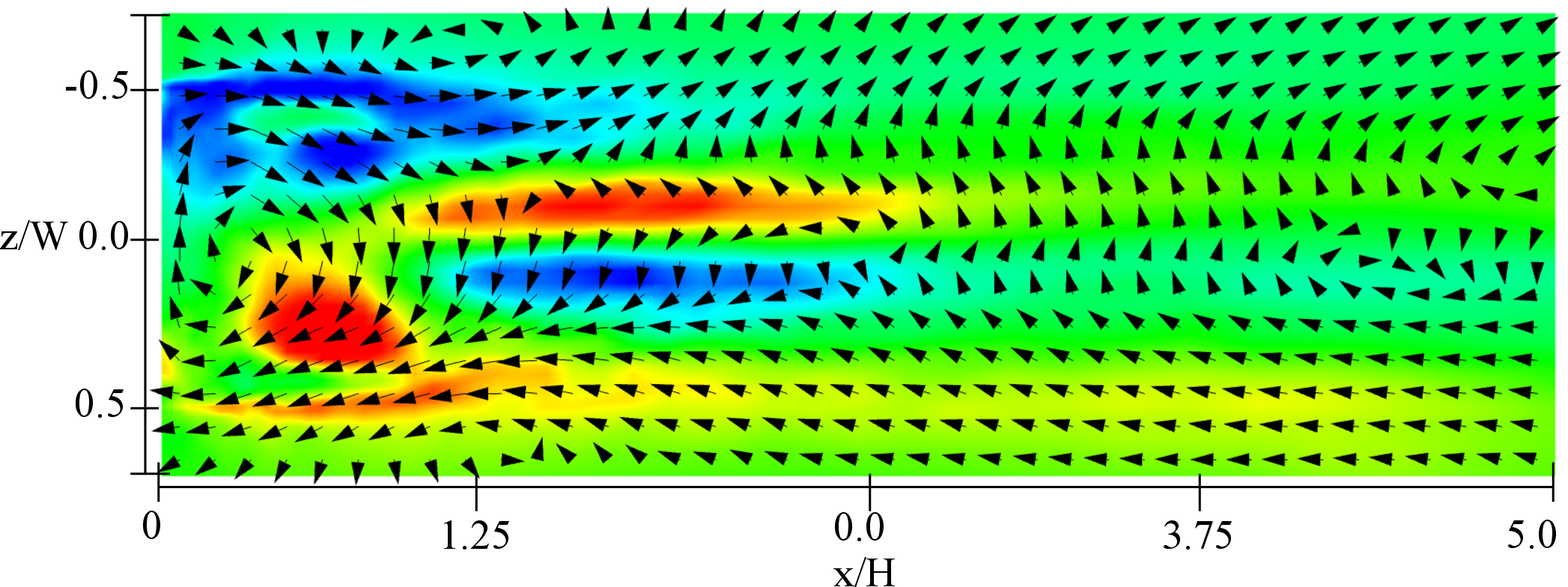}}
 \end{minipage}
 \caption{
 Simplified cluster transitions (analogous to figure~\ref{Fig:Figure16}).
The arrows indicate possible transitions above a certain threshold.
This threshold is chosen small enough to guarantee full connectivity.
The graph  highlights three cyclic groups associated with 
the two asymmetric meta-stable states
and a transition region between them. 
Two branching clusters, $k=5$ and $k=8$, connect these groups.
Mean velocity fields of the three cluster groups of Ahmed body are shown below:
 (b) bi-modal group 'B2' of clusters $k=8,9,10$,
 (c) symmetric group 'T' of clusters $k=1,2,3,4$,
 (d) bi-modal group 'B1' of clusters $k=5,6,7$.
The visualisation of the mean flows corresponds to those in figure~\ref{Fig:Figure27}.}
 \label{Fig:Figure28}
\end{figure}

An intuitive picture of the cluster transitions and the corresponding states
is provided in figure~\ref{Fig:Figure28}.
The simplified cluster transitions
determined as for the mixing layer in \S~\ref{Sec:MixingLayerModel}
and neglecting inner-cluster transitions 
are displayed in figure~\ref{Fig:Figure28}\;(a).
The large cluster group 'T'
connects the two semi-modal states 
characterised by 'B1' and 'B2'.
Clusters $k=5$ and $k=8$ have a special role
and serve as branchig clusters.
Apparently, the plots of the mean of each cluster group (figure~\ref{Fig:Figure28}\;(b-d))
confirms that 'B1' and 'B2' represent the two asymmetric states
while 'T' represents a low-amplitude oscillation around the symmetric base flow (figure~\ref{Fig:Figure28}\;(c)).

\begin{figure}
 \psfrag{1-cD}[cc][][1][0]{\colorbox{white}{$1-c_D$}\hspace{0.07cm}}
 \psfrag{2-cL}[cc][][1][0]{\colorbox{white}{$2-c_L$}\hspace{0.05cm}}
 \psfrag{3-cS}[cc][][1][0]{\colorbox{white}{$3-c_S$}\hspace{0.04cm}}
 \psfrag{alpha1}[cc][][1][0]{$\alpha_1$}
 \psfrag{alpha2}[cc][][1][-90]{$\alpha_2$}
 \psfrag{0}[cc][][1][0]{$0$}
 \psfrag{0.2}[cc][][1][0]{$0.2$}
 \psfrag{0.4}[cc][][1][0]{$0.4$}
 \psfrag{0.6}[cc][][1][0]{$0.6$}
 \psfrag{-0.2}[cc][][1][0]{$-0.2$\hspace{0.2cm}}
 \psfrag{-0.4}[cc][][1][0]{$-0.4$\hspace{0.2cm}}
 \psfrag{-0.6}[cc][][1][0]{$-0.6$\hspace{0.2cm}}
 \psfrag{c1}[cc][][1][0]{\colorbox{white}{$\boldsymbol{c}_1$}}
 \psfrag{c2}[cc][][1][0]{\colorbox{white}{$\boldsymbol{c}_2$}}
 \psfrag{c3}[cc][][1][0]{\colorbox{white}{$\boldsymbol{c}_3$}}
 \psfrag{c4}[cc][][1][0]{\colorbox{white}{$\boldsymbol{c}_4$}}
 \psfrag{c5}[cc][][1][0]{\colorbox{white}{$\boldsymbol{c}_5$}}
 \psfrag{c6}[cc][][1][0]{\colorbox{white}{$\boldsymbol{c}_6$}}
 \psfrag{c7}[cc][][1][0]{\colorbox{white}{$\boldsymbol{c}_7$}}
 \psfrag{c8}[cc][][1][0]{\colorbox{white}{$\boldsymbol{c}_8$}}
 \psfrag{c9}[cc][][1][0]{\colorbox{white}{$\boldsymbol{c}_9$}}
 \psfrag{c10}[cc][][1][0]{\colorbox{white}{$\boldsymbol{c}_{10}$}}
 \centering
 \includegraphics[scale = 1]{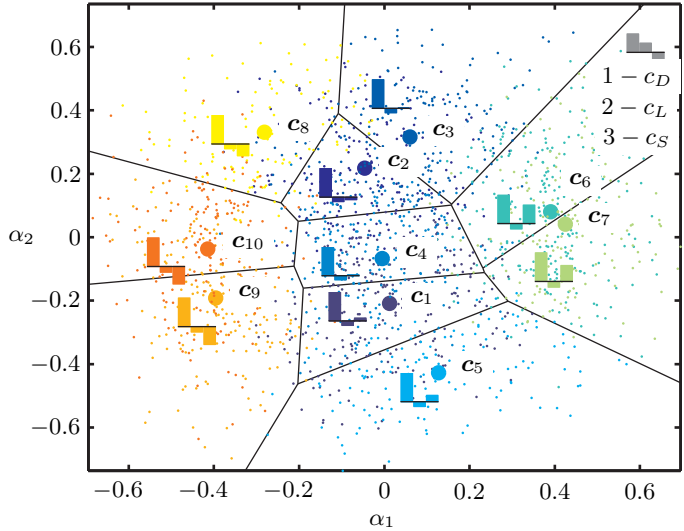}
 \caption{
 Cluster forces of the Ahmed body: 
 Visualisation of the Voronoi diagram of the clusters, 
 the cluster centroids (coloured bullets),
 the data points (coloured dots), 
 and bar plots of the mean force-related coefficients of each cluster
 in the space of the first two POD modes $\alpha_1$ and $\alpha_2$ 
 associated with the centroids
 (as explained in appendix~\ref{Sec:AppD:Visualisation}).
 The coefficients are normalised with respect to the drag coefficient.
 The side coefficient is $\times10$ enlarged for visualisation purposes.
 The colours indicate the different clusters.
 The clusters associated with the largest magnitude of the side force correspond to the 
 outer clusters $k=5,6,7$ and $k=8,9,10$, respectively, 
 which represent the two asymmetric flow states.}
 \label{Fig:Figure29}
\end{figure}

A further analysis of the forces reveals that 
the two semi-modal states are associated with an increase in the magnitude of the side force
which was also found by \cite{Grandemange2013jfm} and \cite{Oesth2013jfm} for the same data.
In figure~\ref{Fig:Figure29} the Voronoi diagram of the clusters,
the cluster centroids (coloured bullets),
and the data points (coloured dots) are displayed as described in appendix~\ref{Sec:AppD:Visualisation}.
The mean forces associated with each cluster are shown as bar plots
of which the first bar corresponds to the drag coefficient $c_D := F_x/(\frac{1}{2}\rho U_{\infty} HW)$, 
the second to the lift cooefficient $c_L := F_y/(\frac{1}{2}\rho U_{\infty} HW)$,
and the third to the side force coefficient $c_S := F_z/(\frac{1}{2}\rho U_{\infty} HW)$.
While the drag and the lift do not show any significant drift,  
the side force coefficient changes signs
when the flow switches from one side to another.
The lowest values in magnitude correspond to the clusters in 'T'
which is consistent since they are tendentially symmetric.

\begin{figure}
 \psfrag{0}[cc][][1][0]{$0$}
 \psfrag{-1}[cc][][1][0]{$-1$}
 \psfrag{-2}[cc][][1][0]{$-2$}
 \psfrag{-3}[cc][][1][0]{$-3$}
 \psfrag{-1.5}[cc][][1][0]{}
 \psfrag{1.5}[cc][][1][0]{}
 \psfrag{-0.5}[cc][][1][0]{}
 \psfrag{0.5}[cc][][1][0]{}
 \psfrag{imag(lambda)}[cc][][1][0]{$\omega$}
 \psfrag{real(lambda)}[cc][][1][0]{$\sigma$}
 \centering
 \includegraphics[scale = 1]{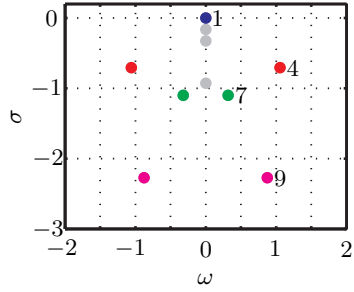}
 \caption{
 Stability analysis of the transition matrix associated with the Ahmed body:
 In contrast to figure~\ref{Fig:Figure19}, 
 the growth rate $\sigma = \Re(\mu)$ and the frequency $\omega = \Im(\mu)$ 
 are shown according to the transformation $\mu = 1/\Delta t\log(\lambda)$.
 The marginal stable eigenvalue $\mu_1$, or $\lambda=1$, 
 has vanishing growth rate and frequency (blue bullet)
 and is associated with the asymptotic probability distribution.
 Other non-oscillatory modes are visualised with grey bullets.
 The frequency corresponding to the oscillatory mode with the smallest damping 
 (red bullets) is $St \approx 0.17$.}
 \label{Fig:Figure30}
\end{figure}

Finally, the spectrum of the CTM displayed in figure~\ref{Fig:Figure30} is analysed.
In contrast to \S~\ref{Sec:LorenzModel} and \S~\ref{Sec:MixingLayerModel}, 
a different visualisation is employed 
displaying the real and imaginary part of $\mu = 1/\Delta t\log(\lambda)$
where $\lambda$ is an eigenvalue of the CTM as defined in \S~\ref{Sec:CROM}. 
This transformation yields the growth rates and frequencies 
of the corresponding probability eigenvectors
which can be directly linked to the growth rates and frequencies of observables,
e.g. velocity field or pressure measurements (for details see App.~\ref{Sec:AppC:Comparison-KoopmanModes}).
Besides the invariant distribution corresponding to $\lambda=1$,
there exist three oscillatory modes.
The frequency of the oscillatory mode with the smallest damping 
is $St = \omega_4/2\pi \approx 0.17$ 
(displayed in red bullets) close to $St=0.2$ of the global wake shedding.
Note that the oscillations are only resolved by $3$ or $4$ clusters.
Better estimations can be obtained by a refined resolution, 
i.e. increasing the number of clusters $K$.

In summary,
CROM distils the two bi-modal states of the base flow
and identifies the main transition mechanism.
These asymmetric quasi-attractors are connected
to the comparably symmetric transition region
solely via the two branching clusters $k=5$ and $k=8$.
All groups are intrinsically periodic and 
the dominant frequency is determined as $St=0.17$.
A larger number of clusters increases the resolution of the bi-modal states 
and the branching clusters
but not change the main mechanism.

\section{Towards CROM-based flow control}
\label{Sec:AppB:TowardsCROMbasedFlowControl}
In this section, a control design is proposed for the Markov model.
The goal is to bring the probability distribution $\boldsymbol{p}$
as close as possibly to the desired distribution $\boldsymbol{p}^{\text{target}}$ 
with the corresponding Markov matrix $\mathcal{P}^{\text{target}}$.
In the following, we drop the symbol $^{cont}$ to simplify the notation.

We start with the Galerkin system
\begin{equation}
\frac{d}{dt} \boldsymbol{a} = \boldsymbol{f} (\boldsymbol{a}) + \boldsymbol{g} (\boldsymbol{a}), 
\label{Eqn:GS+CLC}
\end{equation}
where $\boldsymbol{f}$ is the propagator of the unforced dynamics
and $\boldsymbol{g}$ the actuation term incorporating the control law.
The corresponding Liouville equation 
for the probability distribution $p(\boldsymbol{a},t)$ reads
\begin{equation}
\frac{\partial}{\partial t} p(\boldsymbol{a},t)
+ \nabla_{\boldsymbol{a}} \cdot 
\left[ p(\boldsymbol{a},t) \left(  \boldsymbol{f}(\boldsymbol{a}) + \boldsymbol{g}(\boldsymbol{a} ) \right) \right]
= 0.
\label{Eqn:LiouvilleEquationForced}
\end{equation}
Let $\boldsymbol{\rho}=\left[\rho_1,\ldots,\rho_K\right]^T$ be the Galerkin coefficients of $p(\boldsymbol{a},t)$ on the basis $\{\Psi_k(\boldsymbol{a})\}_{k=1}^K$, the Markov model 
can be derived in complete analogy to \S~\ref{Sec:UlamGalerkinMethod}
\begin{equation}
\frac{d}{dt} \boldsymbol{\rho} = \left [ \mathcal{P}_F + \mathcal{P}_G \right] \boldsymbol{\rho},
\label{Eqn:MarkovModelForced}
\end{equation}
where $\mathcal{P}_F =(\mathcal{P}^F_{jk})$ is the transition matrix 
for the unactuated propagator $\boldsymbol{f}$ 
while $\mathcal{P}_G =(\mathcal{P}^G_{jk})$ represents 
the closed-loop actuation term $\boldsymbol{g}$.
Note that $\sum_{k=1}^K |\Omega_k|^{1/2}\,\rho_k = 1$ is guaranteed by
\begin{subequations}
\begin{eqnarray}
\sum\limits_{j=1}^K \mathcal{P}^F_{jk} &=&0 \quad \forall k=1,\ldots, K,\\
\sum\limits_{j=1}^K \mathcal{P}^G_{jk} &=&0 \quad \forall k=1,\ldots, K.
\end{eqnarray}
\end{subequations}

The actuation term comprises 
a fixed actuation hardware 
and a free control law to be optimised.
For simplicity,
we assume a single volume force actuation.
Then, the Galerkin system \eqref{Eqn:GS+CLC} 
has the actuation term $\boldsymbol{g} = \slsB b$,
where $b$ is the actuation command 
and $\slsB$ a column vector representing the shape of the volume force.
Again, for simplicity, 
the control law is searched in form of a linear full-state feedback ansatz, i.e.
\begin{equation}
b =  \slsK \boldsymbol{a},
\label{Eqn:LCL}
\end{equation}
where $\slsK$ is a row vector.
Summarising,
\begin{equation}
 \boldsymbol{g} = \slsB \slsK \boldsymbol{a}.
\label{Eqn:GS+VF}
\end{equation}
For each cluster $k=1,\ldots,K$, 
the actuation term can be estimated 
by the corresponding centroid $\boldsymbol{a} = \boldsymbol{c}_k$,
i.e.\ $\boldsymbol{g}_k=\slsB \slsK \boldsymbol{c}_k$.
Following, the derivation of Markov model \eqref{Eqn:LiouvilleVFM}
the corresponding actuation transition matrix reads
\begin{equation}
\mathcal{P}_G = \mathcal{P}_B \> \slsK,
\end{equation}
where $\mathcal{P}_B$ corresponds to the volume force.

The solution  
of \eqref{Eqn:MarkovModelForced}
should be as close as possible to the post-transient goal distribution, i.e.
\begin{equation}
\boldsymbol\rho(t) = \mathrm{e}^{\mathcal{P}_F t}\boldsymbol\rho_0 
+\int\limits_0^t\mathrm{d}\tau\> \mathrm{e}^{\mathcal{P}_F (t-\tau)} \mathcal{P}_B \slsK \boldsymbol{\rho}(\tau)
\approx \boldsymbol\rho^{\text{target}} 
\label{Eqn:ControlEquation}
\end{equation}
This implies that the desired attractor shall be reached 
for arbitrary initial conditions.
The corresponding optimality condition reads
\begin{equation}
\Vert \boldsymbol{\rho}(t)
- \boldsymbol{\rho}^{\text{target}} \Vert  \overset{!}{=} \hbox{min},
\label{Eqn:CROM+Control}
\end{equation}
where $\Vert\cdot\Vert$ may denote any suitable matrix norm.

The control design task is to determine 
$\slsK$ from this optimality condition.
A small (large) time horizon $t$ 
corresponds to an aggressive (conservative) control law.

It should be noted that this linear control law 
takes into account the nonlinear actuation dynamics
resolved by the clustered state space.
In principle, the CROM-based control strategy
can be expected to exploit strong nonlinearities,
like frequency cross-talk.
We actively pursue this direction
for our experimental flow control plants.

\section{Perron-Frobenius and Koopman operators}
\label{Sec:AppC:Comparison-KoopmanModes}
In this section, 
relations of the Liouville equation
with the Perron-Frobenius 
and Koopman operators are elaborated.
Intriguingly, 
these are linear operators
although the system dynamics is nonlinear
which explains their 
important role in dynamical systems theory.
If the system is noisy, 
the Perron-Frobenius operator is replaced by the Fokker-Planck operator.
We refer to \cite{Lasota1994book} for a mathematically rigorous treatment, 
and give here a more intuitive picture.

The Perron-Frobenius operator describes the temporal evolution
of a probability density.
Let $\hat{\slsL}$ be the Liouville operator as in \eqref{Eqn:LiouvilleEquation}, 
the solution $p(\boldsymbol{a}, t)$ of the Liouville equation 
\begin{equation}
\frac{\partial}{\partial t} p(\boldsymbol{a},t)
= \hat{\slsL}p(\boldsymbol{a},t)
\label{Eqn:LiouvilleEquationapp}
\end{equation}
can be written formally as 
$p(\boldsymbol{a}, t) = \hat{\slsP}_t\,p_0(\boldsymbol{a})$
where $p_0(\boldsymbol{a})$ is the initial probability distribution.
The Perron-Frobenius operator $\hat{\slsP}_t$ maps the PDF $p(\boldsymbol{a},t)$ forward in time.
Since 
$\hat{\slsP}_t(\alpha p_1 + \beta p_2) = \alpha\hat{\slsP}_t p_1 + \beta\hat{\slsP}_t p_2$
for any functions $p_1$, $p_2$ and scalars $\alpha$, $\beta$,
this operator is linear.
From \eqref{Eqn:LiouvilleEquationapp}, 
we can infer that the Perron-Frobenius operator is also given by
$\hat{\slsP}_t=\exp(t\hat{\slsL})$ 
where
$\hat{\slsL}$ is the Liouville operator defined by
$\hat{\slsL}\,p:=-\nabla_{\boldsymbol{a}}\cdot(\boldsymbol{f}p)$ 
with the system dynamics $\boldsymbol{f}$ as in \eqref{Eqn:GalerkinSystem}.
The PDF at time $t$ can also be written as an integral \citep{PhysRevE.51.74}
\begin{equation}
\label{Eqn:ProbabilityDensity_Solution}
 p(\boldsymbol{a},t) = \hat{\slsP}_t\,p_0(\boldsymbol{a}) =  \int \mathrm d\boldsymbol{a}_0 \,
 \delta\left( \boldsymbol{a}-\boldsymbol\Phi_t\left( \boldsymbol{a}_0 \right)\right)\,p_0(\boldsymbol{a}_0) 
\end{equation}
where $\boldsymbol\Phi_t$ is the flow map associated {with} the Galerkin system \eqref{Eqn:GalerkinSystem}.
Analogous to the Perron-Frobenius operator for a PDF, 
the flow map $\boldsymbol\Phi_t$ 
shifts the state variable $\boldsymbol{a}$
forward in time with
$\boldsymbol{a}(t) = \boldsymbol\Phi_t\,\boldsymbol{a}_0$ 
where $\boldsymbol{a}_0:=\boldsymbol{a}(0)$ is the initial condition.
The kernel $\delta\left( \boldsymbol{a}-\boldsymbol\Phi_t\left( \boldsymbol{a}_0 \right)\right)$ 
of the Perron-Frobenius operator 
can be interpreted as a
conditional probability density for the trajectory 
to be at point $\boldsymbol{a}$
if it had been initially at point $\boldsymbol{a}_0$. 
A finite-rank approximation of the Perron-Frobenious operator 
can be obtained from the Ulam-Galerkin method which is described in 
\S~\ref{Sec:UlamGalerkinMethod}.

In general, one is interested in observables of a system like e.g. 
measurements of the pressure or the velocity field.
The average or the expected value of a scalar observable $g$ at time $t$ is given by
\begin{equation}
\label{Eqn:ObservableDefinition}
 \langle g\rangle(t) 
 = \int\mathrm d\boldsymbol{a} \, g(\boldsymbol{a}) p(\boldsymbol{a},t)
 = \int\int\mathrm d\boldsymbol{a}\,\mathrm d\boldsymbol{a}_0 \> g(\boldsymbol{a}) 
 \delta\left( \boldsymbol{a}-\boldsymbol\Phi_t\left( \boldsymbol{a}_0 
\right)\right) p_0(\boldsymbol{a}_0).
\end{equation}

This expression and the definition \eqref{Eqn:ProbabilityDensity_Solution} 
of $\hat{\slsP}_t$ can be exploited \citep[see][for a proof]{PhysRevE.51.74} 
to define the Koopman operator, denoted by $\hat{\slsK}_t$, which is
adjoint to the Perron-Frobenius operator according to
$$
\langle g\rangle(t) = \langle g,\,\hat{\slsP}_t p_0 \rangle = \langle \hat{\slsK}_t g,\,p_0\rangle
$$
where the inner product is defined by $\langle g,f\rangle:= \int\mathrm d\boldsymbol{a} \,g(\boldsymbol{a})\,f(\boldsymbol{a})$
for two functions $f$ and $g$.
The observable $g$ at time $t$ is then given by
\begin{equation}
 g\left(\boldsymbol{a}(t)\right) 
 = \hat{\slsK}_t g(\boldsymbol{a}_0) 
 = \int \mathrm d\boldsymbol{a}\,  \delta\left( \boldsymbol{a}-\boldsymbol\Phi_t\left( \boldsymbol{a}_0 \right)\right) g(\boldsymbol{a}).
\end{equation}
Thus, the time evolution of the observable is ruled by the Koopman operator. 
The corresponding adjoint Liouville equation reads
\begin{equation}
\frac{\partial}{\partial t} g(\boldsymbol{a})
= \left(\boldsymbol{f}\cdot \nabla_{\boldsymbol{a}}\right)\, g(\boldsymbol{a})
= \hat{\slsL}^\dagger g(\boldsymbol{a}),
\label{Eqn:AdjointLiouvilleEquation}
\end{equation}
where $\hat{\slsL}^\dagger$ is the adjoint Liouville operator. 
In general, the Liouville operator and its adjoint are related by 
$\hat{\slsL}+\hat{\slsL}^\dagger=-\nabla\cdot\boldsymbol{f}$ 
where $\nabla\cdot\boldsymbol{f}$ is not vanishing for dissipative systems.

In recent years, 
the spectral analysis of complex nonlinear flows 
has received increasing attention.
\citet{Rowley2009jfm}
determined 
the eigenfunctions and eigenvalues of the Koopman operator $\hat{\slsK}_t$ 
from only an ensemble of observables
 via the Dynamic Mode Decomposition (DMD) 
 which was introduced in \citet{Schmid2010jfm}.

More recently, \citet{Bagheri_JFM2013} 
studied the conditions under which the DMD algorithm 
approximates the Koopman modes. 
For this purpose, 
he considered the first Hopf bifurcation of the flow past a circular cylinder and 
constructed analytically the Koopman modes. 
This analytical approach is clearly unthinkable 
for a complex dynamical system, 
and new approaches must be 
considered where CROM can play an important role.
Indeed, the Koopman analysis and CROM are strongly connected. 
As discussed in \S~\ref{Sec:UlamGalerkinMethod}, 
CROM can be interpreted as a finite-rank approximation of the 
Perron-Frobenius operator which is adjoint to the Koopman operator.

\section{Visualisation of the cluster topology}
\label{Sec:AppD:Visualisation}
The visualisation of trajectories in the state space can give an intuitive
picture of the underlying dynamics and the state space structure.
It can become a challenging and tedious task if one considers
high-dimensional data like a velocity snapshot ensemble or POD coefficients
and wants to keep certain properties of the data.

In the case of CROM,
the two-dimensional visualisation of the cluster arrangement in the state space
can contribute to a better understanding of the dynamics on the attractor.
A simple method that optimally preserves the centroids' pointwise distances 
in a least-mean-square-error sense employs 
the proper orthogonal decomposition as mentioned in \S~\ref{Sec:CROM}.
Instead of the velocity snapshots we consider here the centroids.

The Matrix $\slsC$ contains the centered centroids $\boldsymbol{c}_k' = \boldsymbol{c}_k - 1/K\,\sum_{k=1}^K \boldsymbol{c}_k$ as columns, i.e.
$\slsC = [\boldsymbol{c}_1' \cdots \boldsymbol{c}_K']$.
The eigenvalues $\nu_k$, $k=1,\ldots,K$, are obtained from a spectral decomposition of the 
covariance matrix $\slsC^T\slsC$,
where the supercscript 'T' denotes the transpose.
They are ordered, $\nu_1 \geq \nu_2 \geq \ldots \geq \nu_k \geq 0$.
The matrix $\slsW_r$ contains the first $r$ eigenvectors of the covariance matrix
in the columns.
Then, a projection of the original centroids $\boldsymbol{c}_k$ onto $\slsW_r$
yields points
\begin{equation}
 \left(\boldsymbol{c}^{r}_k\right)^T = \left(\boldsymbol{c}_k\right)^T \slsW_r, \quad k=1,\ldots,K,
\end{equation}
in $r$ dimensions where the distances between the points $\boldsymbol{c}^r_k$ approximate 
the distances between the centroids $\boldsymbol{c}_k$. 
Analogously, the projection of the snapshots is given by 
\begin{equation}
 \left(\boldsymbol{u}^{m,r}\right)^T = \left(\boldsymbol{u}^{m}\right)^T \slsW_r.
\end{equation}
For visualisation purposes in \S~\ref{Sec:AppA:ExampleBroadbandTurbulenceSimpleStructure}, 
we choose $r=2$.
The POD mode amplitudes associated with the cluster centroids are denoted by $\alpha_k$.

In a more general context, for a given distance matrix according to a (non-Euclidean) distance metric,
multidimensional scaling (MDS) \citep{Mardia1979book,Cox2001book}
aims to find corresponding points in a low-dimensional subspace so that 
the distances between the points are preserved. This is referred to as classical scaling. 
The solution can vary in terms of a translation, a rotation, and reflections.
In the case where the distance is measured via the Euclidean metric, this method coincides with 
the proper orthogonal decomposition, and the mean is at the origin and the axes are the POD eigenvectors \citep{Cox2001book}.

\end{document}